\documentclass[twoside]{bourlarge}
\usepackage{votrenom_h}

\usepackage{amssymb}
\usepackage{amsmath}
\usepackage{graphicx}
\usepackage{bm}

\def\dd{{\rm d}}
\def\etal{{\em et al.}}
\newcommand{\ob}{$\Omega_{\mathrm{b}}$}

\newcommand{\deu}{D}
\newcommand{\tro}{$^3$He}
\newcommand{\qua}{$^4$He}
\newcommand{\six}{$^{6}$Li}
\newcommand{\sep}{$^{7}$Li}

\newcommand{\neu}{$^{9}$Be}
\newcommand{\dix}{$^{10}$B}
\newcommand{\onz}{$^{11}$B}

 \newcommand{\Hconf}{\mathcal{H}}
 \newcommand{\deltanewt}{{\delta^{\mathrm{N}}}}
 \newcommand{\deltaflat}{{\delta^{\mathrm{F}}}}
 \newcommand{\deltacom}{{\delta^{^\mathrm{C}}}}

\begin{document}
\Large

\title{The big-bang theory: construction, evolution and status}

\author{Jean-Philippe {\sc Uzan} \\
Institut d'Astrophysique de Paris\\ UMR 7095 du CNRS,\\
98 bis, bd Arago, 75014 Paris.\\}

\maketitle

\begin{abstract}
Over the past century, rooted in the theory of general relativity, cosmology has developed a very successful physical model of the universe: the {\em big-bang model}. Its construction followed different stages to incorporate nuclear processes, the understanding of the matter present in the universe, a description of the early universe and of the large scale structure. This model has been confronted to a variety of observations that allow one to reconstruct its expansion history, its thermal history and the structuration of matter. Hence, what we refer to as the big-bang model today is radically different from what one may have had in mind a century ago. This construction changed our vision of the universe, both on observable scales and for the universe as a whole. It offers in particular physical models for the origins of the atomic nuclei, of matter and of the large scale structure. This text summarizes the main steps of the construction of the model, linking its main predictions to the observations that back them up. It also discusses its weaknesses, the open questions and problems, among which the need for a dark sector including dark matter and dark energy.
\end{abstract}
%\pacs{Valid PACS appear here}

\section{Introduction}

%-----------------------------------------------------------------
\subsection{From General Relativity to cosmology}

A cosmological model is a mathematical representation of our universe that is based on the laws of nature that have been validated locally in our Solar system and on their extrapolations (see~Refs.~\cite{gfrphilo,cosmomodel,jpu_cup} for a detailed discussion). It thus seats at the crossroad between theoretical physics and astronomy.  Its basic enterprise is thus to use tested physical laws to understand the properties and evolution of our universe and of the matter and the astrophysical objects it contains. 

Cosmology is however peculiar among sciences at least on two foundational aspects. The {\em uniqueness of the universe} limits the standard scientific method of comparing similar objects in order to find regularities and to test for reproductibility; indeed this limitation depends on the question that is asked. In particular, this will tend to blur many discussions on chance and necessity. Its {\em historical dimension} forces us to use abduction\footnote{Abduction is a form of inference which goes from an observation to a theory, ideally looking for the simplest and most likely explanation. In this reasoning, unlike with deduction, the premises do not guarantee the conclusion, so that it can be thought as  ``inference to the best explanation".} together with deduction (and sometime induction) to reconstruct the most probable cosmological scenario\footnote{A property cosmology shares with Darwinian evolution.}. One thus needs to reconstruct the conditions in the primordial universe to fit best what is observed at different epochs, given a set of physical laws. Again the distinction between laws and initial conditions may also be subtle.This means that cosmology also involves, whether we like it or not, some philosophical issues~\cite{gfrverif}. 

In particular, one carefully needs to distinguish {\em physical cosmology} from the {\em Cosmology} that aims to propose a global picture of the universe~\cite{gfrphilo}.  The former has tremendously progressed during the past decades, both from a theory and an observation point of view. Its goal is to relate the predictions of a physical theory of the universe to actual observations. It is thus mostly limited to our {\em observable universe}. The latter is aiming at answering broader questions on the universe as a whole, such as questions on origins or its finiteness but also on the apparent fine-tuning of the laws of nature for complexity to emerge or the universe to host a viable form of life. The boundary between these two approaches is ill-defined and moving, particularly when it comes to recent developments such as inflation or the multiverse debate. They are related to the two notions, the universe, i.e. the ensemble of all what exist, and our observable universe. Both have grown due to the progresses of our theories, that allow us to conceptualize new continents, and of the technologies, that have extended the domain of what we can observe and test.

Indeed the physical cosmology sets very strong passive constraints on Cosmology. It is then important to evaluate to which extent our observable universe is representative of the universe as a whole, a question whose answer depends drastically of what is meant by ``universe as a whole''.  Both approaches are legitimate and the general public is mostly interested by the second. This is why we have the moral duty to state to which of those approaches we are referring to when we talk about cosmology.

While a topic of interest for many centuries -- since any civilization needs to be structured by an anthropology and a cosmology, through mythology or science -- we can safely declare~\cite{jeref} that scientific cosmology was born with Albert Einstein's general relativity a century ago. His theory of gravitation made the geometry of spacetime dynamical physical fields, $g_{\mu\nu}$, that need to be determined by solving equations known as Einstein field equations,
\begin{equation}\label{eq.efe}
G_{\mu\nu}[g_{\alpha\beta}]=\frac{8\pi G}{c^4}T_{\mu\nu},
\end{equation}
where the stress-energy tensor $T_{\mu\nu}$ charaterizes the matter distribution. From this point of view, the cosmological question can be phrased as {\em What are the spacetime geometries and topologies that correspond to our universe?}\\

This already sets limitations on how well we can answer this question. First, from a pure mathematical perspective, the Einstein equations~(\ref{eq.efe}) cannot be solved in their full generality. They represent 10 coupled and non-linear partial differential equations for 10 functions of 4 variables and there is, at least for now, no general procedure to solve such a system. This concerns only the structure of the left-hand-side of Eq.~(\ref{eq.efe}). This explains the huge mathematical literature on the existence and stability of the solution of these equations.

%%----------------------------------------------------
\begin{figure}[h!]
\centering
\includegraphics[width=.9\columnwidth]{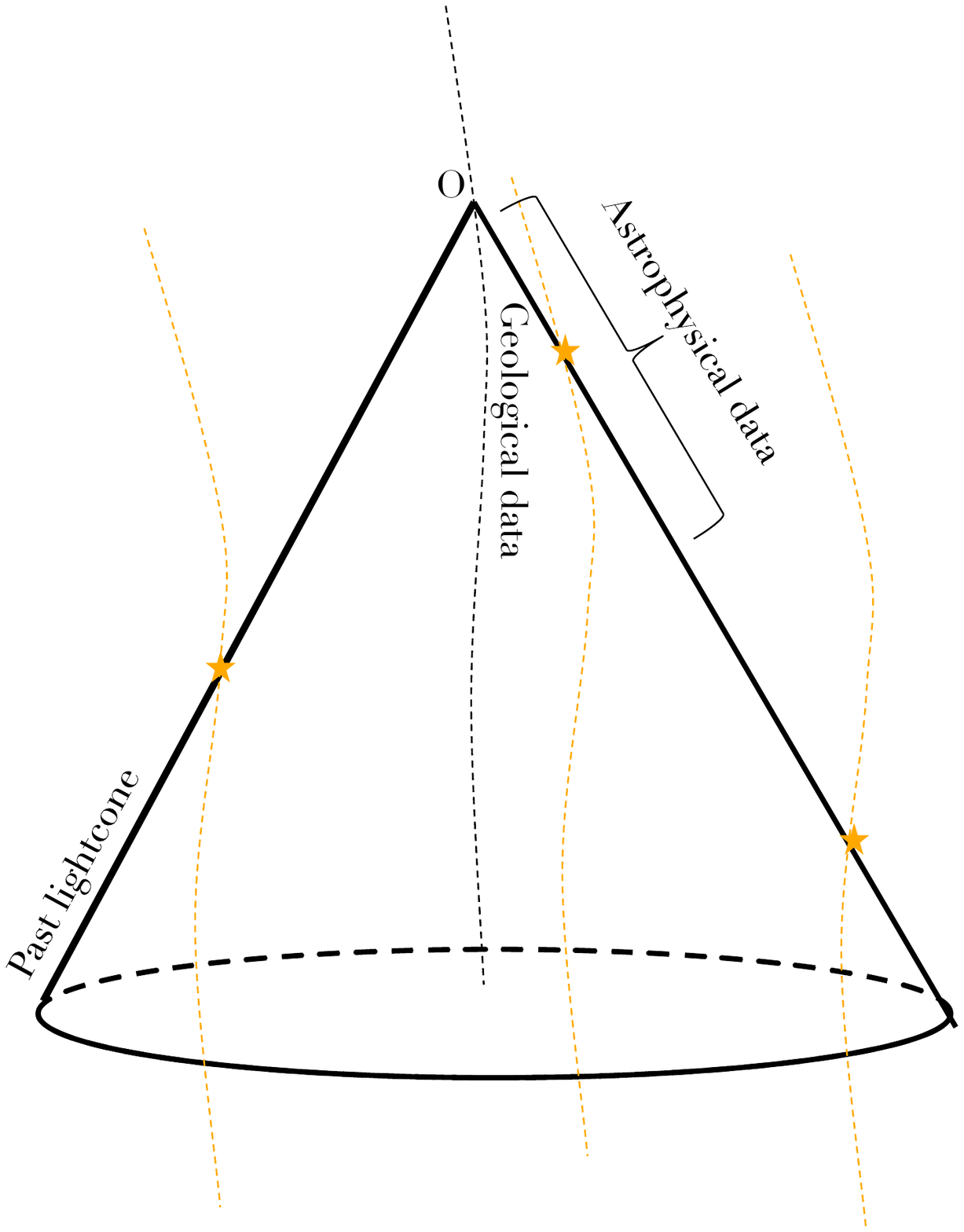}
\vskip-.5cm
 \caption{\footnotesize Astrophysical data are mostly located on our past lightcone and fade away with distance from us so that we have access to a portion of a 3-dimensional null hypersurface -- an object can be observed only when its worldline (dashed lines) intersects our past lightcone. Some geological data can be extracted on our Solar system neighborhood. It is important to keep in mind that the interpretation of the observations is not independent of the spacetime structure, e.g. assigning distances. We are thus looking for compatibility between a universe model and these observations.} 
\label{fig0}
\end{figure}
%%----------------------------------------------------

Another limitation arises from the source term in its right-hand-side. In order to solve these equations, one needs to have a good description of the matter content in the universe. Even with perfect data, the fact that (almost)\footnote{We also collect information in the Solar neighborhood and of high energy cosmic rays.} all the information we can extract from the universe is under the form of electromagnetic signal implies that observations are located on our past lightcone, that is on a 3-dimensional null hypersurface (see Fig.~\ref{fig0}). It can be demonstrated (see e.g. Ref.~\cite{ellisphysrep} and Ref.~\cite{dunsbyetal} for a concrete of 2 different cosmological spacetimes which enjoy the same lightcone observations) that the 4-dimensional metric cannot be reconstructed from this information alone. Indeed a further limitation arises from the fact that there is no such thing as perfect observations. Galaxy catalogs are limited in magnitude or redshift, evolution effects have to be taken into account, some components of matter (such a cold diffuse gas or dark matter) cannot be observed electromagnetically. We do not observe the whole matter distribution but rather classes of particular objects (stars, galaxies,...) and we need to deal with the variations in the properties of these individual objects and evolution effects. A difficult task is to quantify how the intrinsic properties of these objects influence our inference of the properties of the universe. 

As a consequence, the cosmological question is replaced by a more modest question, the one of the construction of a good cosmological model, that can be phrased as {\em Can we determine metrics that are good approximations of our universe?} This means that we need to find a guide (such as symmetries) to exhibit some simplified solutions for the metrics that offer a good description on the universe on large scales. From a mathematical point of view, many such solutions are known~\cite{solexact}. Indeed our actual universe has no symmetry at all and these solutions have to be thought as a description of the universe smoothed on ``some'' scale, and we should not expect them to describe our spacetime from stellar scales to the Hubble scale.\\

The first relativistic cosmological model~\cite{Ei17}  was constructed by Einstein in 1917 and it can be considered as the birthdate of modern cosmology. As we shall see, most models rely on some hypotheses that are difficult to test on the size of the observable universe. The contemporary cosmological model is often referred to as the {\em big-bang} model and Section~\ref{section2} describes its construction and structure. It is now complemented by a description of the early universe that we detail in Section~\ref{section3}, that offers both a model for the origin of the large scale structure but also a new picture of the universe as a whole. As any scientific model, it has to be compared to observation and a large activity in cosmology is devoted to understanding of how the universe looks to an observer inside this universe. Section~\ref{section3} also sketches this theoretical activity and summarizes the observational landscape, unfortunately too wide to be described completely. This construction, while in agreement with all observation, relies on very simple, some may say crude, assumptions, which lead to the question {\em why does it works so well?} that we shall address in Section~\ref{section5}.

To summarize the main methodological peculiarities of cosmology, we have to keep in mind, (1) the uniqueness of the universe, (2) its historical dimension (hence the necessity of abduction), (3) the fact that we only reconstruct the most probable history (and then have to quantify its credence) that is backed up by the consistency of different facts (to be contrasted with a explanation designed to explain an isolated phenomena), (4) the need for a large extrapolation of the laws of nature, and (5) the existence of a (unspecified) smoothing scale. 

Indeed, as for any model in physics, our model cannot explain its own ontology and opens limiting questions, such as its origin. The fact that we cannot answer them within the model is indeed no flaw and cannot be taken as an argument against the model. They just trivially show that it needs to be extended.

%-----------------------------------------------------------------
\subsection{Hypotheses}

The construction of a cosmological model depends on our knowledge of microphysics as well as on {\em a priori} hypotheses on the geometry of the spacetime describing our universe.

{\em Theoretical physics} describes the fundamental components of nature and their interactions. These laws can be probed locally by experiments. They need to be extrapolated to construct a cosmological model. On the one hand, any new idea or discovery will naturally call for an extension of the cosmological model (e.g. introducing massive neutrinos in cosmology is now mandatory). On the other hand, cosmology can help constraining extrapolations of the established reference theories to regimes that cannot be accessed locally. As explained above, the knowledge of the laws of microphysics is not sufficient to construct a physical representation of the universe. These are reasons for the need of extra-hypotheses, that we call cosmological hypotheses.

{\em Astronomy} confronts us with phenomena that we have to understand and explain consistently. This often requires the introduction of hypotheses beyond those of the physical theories in order to ``{\em save the phenomena}''~\cite{duhem}, as is actually the case with the dark sector of our cosmological model~\cite{jpu-general relativityG}. Needless to remind that even if a cosmological model is in agreement with all observations, whatever their accuracy, it does not prove that it is the ``correct'' model of the universe, in the sense that it is the correct cosmological extrapolation and solution of the local physical laws. 

When confronted with an inconsistency in the model, one can either invoke the need for new physics, i.e. a modification of the laws of physics we have extrapolated in a regime outside of the domain of validity that has been established so far (e.g. large cosmological distance, low curvature regimes etc.), or have a more conservative attitude concerning fundamental physics and modify the cosmological hypotheses.\\

Let us start by reminding that the construction of any cosmological model relies on 4 main hypotheses (see Ref.~\cite{jpu_cup} for a detailed description),
\begin{itemize}
  \item[(H1)] a theory of gravity,
  \item[(H2)] a description of the matter contained in the universe and their non-gravitational interactions,
  \item[(H3)] symmetry hypothesis,
  \item[(H4)] a hypothesis on the global structure, i.e. the topology, of the universe.
\end{itemize}
These hypotheses are indeed not on the same footing since H1 and H2 refer to the local (fundamental) physical theories. These two hypotheses are however not sufficient to solve the field equations and we must make an assumption on the symmetries (H3) of the solutions describing our universe on large scales while H4 is an assumption on some global properties of these cosmological solutions, with same local geometry.

\subsubsection{Gravity}

Our reference cosmological model first assumes that gravity is well-described by general relativity (H1). This theory is well-tested on many scales and we have no reason to doubt it today~\cite{will,dubook}. It follows that we shall assume that the gravitational sector is described by the Einstein-Hilbert action
\begin{equation}
 S=\frac{1}{16\pi G}\int(R-2\Lambda) \sqrt{-g}\dd^4x,
\end{equation}
where a cosmological constant $\Lambda$ has been included.

Indeed, we cannot exclude that it does not properly describe gravity on large scales and there exists a large variety of theories (e.g. scalar-tensor theories, massive gravity, etc.) that can significantly differ from general relativity in the early universe while being compatible with its predictions today. This means that we will have to design tests of general relativity on astrophysical scales~\cite{jpu-testRGcosmo,ub2001}. Indeed, from a theoretical point of view, we know that general relativity needs to be extended to a theory of quantum gravity. It is however difficult, on very general grounds, to determine if that would imply that there exist an ``intermediate'' theory of gravity that differs from general relativity on energy and distance scales that are relevant for the cosmological model. Indeed, there exist classes of theories, such as the scalar-tensor theories of gravity, that can be dynamically attracted ~\cite{dn} toward general cosmology during the cosmic evolution. Hence all the cosmological test of general relativity complement those on Solar system scales.

\subsubsection{Non-gravitational sector}

Einstein equivalence principle, as the heart of general relativity, also implies that the laws of non-gravitational physics validated locally can be extrapolated. In particular the constants of nature shall remain constant, a prediction that can also be tested on astrophysical scales~\cite{uzancst1,uzancst2}. Our cosmological model assumes (H2) that the matter and non-gravitational interactions are described by the standard model of particle physics. As will be discussed later, but this is no breaking news, modern cosmology requires the universe to contain some dark matter (DM) and a non-vanishing cosmological constant ($\Lambda$). Their existence is inferred from cosmological observations assuming the validity of general relativity (e.g. flat rotation curves, large scale structure, dynamics of galaxy clusters for dark matter, accelerated cosmic expansion for the cosmological constant; see chapters 7 and 12 of Ref.~\cite{jpu-book}). Dark matter sets many questions on the standard model of particle physics and its possible extensions since the physical nature of this new field has to be determined and integrated consistently in the model. The cosmological constant problem is argued to be a sign of a multiverse, indeed a very controversial statement. If solved then one needs to infer some {\em dark energy} to be consistently included.

We thus assume that the action of the non-gravitational sector is of the form
\begin{equation}
 S=\int{\cal L}(\psi,g_{\mu\nu})\sqrt{-g}\dd^4x,
\end{equation}
in which all the matter fields, $\psi$, are universally coupled to the spacetime metric. 

Note that H2 also involves an extra-assumption since what will be required by the Einstein equations is the effective stress-energy tensor averaged on cosmological scales. It thus implicitly refers to a, usually not explicited, averaging procedure~\cite{averaging}. On large scale, matter is thus described by a mixture of pressureless matter ($P=0$) and radiation ($P=\rho/3$).

\subsubsection{Copernican principle}

Let us now turn the cosmological hypotheses. In order to simplify the expected form of our world model, one first takes into account that observations, such as the cosmic microwave background or the distribution of galaxies, look isotropic around us. It follows that we may expect the metric to enjoy a local rotational symmetry and thus to be of the form
\begin{equation}
 \dd s^2 = -A^2(t,r)\dd t^2+B^2(t,r)\left[\dd r^2 + R^2(t,r)\dd\Omega^2\right].
\end{equation}
We are left with two possibilities. Either our universe is spherically symmetric and we are located close to its center or it has a higher symmetry and is also spatially homogeneous. Since we observe the universe from a single event, this cannot be decided observationally. It is thus postulated that we do not stand in a particular place of the universe, or equivalently that we can consider ourselves as a typical observer. This {\em Copernican principle} has strong implications since it implies that the universe is, at least on the size of the observable universe, spatially homogeneous and isotropic. Its validity can be tested~\cite{testCP} but no such test did actually exist before 2008. It is often distinguished from the {\em cosmological principle} that states that the universe is spatially homogeneous and isotropic. This latter statement makes an assumption on the universe on scales that cannot be observed~\cite{gfrphilo}. From a technical point of view, it can be shown that it implies that the metric of the universe reduces to the Friedmann-Lema\^{\i}tre form (see e.g. chapter~3 of Ref.~\cite{jpu-book})
\begin{equation}\label{eq.fl}
 \dd s^2 = -\dd t^2+a^2(t)\left[\dd \chi^2 + f_K^2(\chi)\dd\Omega^2\right]\equiv g_{\mu\nu}\dd x^\mu\dd x^\nu,
\end{equation}
where the scale factor $a$ is a function of the cosmic time $t$. Because of the spatial homogeneity and isotropy, there exists a preferred slicing $\Sigma_t$ of the spacetime that allows one to define this notion of cosmic time, i.e. $\Sigma_t$ are constant $t$ hypersurfaces. One can introduce the family of observers with worldlines orthogonal to $\Sigma_t$ and actually show that they are following comoving geodesics. In terms of the tangent vector to their worldline, $u^\mu=\delta^\mu_0$, the metric~(\ref{eq.fl}) takes the form
\begin{equation}\label{eq.fl2}
 \dd s^2 = -(u_\mu \dd x^\mu)^2+(g_{\mu\nu}+u_\mu u_\nu)\dd x^\mu\dd x^\nu,
\end{equation}
which clearly shows that the cosmic time $t$ is the proper time measured by these fundamental observers. As a second consequence, this symmetry implies that the most general form of the stress-energy tensor is the one of a perfect fluid
\begin{equation}\label{eq.tmunu}
 T_{\mu\nu} =\rho u_\mu u_\nu + P (g_{\mu\nu}+u_\mu u_\nu),
\end{equation}
with $\rho$ and $P$, the energy density and isotropic pressure measured by the fundamental observers.

Now the 3-dimensional spatial hypersurfaces $\Sigma_t$ are homogeneous and isotropic, which means that they are maximally symmetric. Their geometry can thus be only the one of either a locally spherical, Euclidean or hyperbolic space, depending on the sign of $K$. The form of $f_K(\chi)$ is thus given by
\begin{eqnarray}
f_K(\chi) =\left\lbrace\begin{matrix}
                  K^{-1/2}\sin\left(\sqrt{K}\chi\right) &K>0\\
                 \chi  &K=0 \\
                   (-K)^{-1/2}\sinh\left(\sqrt{-K}\chi\right) &K<0
                   \end{matrix}\right.,
\end{eqnarray}
$\chi$ being the comoving radial coordinate. The causal structure of this class of spacetimes is discussed in details in Ref.~\cite{eu_cras}.

\subsubsection{Topology}

The Copernican principle allowed us to fix the general form of the metric. Still, a freedom remains on the topology of the spatial sections~\cite{toporevue}. It has to be compatible with the geometry. 

In the case of a multiply connected universe, one can visualize space as the quotient $X / \Gamma$ of a simply connected space $X$ (which is just he {Euclidean space} {$E^3$}, the {hypersphere} {$S^3$} or the {3-hyperboloid} {$H^3$},  $\Gamma$ being a discrete and fixed point free symmetry group of $X$.  This holonomy group $\Gamma$ changes the boundary conditions on all the functions defined on the spatial sections, which subsequently need to be $\Gamma$-periodic. Hence, the topology leaves the local physics unchanged while modifying the boundary conditions on all the fields. Given a field $\phi({\bm x},t)$ living on $X$, one can construct a field $\overline\phi({\bm x},t)$ leaving on $X / \Gamma$ by projection as
\begin{equation}
 \overline\phi({\bm x},t)=\frac{1}{\vert\Gamma\vert}\sum_{g\in\Gamma}\phi(g({\bm x}),t)
\end{equation}
since then, for all $g$, $\bar\phi(g({\bm x}),t)=\bar\phi({\bm x},t)$. It follows that any $\Gamma$-periodic function of $L^2(X)$ can be identified to a function of $L^2(X/\Gamma)$.

In the standard model, it is assumed that the spatial sections are simply connected. The observational signature of a spatial topology decreases when the size of the universe becomes larger than the Hubble radius~\cite{topogen}. Its effects on the CMB anisotropy have been extensively studied~\cite{topocmb} to conclude that a space with typical size larger than 20\% of the Hubble radius today cannot be observationally distinguished from a infinite space~\cite{topolimite}. From a theoretical point of view, inflation predicts that the universe is expected to be extremely larger than the Hubble radius (see below).

%%%%%%%%%%%%%%%%%%%%%%%%%%%%%%%%%%%%%
\section{The construction of the hot big-bang model}\label{section2}

The name of the theory, {\it big-bang}, was coined by one of its opponent, Fred Hoyle, during a BBC broadcast on March $28^{\rm th}$ 1948 for one was referred to the model of dynamical evolution before then. It became very media and one has to be aware that there exist many versions of this big-bang model and that it has tremendously evolved during the past century.

This section summarizes the main evolutions of the formulation of this model that is today in agreement with all observations, at the expense of introducing 6 free parameters. This version has now been adopted as the standard model for cosmology, used in the analysis and interpretation of all observational data.

\subsection{General overview}

The first era of relativistic cosmology, started in 1917 with the seminal paper by Einstein~\cite{Ei17} in which he constructed, at the expense of the introduction of a cosmological constant, a static solution of its equations in which space enjoys the topology of a 3-sphere. 

This paved the way to the derivations of exact solutions of the Einstein equations that offer possible world-models. Alexandr Friedmann and independently Georges Lema\^{\i}tre~\cite{lemaitre27} developed the first dynamical models~\cite{fried22}, hence discovering the cosmic expansion as a prediction of the equations of general relativity. 

An important step was provided by Lema\^{\i}tre who connected the theoretical prediction of an expanding universe to observation by linking it to the redshifts of electromagnetic spectra, and thus of observed galaxies. This was later confirmed by the observations by Edwin Hubble~\cite{huublelaw} whose {\em Hubble law}, relating the recession velocity of a galaxy to its distance from us, confirms the cosmological expansion. The law of expansion derives from the Einstein equations and thus relates the cosmic expansion rate, $H$, to the matter content of the universe, offering the possibility to ``weight the universe''. This solution of a spatially homogeneous and isotropic expanding spacetime, referred to as Friedmann-Lema\^{\i}tre (FL) universe, serves as the reference background spacetime for most later developments of cosmology. It relies on the so-called Copernican principle stating that we do not seat in a particular place in the universe, and introduced by Einstein~\cite{Ei17}.\\

In a second era, starting in 1948, the properties of atomic and nuclear processes in an expanding universe were investigated (see e.g. Ref.~\cite{tolman} for an early texbook). This allowed Ralph Alpher, Hans Bethe and George Gamow~\cite{abg48} to predict the existence and estimate the temperature of a cosmic microwave background (CMB) radiation and understand the synthesis of light nuclei, the big-bang nucleosynthesis (BBN), in the early universe. Both have led to theoretical developments compared successfully to observation. It was understood that the universe is filled with a thermal bath with a black-body spectrum, the temperature of which decreases with the expansion of the universe. The universe cools down and has a thermal history, and more important it was concluded that it emerges from a hot and dense phase at thermal equilibrium (see e.g. Ref~\cite{jpu-book} for the details).  This model has however several problems, such as the fact that the universe is spatially extremely close to Euclidean ({\em flatness problem}), the fact that it has an {\em initial spacelike singularity} (known as big-bang) and the fact that thermal equilibrium, homogeneity and isotropy are set as initial conditions and not explained ({\em horizon problem}). It is also too idealized since it describes no structure, i.e. does not account for the inhomogeneities of the matter, which is obviously distributed in galaxies, clusters and voids. The resolution of the naturalness of the initial conditions was solved by the postulate~\cite{guth} of the existence of a primordial accelerated expansion phase, called {\em inflation}.\\

The third and fourth periods were triggered by an analysis of the growth of the density inhomogeneities by Lifshitz~\cite{landau}, opening the understanding of the evolution of the large scale structure of the universe, that is of the distribution of the galaxies in cluster, filaments and voids. Technically, it opens the way to the theory of cosmological perturbations~\cite{pertcosmo,pertbardeen,pertstewart} in which one considers the FL spacetime as a background spacetime the geometry and matter content of which are perturbed. The evolution of these perturbations can be derived from the Einstein equations. For the mechanism studied by Evgeny Lifshitz to be efficient, one needed initial density fluctuations large enough so that their growth in a expanding universe could lead to non-linear structure at least today. In particular, they cannot be thermal fluctuations. This motivated the question of the understanding of the origin and nature (amplitude, statistical distribution) of the initial density fluctuations, which turned out to be the second success of the inflationary theory which can be considered as the onset as the third era, the one of {\em primordial cosmology}. From a theory point of view, the origin of the density fluctuation lies into the quantum properties of matter~\cite{chibi}. From an observational point of view, the predictions of inflation can be related to the distribution of the large scale structure of the universe, in particular in the anisotropy of the temperature of the cosmic microwave background~\cite{cmbth}. This makes  the study of inflation an extremely interesting field since it requires to deal with both general relativity and quantum mechanics and has some observational imprints. It could thus be a window to a better understanding of quantum gravity.\\

The observational developments and the progresses in the theoretical of the understanding of the growth of the large scale structure led to the conclusion that
\begin{itemize}
\item there may exist a fairly substantial amount of non relativistic dark matter, or cold dark matter (hence the acronym CDM)
\item there shall exist a non-vanishing cosmological constant ($\Lambda$). 
\end{itemize}
This led to the formulation of the $\Lambda$CDM model~\cite{LCDM}  by Jeremy Ostriker and Paul Steinhardt in 1995. The community was reluctant to adopt this model until the results of the analysis of the Hubble diagram of type~Ia supernovae in 1999~\cite{sn1a}. This $\Lambda$CDM model is in agreement with all the existing observations of the large survey (galaxy catalogs, CMB, weak lensing, Hubble diagram etc.) and its parameters are measured with increasing accuracy. This has opened the era of {\em observational cosmology} with the open question of the physical nature of the dark sector. 

This is often advertised as {\em precision cosmology}, mostly because of the increase of the quality of the observations, which allow one to derive sharp constraints on the cosmological parameters. One has however to be aware, that these parameters are defined within a very specific model and require many theoretical developments (and approximations) to compare the predictions of the model to the data. Both set a limit a on the accuracy of the interpretation of the data; see e.g. Refs.~\cite{fleury1,fleury2} for an example of the influence of the small scale structure of the universe on the accuracy of the inference of the cosmological parameters. And indeed, measuring these parameters with a higher accuracy often does not shed more light on their physical nature.\\

\begin{figure}[htb!]
\centering
\includegraphics[width=.9\columnwidth]{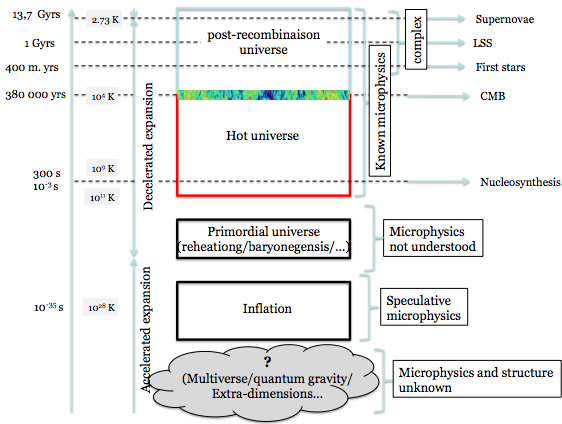}
 \caption{\footnotesize The standard history of our universe. The local universe provides observations on phenomena from big-bang nucleosynthesis to today spanning a range between $10^{-3}$~s to $13.7$~Gyrs. One major transition is the equality which separates the universe in two era: a matter dominated era during which large scale structure can grow and a radiation dominated era during which the radiation pressure plays a central role, in particular in allowing for acoustic waves. Equality is followed by recombination, which can be observed through the CMB anisotropies. For temperatures larger that $10^{11}$~K, the microphysics is less understood and more speculative. Many phenomena such as baryogenesis and reheating still need to be understood in better details. The whole history of our universe appears as a parenthesis of decelerated expansion, during which complex structures can form, in between two periods of accelerated expansion, which do not allow for this complex structures to either appear or even survive. From Ref.~\cite{cosmomodel}.} 
\label{fig1}
\end{figure}

The standard history of our universe, according to this model, is summarized in Fig.~\ref{fig1}. Interestingly, the evolution of the universe and of its structures spans a period ranging from about 1 second after the big-bang to today. This is made possible by the fact that (1) the relevant microphysics is well-known in the range of energies reached by the thermal bath during that period ($<100$~MeV typically) so that it involves no speculative physics, and (2) most of the observables can be described by a linear perturbation theory, which technically simplifies the analysis. 

This description is in agreement with all observations performed so far (big-bang nucleosynthesis abundances, cosmic microwave background temperature and polarisation anisotropies, distribution of galaxies and galaxy clusters given by large catalogs and weak lensing observations, supernovae data and their implication for the Hubble diagram). 

This short summary shows that today inflation is a cornerstone of the standard cosmological model and emphasizes its roles in the development and architecture of the model,
\begin{enumerate}
\item it was postulated in order to explain the required fine-tuning of the initial conditions of the hot big-bang model,
\item it provides a mechanism for the origin of the large scale structure,
\item it gives a new and unexpected vision of the universe on large scale,
\item it connects, in principle~\cite{uee}, cosmology to high energy physics.
\end{enumerate}
This construction is the endpoint of about one century of theoretical and observational developments, that we will now detail.

%%----------------------------------------------------
%%----------------------------------------------------

\subsection{Relativistic cosmology}

\subsubsection{Einstein static universe (1917)}

The Einstein static universe is a static homogeneous and isotropic universe with compact spatial sections, hence enjoying the topology of a 3-sphere. It is thus characterized by 3 quantities: the radius of the 3-sphere, the matter density and the cosmological constant, that is mandatory to ensure staticity. They are related by
\begin{equation}
 \Lambda=\frac{1}{R^2}=4\pi G\rho.
\end{equation}
so that the volume of the universe is $2\pi^2R^3$.

\subsubsection{de Sitter universe (1917)}

The same year, Wilhelm de Sitter constructed a solution that is not flat despite the fact that it contains no matter, but at the expense of having a cosmological constant. This solution was important in the discussion related to the Machs principle and today it plays an important role in inflation.

\begin{figure}[t!]
\begin{center}
\resizebox{11cm}{!}{\includegraphics[clip=true]{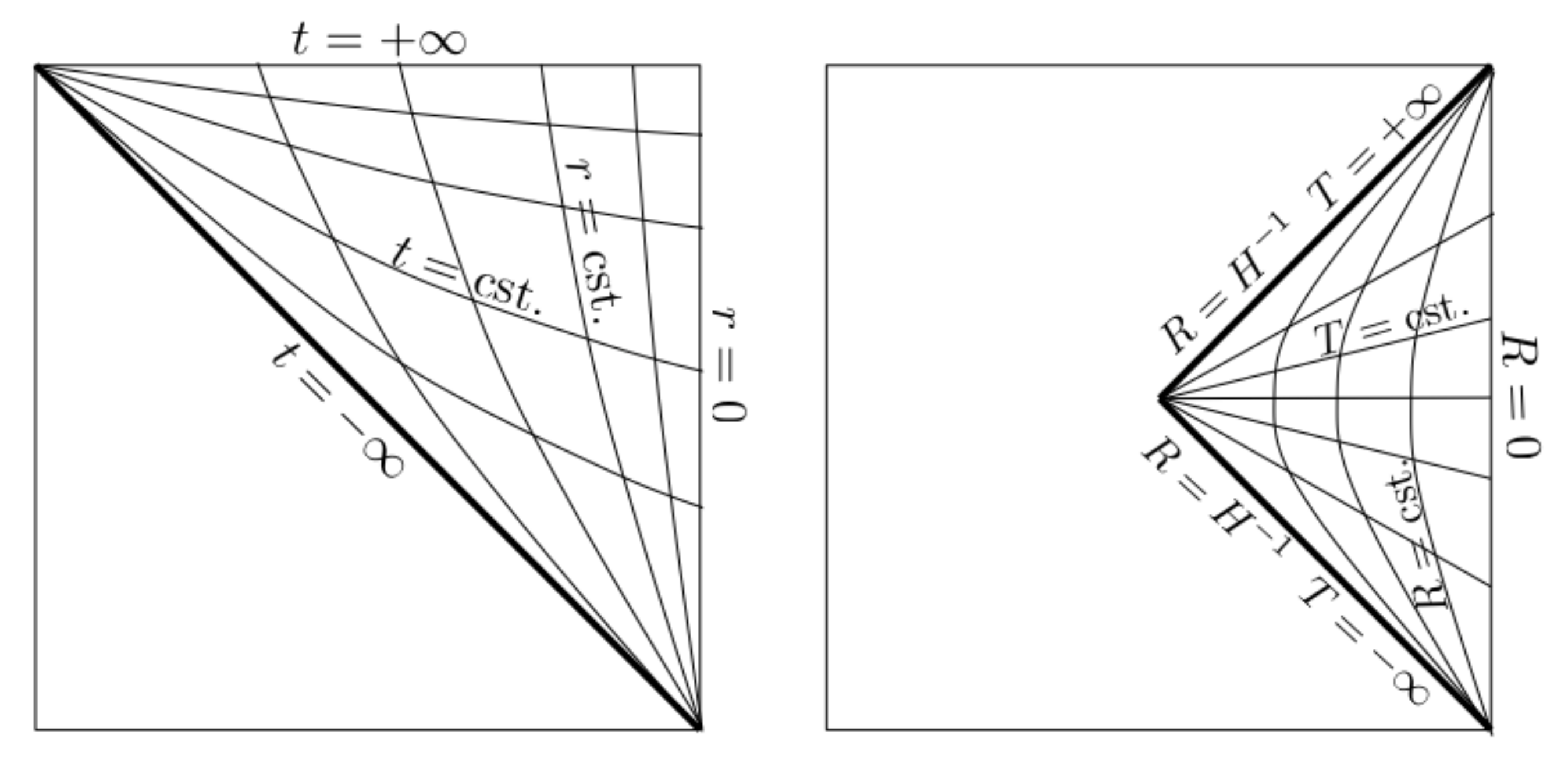}}
\end{center}
 \caption{\footnotesize Penrose diagrams of de Sitter space in the flat (left) and static (right) slicings that each cover only part of the whole de Sitter space, and that are both geodesically incomplete. From Ref.~\cite{eu_cras}.}
  \label{fig4b}
\end{figure}

The de Sitter spacetime is a maximally symmetric spacetime and has the structure of a 4-dimensional hyperboloid. It enjoys many slicings, only one of them being geodesically complete. These different representations, corresponding to different choices of the family of fundamental observers,  are  \cite{eu_cras,HawEll73}:
\begin{enumerate}
 \item the spherical slicing in which the metric has a FL form~(\ref{eq.fl}) with $K=+1$ spatial sections and scale factor $a\propto\cosh Ht$ with $H=\sqrt{\Lambda/3}$ constant. It is the only geodesically complete representation;
 \item the flat slicing in which the metric has a FL form~(\ref{eq.fl})  with Euclidean spatial section, in which case $a\propto \exp Ht$;
 \item the hyperbolic slicing in which the metric has a FL form~(\ref{eq.fl})  with $K=-1$ spatial sections and scale factor $a\propto\sinh Ht$;
 \item the static slicing in which the metric takes the form $ds^2=-(1-H^2R^2)dT^2+dR^2/(1-H^2R^2)-r^2d\Omega^2$.
\end{enumerate}
In each of the last three cases, the coordinate patch used covers only part of the de Sitter hyperboloid. Fig.~\ref{fig4b} gives their Penrose diagrams to illustrate their causal structure (see Refs.~\cite{dubook,eu_cras,HawEll73} for the interpretation of these diagrams).

\subsubsection{Dynamical models (1922-...)}

\begin{figure}[htb!]
\centering
\includegraphics[width=.6\columnwidth]{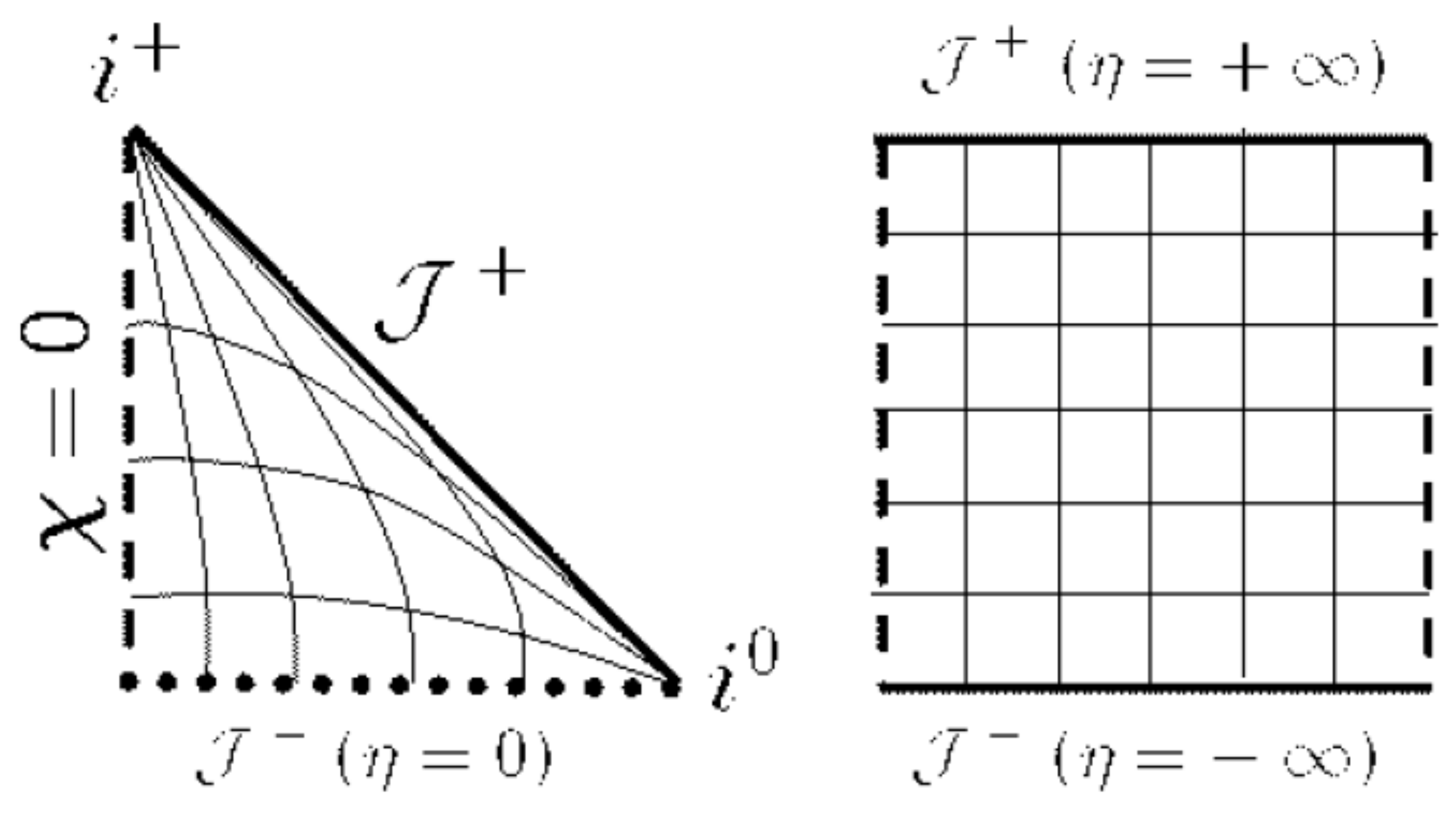}
 \caption{\footnotesize Conformal diagram of the Friedmann-Lema\^{\i}tre spacetimes with Euclidean spatial sections with $\Lambda= 0$ and $P > $0 (left) and de Sitter space in the spherical slicing in which it is geodesically complete (right) -- dashed line corresponds to $\chi=0$ and $\chi=\pi$; From Ref.~\cite{jpu-book}.} 
\label{fig-dyn2}
\end{figure}

\begin{figure}[htb!]
\centering
\includegraphics[width=.8\columnwidth]{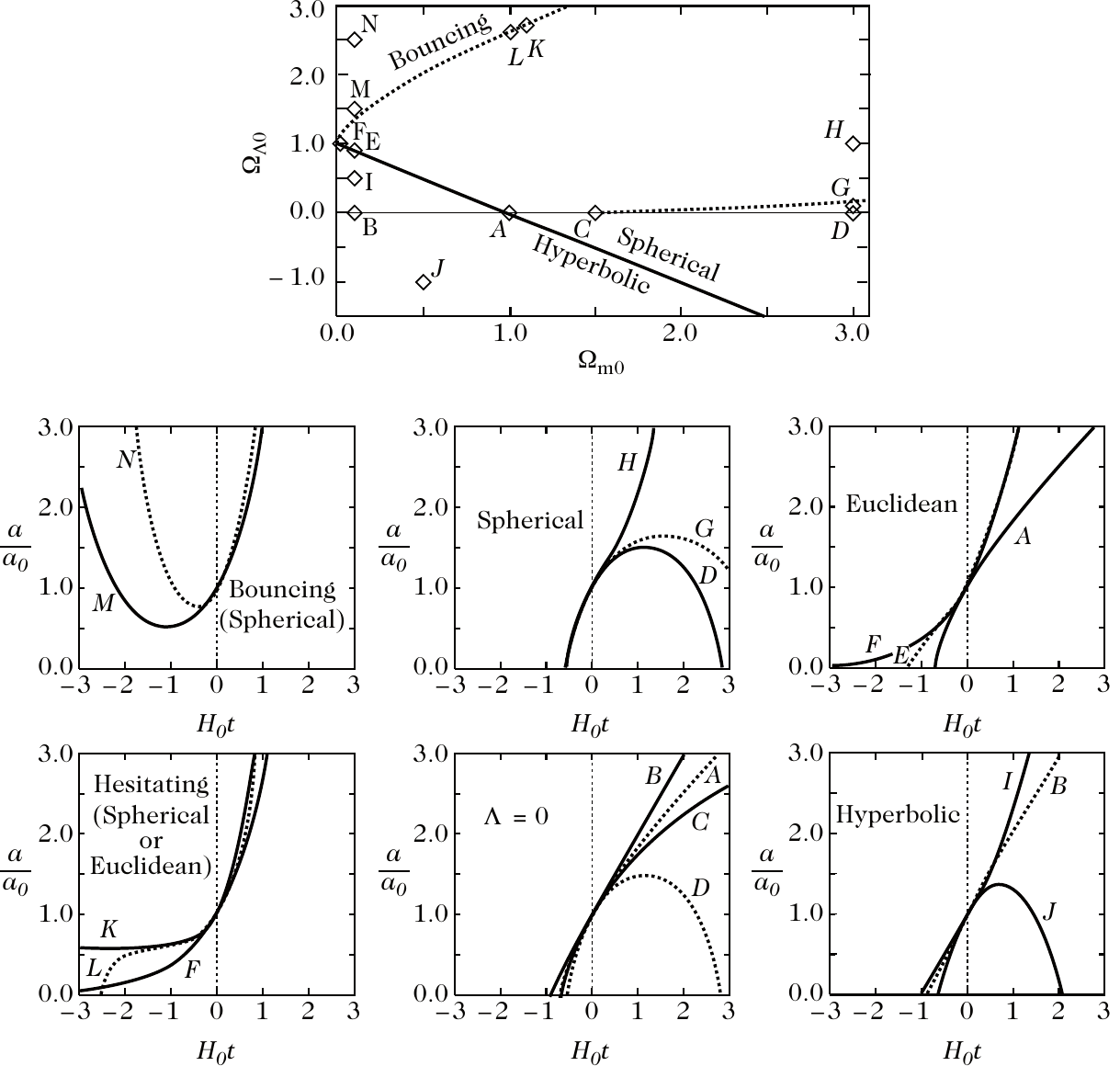}
 \caption{\footnotesize Depending on the values of the cosmological parameters (upper panel), the scale
factor can have very different evolutions. It includes bouncing solutions with no big-bang (i.e. no initial singularity), hesitating universes (with a limiting case in which the universe is asymptotically initially Einstein static), collapsing universes. The expansion can be accelerating or decelerating. From Ref.~\cite{jpu-book}.} 
\label{fig-dyn}
\end{figure}

Starting from the previous assumptions, the spacetime geometry is described by the Friedmann-Lema\^{\i}tre metric~(\ref{eq.fl}). The Einstein equations with the stress-energy tensor~(\ref{eq.tmunu}) reduce to the Friedmann equations
\begin{eqnarray}
 H^2&=&\frac{8\pi G}{3}\rho -\frac{K}{a^2}+\frac{\Lambda}{3},\label{eq.fried1}\\
 \frac{\ddot a}{a}&=& -\frac{4\pi G}{3}(\rho+3P)+\frac{\Lambda}{3},\label{eq.fried2}
\end{eqnarray}
together with the conservation equation ($\nabla_\mu T^{\mu\nu}=0$)
\begin{equation}\label{eq.cons}
\dot\rho+3H(\rho+P)=0.
\end{equation}
This gives 2 independent equations for 3 variables ($a,\rho,P$) that requires the choice of an equation of state
\begin{equation}\label{eq.state}
 P=w\rho
\end{equation}
to be integrated. It is convenient to use the conformal time defined by $\dd t =a(\eta)\dd\eta$ and the normalized density parameters 
\begin{equation}
\Omega_i=8\pi G\rho_i/{3H_0^2},\qquad
\Omega_\Lambda=\Lambda/3H_0^2,
\qquad\Omega_K=-K/a_0^2H_0^2,
\end{equation}
that satisfy, from Eq.~(\ref{eq.fried1}), $\sum_i\Omega_i+\Omega_\Lambda+\Omega_K=1$, so that the Friedmann equation takes the form
\begin{equation}
 \frac{H^2}{H_0^2}= \sum_i\Omega_i (1+z)^{3(1+w_i)}+\Omega_K(1+z)^2+\Omega_\Lambda
\end{equation}
where the redshift $z$ has been defined as $1+z=a_0/a$. The Penrose diagram a FL spacetime with $\Lambda=0$ is presented in Fig.~\ref{fig-dyn2} and the solutions of the Friedmann equations for different sets of cosmological parmaters are depicted on Fig.~\ref{fig-dyn}.

%%----------------------------------------------------
%%----------------------------------------------------

It is obvious from Eqs.~(\ref{eq.fried1}-\ref{eq.fried2}) that the Einstein static solution and de Sitter solution are particular cases of this more general class of solutions. The general cosmic expansion history can be determined from these equations and are depicted in Fig.~\ref{fig-dyn}. A first property of the dynamics of these models can be obtained easily and without solving the Friedmann equations, by performing a Taylor expansion of $a(t)$ round $t=t_0$ (today). It gives $a(t)=a_0[1+H_0(t-t_0)-\frac12q_0H_0^2(t-t_0)^2+\ldots]$ where
\begin{equation}
 H_0= \left.\frac{\dot a}{a}\right\vert_{t=t_0},\qquad
 q_0= -\left.\frac{\ddot a}{aH^2}\right\vert_{t=t_0}.
\end{equation}
From the Friedmann equations, one deduces that $q_0=\frac12\sum(1+3w_i)\Omega_{i 0}-\Omega_{\Lambda0}$ so that the cosmic expansion cannot be accelerated unless there is a cosmological constant.

%%----------------------------------------------------
%%----------------------------------------------------

In that description, the model can already be compared to some observations. 
\begin{enumerate}
\item{\em Hubble law}. The first prediction of the model concerns the recession velocity of distant galaxies. Two galaxies with comoving coordinates $0$ and ${\bm x}$ have a physical separation ${\bm r}(t)=a(t){\bm x}$ so that their relative physical velocity is
$$
\dot{\bm r} = H{\bm r},
$$
as first established by Lema\^{\i}tre~\cite{lemaitre27}. This gives the first observational pillar of the model and the order of magnitudes of the Hubble time and radius are obtained by expressing the current value of the Hubble parameter as $H_0 = 100\,h\, \hbox{km}\cdot\hbox{s}^{-1}\cdot\hbox{Mpc}^{-1}$ with $h$ typically of the of order $0.7$ so that the present Hubble distance and time are 
\begin{eqnarray}
 D_{{\rm H}_0}&=&9.26\,h^{-1}\times10^{25}\sim3000\,h^{-1}\,\hbox{Mpc} ,\\
 t_{{\rm H}_0}&=&9.78\,h^{-1}\times10^{9}\,\hbox{years}\ .
\end{eqnarray}
\item{\em Hubble diagram}. Hubble measurements~\cite{huublelaw} aimed at mesuring independenly distances and velocities. Today, we construct observationally a Hubble diagram that represents the distance in terms of the redshift. One needs to be careful when defining the distance since one needs to distinguish angular and luminosity distances. The luminosity distance can be shown to be given by
\begin{equation}\label{eq.20}
 D_L(z)=a_0(1+z)f_K\left(\frac{1}{a_0H_0}\int_0^z\frac{\dd z'}{\mathbb{E}(z')} \right)
\end{equation}
with $\mathbb{E}\equiv H/H_0$. The use of standard candels, such as type~Ia supernovae allow to constrain the parameter of the models and measured the Hubble constant (see Fig.~\ref{fig-hd}).
%%----------------------------------------------------
\begin{figure}[htb!]
\centering
\includegraphics[width=.41\columnwidth]{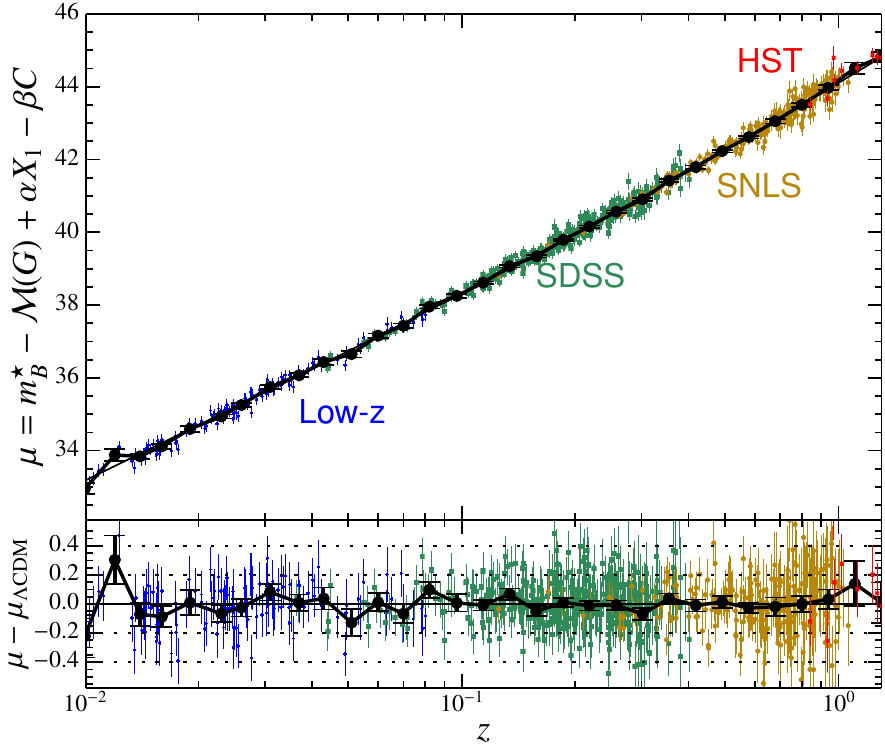}$\,$
\includegraphics[width=.49\columnwidth]{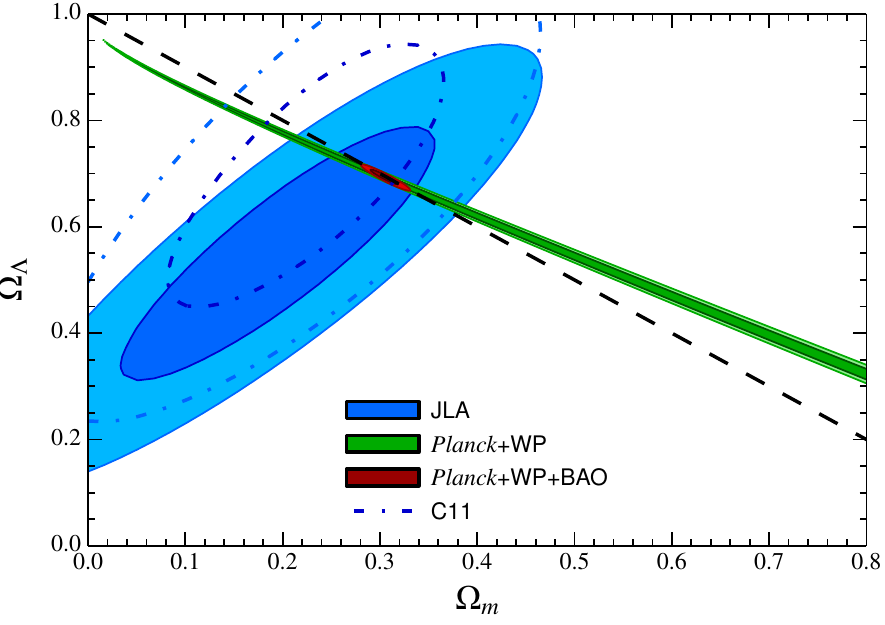}
 \caption{\footnotesize (left) A Hubble diagram obtained by the joint lightcurve analysis of 740 SNeIa from four different samples: Low-$z$, SDSS-II, SNLS3, and HST. The top panel depicts the Hubble diagram itself with the CDM best fit (black line); the bottom panel shows the residuals. (right) Constraints on the cosmological parameters obtained from this Hubble (together with other observations: CMB (green), and CMB+BAO (red). The dot-dashed contours corresponds to the constraints from earlier SN data. From Ref.~\cite{sn40}.} 
\label{fig-hd}
\end{figure}
%%----------------------------------------------------
\item{\em Age of the universe}. Since $\dd t=\dd a/aH$, the dynamical age of the universe is given as
\begin{equation}
t_0   = t_{{\rm H}_0}\int_0^\infty\frac{\dd z}{(1+z)\mathbb{E}(z)}.
\end{equation}
A constraint on the cosmological parameters can also be obtained from the measurements of the age of the universe. A lower bound on this age can then be set that should be compatible with the dynamical age of the universe computed from the Friedmann equations, and it needs to be larger than the age of its oldest objects (see Chap~4 of Ref.~\cite{jpu-book}).
\end{enumerate}

During the development of this first formulation of the big-bang model, there is almost no discussion on the physical nature of the matter content (it is modeled as a matter \& radiation fluid). The debate was primarily focused on the expansion, challenged by the steady state model. In the early stages (20ies), the main debate concerned the extragalactic nature of nebulae (the Shapley-Curtis debate). The model depends on only 4 independent parameters: the Hubble constant $H_0$ the density parameters for the matter and radiation, $\Omega_{\rm m}$ and $\Omega_{\rm r}$, the spatial curvature, $\Omega_K$, and the cosmological constant $\Omega_{\Lambda}$ so that the program of observational cosmology reduced mostly to measuring the mean density of the universe. From these parameters one can infer an estimate of the age of the universe, which was an important input in the parallel debate on the legitimity of a non-vanishing cosmological constant. The Copernican principle was also challenged by the derivation of non-isotropic solutions, such as the Bianchi~\cite{bianchisol} family, and non-inhomogeneous solutions, such as the Lema\^{\i}tre-Tolman-Bondi~\cite{LTBsol} spacetime.

\subsection{The hot big-bang model}

As challenged by Georges Lema\^{\i}tre~\cite{lemaitre31} in 1931, ``{\em une cosmogonie vraiment compl\`ete devrait expliquer les atomes comme les soleils.}'' This is mostly what the hot big-bang model will achieved.

This next evolution takes into account a better description of the matter content of the universe. Since for radiation $\rho\propto a^4$ while for pressureless matter $\rho\propto a^3$, it can be concluded that the universe was dominated by radiation. The density of radiation today is mostly determined by the temperature of the cosmic microwave background (see below) so that equality takes place at a redshift
\begin{equation}\label{4.zeq}
 z_{\rm eq} \simeq 3612\, \Theta_{2.7}^{-4}
 \,\left(\frac{\Omega_{\rm m0} h^2}{0.15}\right)\ ,
\end{equation}
obtained by equating the matter and radiation energy densities and where $\Theta_{2.7}\equiv T_{\rm CMB}/2.725\,{\rm m}\rm{K}$. Since the temperature scales as $(1+z)$, the temperature at which the matter and radiation densities were equal is $T_{\rm eq} =
T_{\rm CMB}(1+z_{\rm eq})$ which is of order
\begin{equation}\label{4.Teq}
 T_{\rm eq} \simeq 5.65\,
 \Theta_{2.7}^{-3}\,\Omega_{\rm m0} h^2\,\mathrm{eV}.
\end{equation}
Above this energy, the matter content of the universe is in a very different form to that of today. As it expands, the photon bath cools down, which implies a {\em thermal history}. In particular, 
\begin{itemize}
\item when the temperature $T$ becomes larger than twice the rest mass $m$ of a charged particle, the energy of a photon is large enough to produce particle-antiparticle pairs. Thus when $T\gg m_{\rm e}$, both electrons and positrons were present in the universe, so that the particle content of the universe changes during its evolution, while it cools down;
\item symmetries can be spontaneously broken;
\item some interactions may be efficient only above a temperature, typically as long as the interaction rate $\Gamma$ is larger than the Hubble expansion rate;
\item the freeze-out of some interaction can lead to the existence of relic particles.
\end{itemize}

\subsubsection{Equilibrium and beyond}

Particles interaction are mainly characterized by a reaction rate $\Gamma$. If this reaction rate is much larger than the Hubble expansion rate, then it can maintain these particles in {\em thermodynamic equilibrium} at a temperature $T$. Particles can thus be treated as perfect Fermi-Dirac and Bose-Einstein gases with distribution\footnote{The distribution function depends {\sl a priori} on $({\bm x},t)$ and $(\bm{p},E)$ but the homogeneity hypothesis implies that it does not depend on ${\bm x}$ and isotropy implies that it is a function of $p^2=\bm{p}^2$. Thus it follows from the cosmological principle that $f({\bm x},t,\bm{p},E) = f(E,t) = f[E,T(t)] $.}
\begin{equation}\label{4.fi}
 F_i(E,T) =
 \frac{g_i}{(2\pi)^3}\frac{1}{\exp\left[(E-\mu_i)/T_i(t)\right]\pm1}
 \equiv\frac{g_i}{(2\pi)^3}f_i(E,T)\ ,
\end{equation}
where $g_i$ is the degeneracy factor, $\mu_i$ is the chemical potential and  $E^2=p^2 + m^2$. The normalisation of $f_i$ is such that $f_i=1$ for the maximum phase space density allowed by the Pauli principle for a fermion. $T_i$ is the temperature associated with the given species and, by symmetry, it is a function of $t$ alone, $T_i(t)$. Interacting species have the same temperature. Among these particles, the universe contains an electrodynamic radiation with black body spectrum (see below). Any species interacting with photons will hence have the same temperature as these photons as long as $\Gamma_i\gg H$. The photon temperature $T_\gamma=T$ will thus be called the {\it temperature of the universe}.

If the cross-section behaves as $\sigma\sim E^p\sim T^p$ (for instance $p=2$ for electroweak interactions) then the reaction rate behaves as $\Gamma\sim n\sigma\sim T^{p+3}$; the Hubble parameter behaves as $H\sim T^2$ in the radiation period. Thus if $p+1>0$, there will always be a temperature below which an interaction decouples while the universe cools down. The interaction is no longer efficient; it is then said to be {\it frozen}, and can no longer keep the equilibrium of the given species with the other components. This property is at the origin of the thermal history of the universe and of the existence of relics. This mechanism, during which an interaction can no longer maintain the equilibrium between various particles because of the cosmic expansion, is called  {\it decoupling}. This criteria of comparing the reaction rate and the rate $H$ is simple; it often gives a correct order of magnitude, but a more detailed description of the decoupling should be based on a microscopic study of the evolution of the distribution function. As an example, consider Compton scattering. Its reaction rate, $\Gamma_{\rm Compton}=n_{\rm e}\sigma_{\rm T}c$, is of order $\Gamma_{\rm Compton}\sim1.4\times 10^{-3}H_0$ today, which means that, statistically, only one photon over 700 interacts with an electron in a Hubble time today. However, at a redshift $z\sim10^3$, the electron density is $10^3$ times larger and the Hubble expansion rate is of order $H\sim H_0\sqrt{\Omega_{{\rm m} 0}(1+z)^3}\sim2\times10^4H_0$ so that $\Gamma_{\rm Compton}\sim80H$. This means that statistically at a redshift $z\sim10^3$ a photon interacts with an electron about 80 times in a Hubble time. This illustrates that backward in times densities and temperature increase and interactions become more and more important.\\

As long as thermal equilibrium holds, one can define thermodynamical quantities such as the number density $n$, energy density $\rho$ and pressure $P$ as
\begin{equation}
 n_i =  \int F_i(\bm{p},T)\dd^3\bm{p}\quad
 \rho_i =  \int F_i(\bm{p},T)E(\bm{p})\dd^3\bm{p}\quad
   P_i =  \int F_i(\bm{p},T)\frac{p^2}{3E}\dd^3\bm{p}.
\end{equation}
For ultra-relativistic particles ($m,\mu\ll T$), the density at a given temperature $T$ is then given by
\begin{equation}
 \rho_{\rm r}(T) = g_*(T) \left(\frac{\pi^2}{30}\right) T^4\ .
\end{equation}
$g_*$ represents the effective number of relativistic degrees of freedom at this temperature,
\begin{equation}
 g_*(T) = \sum_{i=\mathrm{bosons}} g_i\left(\frac{T_i}{T}\right)^4
 + \frac{7}{8}\sum_{i=\mathrm{fermions}}
 g_i\left(\frac{T_i}{T}\right)^4\ .
\end{equation}
The factor $7/8$ arises from the difference between the Fermi and Bose distributions. In the radiation era, the Friedmann equation then takes the simple form
\begin{equation}\label{4.Hexp}
 H^2 = \frac{8\pi G}{3}\left(\frac{\pi^2}{30}\right)g_* T^4\ .
\end{equation}
Numerically, this amounts to
\begin{equation}\label{4.20}
 H(T) \cong1.66g_*^{1/2}\frac{T^2}{M_p}\ , \qquad
 t(T) \cong 0.3g_*^{-1/2}\frac{M_p}{T^2}\sim
 2.42\, g_*^{-1/2}\left(\frac{T}{1\,\mathrm{MeV}}\right)^{-2}\,\mathrm{s}\ .
\end{equation}

In order to follow the evolution of the matter content of the universe, it is convenient to have conserved quantities such as the entropy. It can be shown~\cite{jpu-book} to be defined as $S=sa^3$ in terms of the entropy density $s$ as
\begin{equation}\label{4.entropie}
 s \equiv \frac{\rho+P-n\mu}{T}.
\end{equation}
It satisfies $\dd(sa^3) = -(\mu/T)\dd(n a^3)$ and his hence constant (i) as long as matter is neither destroyed nor created, since then $n a^3$ is constant, or (ii) for non-degenerate relativistic matter, $\mu/T\ll1$. In the cases relevant for cosmology, $\dd(s a^3) = 0$. It can be expressed in terms of the temperature of the photon bath as
\begin{equation}\label{4.sexp}
 s=\frac{2\pi^2}{45}q_* T^3\ ,\qquad{\rm with}\qquad
 q_*(T) = \sum_{i=\mathrm{bosons}} g_i\left(\frac{T_i}{T}\right)^3
 + \frac{7}{8}\sum_{i=\mathrm{fermions}}
 g_i\left(\frac{T_i}{T}\right)^3\ .
\end{equation}
If all relativistic particles are at the same temperature, $T_i=T$, then  $q_*=g_*$. Note also that $s=q_*\pi^4/45\zeta(3) n_\gamma\sim 1.8q_*n_\gamma$, so that the photon number density gives a measure of the entropy.

The standard example of the use of entropy is the determination of the temperature of the cosmic neutrino background. Neutrinos are in equilibrium with the cosmic plasma as long as the reactions $\nu+\bar\nu\longleftrightarrow e+\bar e$ and $\nu+e\longleftrightarrow\nu+e$ can keep them coupled. Since neutrinos are not charged, they do not interact directly with photons. The cross-section of weak interactions is given  by $\sigma\sim G_{\rm F}^2E^2\propto G_{\rm F}^2T^2$ as long as the energy of the neutrinos is in the range $m_{\rm e}\ll E\ll m_{\rm W}$. The interaction rate is thus of the order of $\Gamma = n\langle\sigma v\rangle\simeq G_{\rm F}^2T^5$. We obtain that $\Gamma \simeq \left(\frac{T}{1\,\mathrm{MeV}}\right)^3 H$. Thus close to $T_{\rm D}\sim 1\,\mathrm{MeV}$, neutrinos
decouple from the cosmic plasma. For $T<T_{\rm D}$, the neutrino temperature decreases as $T_\nu\propto a^{-1}$ and remains equal to the photon temperature.

Slightly after decoupling, the temperature becomes smaller than $m_{\rm e}$. Between  $T_{\rm D}$ and $T=m_{\rm e}$ there are 4 fermionic states ($e^-$, $e^+$, each having $g_e=2$) and 2 bosonic states (photons with $g_\gamma=2$) in thermal equilibrium with the photons. We thus have that $q_{\gamma}(T> m_{\rm e})=\frac{11}{2}$ while for $T< m_{\rm e}$ only the photons contribute to $q_{\gamma}$ and hence $q_{\gamma}(T<m_{\rm e})= 2$. The conservation of entropy implies that after $\bar{e}-e$ annihilation, the temperatures of the neutrinos and the photons are related by
\begin{equation}
 T_\gamma = \left(\frac{11}{4}\right)^{1/3}T_\nu\ .
\end{equation}
Thus the temperature of the universe is increased by about 40\% compared to the neutrino temperature during the annihilation. Since $n_\nu=(3/11)n_\gamma$, there must exist a cosmic background of neutrinos with a density of 112 neutrinos per cubic centimeter
and per family, with a temperature of around 1.95~K today.\\

The evolution of any decoupled species can thus easily be described. However, the description of the decoupling, or of a freeze-out of an interaction, is a more complex problem which requires to go beyond the equilibrium description.

The evolution of the distribution function is obtained from the Boltzmann equation $L[f] = C[f]$, where $C$ describes the collisions and  $L=\dd/\dd s$ is the Liouville operator, with $s$ the length along a worldline. The operator $L$ is a function of eight variables taking the explicit form
\begin{equation}
 L[f] = p^\alpha\frac{\partial}{\partial x^\alpha}
 -\Gamma_{\beta\gamma}^\alpha p^\beta p^\gamma
 \frac{\partial}{\partial p^\alpha}\ .
\end{equation}
In a homogeneous and isotropic space-time, $f$ is only function of the energy and time, $f(E,t)$, only so that
\begin{equation}\label{4.eqX451}
 L[f] = E\frac{\partial f}{\partial t} - H p^2\frac{\partial f}{\partial  E}\ .
\end{equation}
Integrating this equation with respect to the momentum  $\bm{p}$, we obtain
\begin{equation}\label{4.nevo}
 \dot n_i + 3 H n_i = \mathcal{C}_i\ ,\qquad
 \mathcal{C}_i=\frac{g_i}{(2\pi)^3}\int C\left[f_i(\bm{p_i},t)\right] \frac{\dd^3\bm{p}_i}{E_i}\ .
\end{equation}
The difficult part lies in the modelling and the evaluation of the collision term. In the simple of an interaction of the form $ i + j \longleftrightarrow k+ l$, the collision term can be decomposed as $\mathcal{C}_i=\mathcal{C}_{kl\rightarrow ij}-
\mathcal{C}_{ij\rightarrow kl}$.  There are thus 3 sources of evolution for the number density $n_i$, namely dilution ($3Hn_i$), creation ($\mathcal{C}_i=\mathcal{C}_{kl\rightarrow ij}$) and destruction ($\mathcal{C}_i=\mathcal{C}_{ij\rightarrow kl}$).

To go further, one needs to specify the interaction. We will consider the nuclear interaction for BBN, the electromagnetic interaction for the CMB and particle annihilation for relics.

\subsubsection{big-bang nucleosynthesis (BBN)}\label{secbbn}

big-bang nucleosynthesis describes the synthesis of light nuclei in the primordial universe. It is considered as the {\em second} pillar of the big-bang model. It is worth reminding that BBN has been essential in the past, first to estimate the baryonic density of the universe, and give an upper limit on the number of neutrino families, as was later confirmed from the measurement of the $Z^0$ width by LEP experiments at CERN.

%%----------------------------------------------------
\begin{figure}[htb!]
\centering
\includegraphics[width=.4\columnwidth]{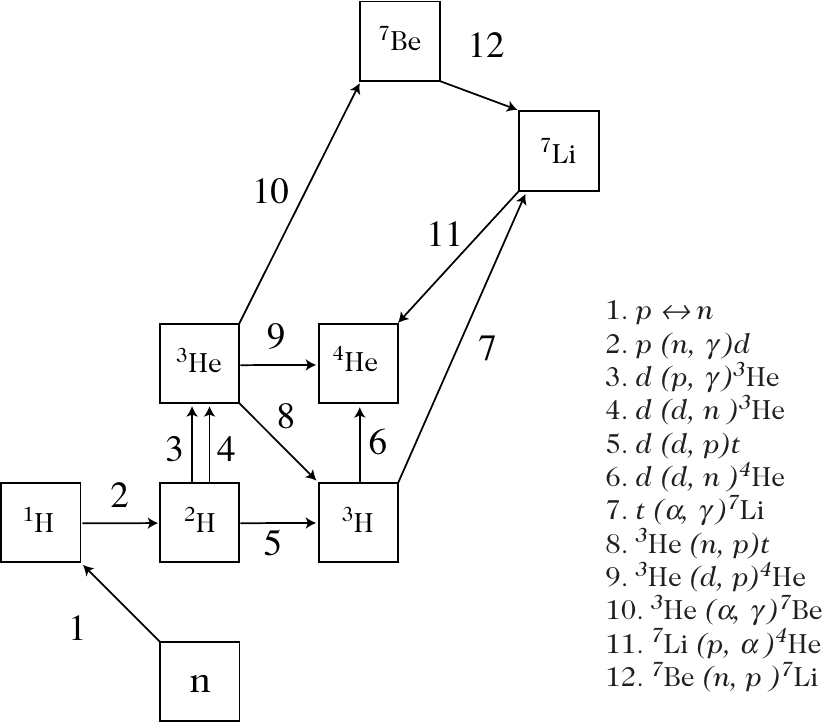}\hskip.5cm
\includegraphics[width=.5\columnwidth]{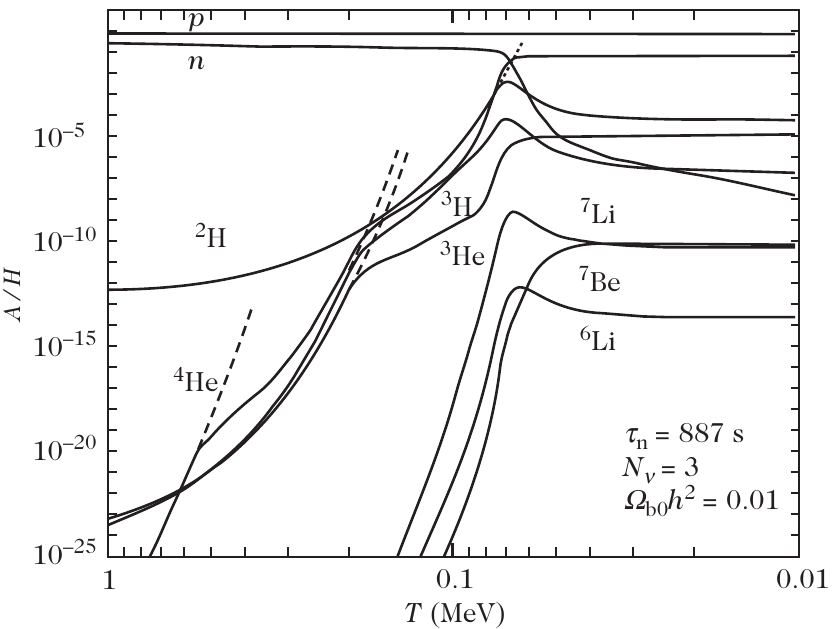}
 \caption{\footnotesize (left) The minimal 12 reactions network needed to compute the abundances up to lithium. (right) The evolution of the abundances of neutron, proton and the lightest elements as a function of temperature (i.e. time). Below 0.01~MeV, the abundances are frozen and can be considered as the primordial abundances. From Ref.~\cite{jpu-book}.} 
\label{fig-bbn}
\end{figure}
%%----------------------------------------------------

\paragraph{Generalities on BBN.}

The standard BBN scenario~\cite{jpu-book,bernstein,4.peeble66,7.bbn1} proceeds in three main
steps:
\begin{enumerate}
\item for $T>1$~MeV, ($t<1$~s) a first stage during which the
neutrons, protons, electrons, positrons an neutrinos are kept in
statistical equilibrium by the (rapid) weak interaction
\begin{eqnarray}\label{bbn0}
&&n\longleftrightarrow p+e^-+\bar\nu_e,\qquad
n+\nu_e\longleftrightarrow  p+e^-,\qquad n+e^+\longleftrightarrow
p+\bar\nu_e.
\end{eqnarray}
As long as statistical equilibrium holds, the neutron to proton
ratio is
\begin{equation}
(n/p)=\hbox{e}^{-Q_{\rm np}/k_{_{\rm B}}T}
\end{equation}
where $Q_{\rm np}\equiv (m_{\rm n}-m_{\rm p})c^2=1.29$~MeV. The
abundance of the other light elements is given
by~\cite{jpu-book}
\begin{eqnarray}
Y_A&=&g_A\left(\frac{\zeta(3)}{\sqrt{\pi}}\right)^{A-1}2^{(3A-5)/2}A^{5/2}
     \left[\frac{k_{_{\rm B}}T}{m_{\rm N}c^2}\right]^{3(A-1)/2}\!\!\!\!\!\!\!\!\!
      \eta^{A-1}Y_{\rm p}^ZY_{\rm n}^{A-Z}\hbox{e}^{B_A/k_{_{\rm B}}T},
\end{eqnarray}
where $g_A$ is the number of degrees of freedom of the nucleus
$_Z^A{\rm X}$, $m_{\rm N}$ is the nucleon mass, $\eta$ the
baryon-photon ratio and $B_A\equiv(Zm_{\rm p}+(A-Z)m_{\rm
n}-m_A)c^2$ the binding energy. \item Around $T\sim0.8$~MeV
($t\sim2$~s), the weak interactions freeze out at a temperature
$T_{\rm f}$ determined by the competition between the weak
interaction rates and the expansion rate of the universe and thus
roughly determined by $\Gamma_{_{\rm w}}(T_{\rm f})\sim H(T_{\rm
f})$ that is
\begin{equation}
G_{\rm F}^2(k_{_{\rm B}}T_{\rm f})^5\sim\sqrt{GN_*}(k_{_{\rm
B}}T_{\rm f})^2
\end{equation}
where $G_{\rm F}$ is the Fermi constant and $N_*$ the number of
relativistic degrees of freedom at $T_{\rm f}$. Below $T_{\rm f}$,
the number of neutrons and protons changes only from the neutron
$\beta$-decay between $T_{\rm f}$ to $T_{\rm N}\sim0.1$~MeV when
$p+n$ reactions proceed faster than their inverse dissociation.
 \item For $0.05$~MeV$<T<0.6$~MeV ($3\,{\rm s}<t<6\,{\rm min}$),
the synthesis of light elements occurs only by two-body reactions.
This requires the deuteron to be synthesized ($p+n\rightarrow D$)
and the photon density must be low enough for the
photo-dissociation to be negligible. This happens roughly when
\begin{equation}\label{n0}
\frac{n_{\rm d}}{n_\gamma}\sim\eta^2\exp(-B_D/T_{\rm N})\sim 1
\end{equation}
with $\eta\sim3\times10^{-10}$. The abundance of $^4{\rm He}$ by
mass, $Y_{\rm p}$, is then well estimated by
\begin{equation}\label{n1}
Y_{\rm p}\simeq2\frac{(n/p)_{\rm N}}{1+(n/p)_{\rm N}}
\end{equation}
with
\begin{equation}\label{n2}
(n/p)_{\rm N}=(n/p)_{\rm f}\exp(-t_{\rm N}/\tau_{\rm n})
\end{equation}
with $t_{\rm N}\propto G^{-1/2}T_{\rm N}^{-2}$ and $\tau_{\rm
n}^{-1}=1.636G_{\rm F}^2(1+3g_A^2)m_{\rm e}^5/(2\pi^3)$, with
$g_A\simeq1.26$ being the axial/vector coupling of the nucleon.
\item The abundances of the light element abundances, $Y_i$, are then
obtained by solving a series of nuclear reactions
$$
 \dot Y_i = J - \Gamma Y_i,
$$
where $J$ and $\Gamma$ are time-dependent source and sink terms (see Fig.~\ref{fig-bbn}).
\item Today, BBN codes include up to 424 nuclear reaction network \cite{Coc12a} with up-to-date nuclear physics. In standard BBN only \deu, \tro, \qua\, and \sep\, are significantly produced as well as traces  of \six, \neu, \dix, \onz\,\ and CNO.  The most recent up-to-date predictions are discussed in Refs.~\cite{Coc13,Coc14}.
\end{enumerate}

%%----------------------------------------------------
\begin{figure}[htb!]
\centering
\vskip-2cm
\includegraphics[width=.4\columnwidth]{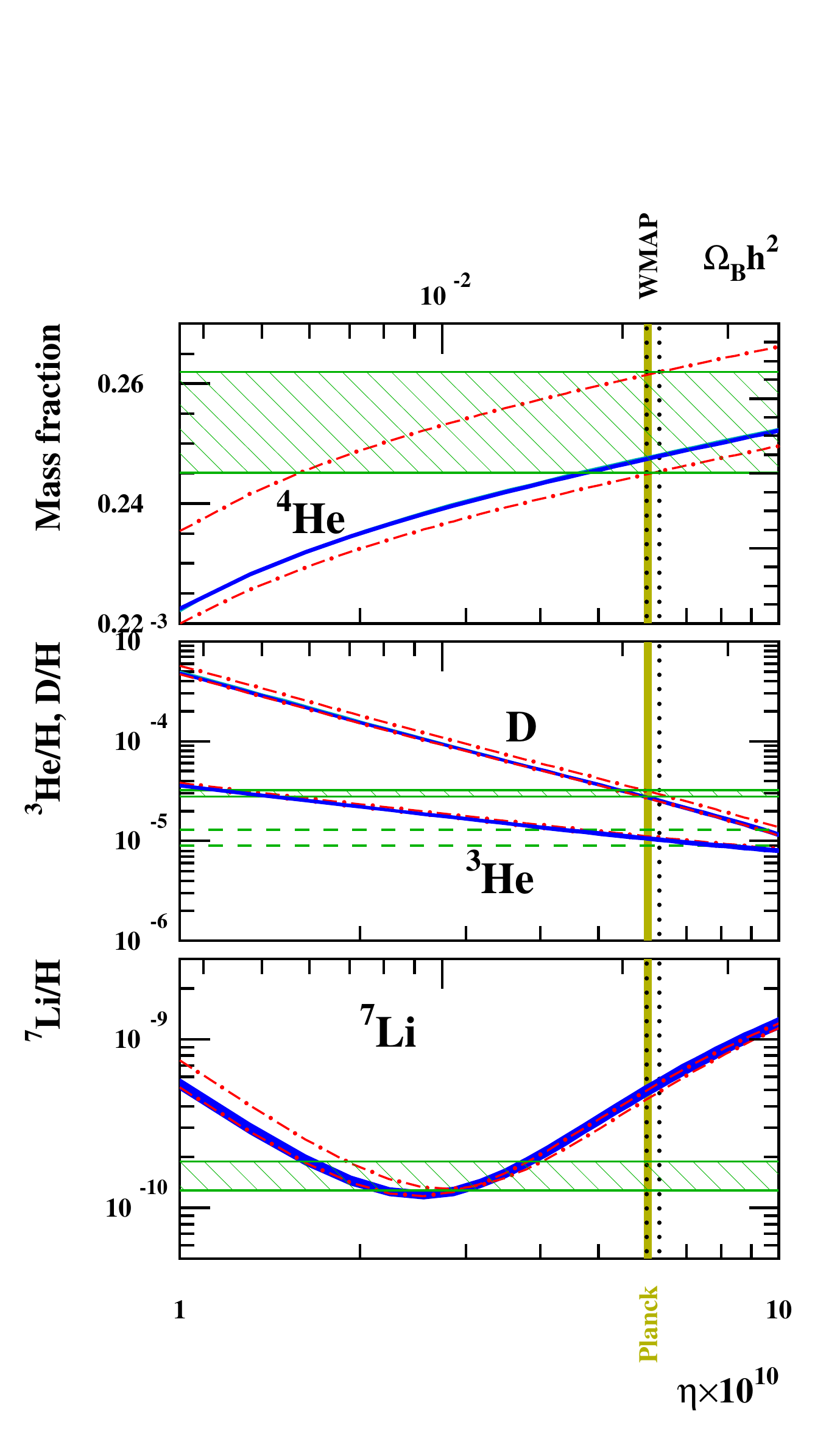}
 \caption{\footnotesize Abundances of \qua\, \deu, \tro\ and \sep\  (blue) as a function of the baryon over photon ratio (bottom) or baryonic density (top). The vertical areas correspond to the WMAP (dot, black) and Planck (solid, yellow) baryonic densities while the horizontal areas (green) represent the adopted observational abundances. The  (red) dot--dashed lines correspond to the extreme values of the {\em effective} neutrino families coming from CMB Planck study, $N_{\rm eff}=(3.02, 3.70)$. From Ref.~\cite{Coc14}.} 
\label{fig-bbn2}
\end{figure}
%%----------------------------------------------------

\paragraph{Observational status.}

These predictions need to be compared to the observation of the abundances of the different nuclei (Fig.~\ref{fig-bbn2}), that we can summarized.

Deuterium is a very fragile isotope, easily destroyed after BBN. Its most primitive abundance is determined from the observation of clouds at high redshift, on the line of sight of distant quasars. Recently, more precise observations of damped Lyman-$\alpha$ systems at high redshift have lead to provide ~\cite{pettini2012, cooke2014}  the mean value
\begin{equation}
{\rm D/H} = (2.53 \pm 0.04) \times 10^{-5}.
 \end{equation}

After BBN, \qua\ is still produced by stars, essentially during the main sequence phase. Its primitive abundance is deduced from observations in H{\sc ii} (ionized hydrogen) regions of compact blue galaxies. The primordial \qua\ mass fraction, $Y_p$, is obtained from the extrapolation to zero metallicity but is affected by systematic uncertainties. Recently, \cite{aver12} have determined that
\begin{equation}
Y_p = 0.2465 \pm 0.0097.
\end{equation}

Contrary to \qua,  \tro\ is both produced and destroyed in stars all along its galactic evolution, so that the evolution of its abundance as a function of time is subject to large uncertainties. Moreover, \tro\ has been observed in our Galaxy \cite{Ban02}, and one only gets a local constraint
\begin{equation}
\hbox{\tro/H}  = (1.1 \pm 0.2) \times 10^{-5}.
\end{equation}
Consequently, the baryometric status of \tro\ is not firmly established \cite{vang03}.

Primitive lithium abundance is deduced from observations of low metallicity stars in the halo of our Galaxy where the lithium abundance is almost independent of metallicity, displaying the so-called Spite plateau \cite{Spite82}. This interpretation assumes that lithium has not been depleted at the surface of these stars, so that the presently observed abundance can be assumed to be equal to the primitive one. The small scatter of values around the Spite plateau is indeed an indication that depletion may not have been very efficient. However, there is a discrepancy between the value {\it{i)}} deduced from these  observed spectroscopic abundances and {\it{ii)}} the  BBN theoretical predictions assuming  \ob is determined by the CMB observations.  Many studies have been devoted to the resolution of this so-called {\it Lithium problem} and many possible ``solutions'', none fully satisfactory, have been proposed. For a detailed analysis see the proceedings of the meeting ``Lithium in the cosmos''~\cite{LiinC}. Note that the idea according to which introducing neutrons during BBN may solve the problem has today been shown~\cite{pospelov} to be generically inconsistent with lithium and deuterium observations. Astronomical observations of these metal poor halo stars~\cite{sbordone10}  have thus led to a relative primordial abundance of
\begin{equation}
  {\rm Li/H}=  (1.58 \pm 0.31) \times 10^{-10}.
\end{equation}  

The origin of the light elements LiBeB, is a crossing point between optical and gamma spectroscopy,  non thermal nucleosynthesis (via spallation with galactic cosmic ray), stellar evolution and big-bang nucleosynthesis. I shall not discuss them in details but just mention that typically, \six/H$\sim10^{-11}$. Beryllium is a fragile nucleus formed in the vicinity of Type II supernovae by non thermal process (spallation). The observations in metal poor stars provide a primitive abundance at very low metallicity of the order of  $ {\rm Be/H}=  3.\times 10^{-14}$ at   [Fe/H] = -4. This observation has to be compared to the typical primordial Be abundance,  $ {\rm Be/H}=  10^{-18}$. Boron has two isotopes:  \dix\   and \onz\ and is also synthesized by non thermal processes. The most recent observations give $ {\rm B/H}=  1.7  \times 10^{-12}$, to be compared to the typical primordial B abundance $ {\rm B/H}=  3  \times 10^{-16}$. For a general review of these light elements, see Ref.~\cite{charbonnel10}.

Finally, CNO elements are observed in the lowest metal poor stars (around [Fe/H]=-5). The observed abundance of CNO is typically [CNO/H]= -4,  relatively to the solar abundance i.e. primordial CNO/H$<10^{-7}$. For a review see Ref.~\cite{frebel13} and references therein.

\paragraph{Discussion}

This shows that BBN is both a historical pillar and lively topic of the big-bang model. It allows one to test physics beyond the standard model. In particular, it can allow us to set strong constraints on the variation of fundamental constants~\cite{uzancst1,bbncte} and deviation from general relativity~\cite{dampich} As discussed it shows one of the major discrepancy of the model, namely the {\em Lithium problem}, that has for now no plausible explanation. Heavier elements such as CNO are now investigated in more details since they influence the evolution of Population~III stars and invovled nuclear reactions the cross-section of which are badly known in the laboratory~\cite{Coc14}.

\subsubsection{Cosmic microwave background}

As long as the temperature of the universe remains large compared to the hydrogen ionisation energy, matter is ionized and photons are then strongly coupled to electrons through Compton scattering. At lower temperatures, the formation of neutral atoms is thermodynamically favored for matter. Compton scattering is then no longer efficient and radiation decouples from matter to
give rise to a fossil radiation: the cosmic microwave background (CMB).

\paragraph{Recombination and decoupling}

As long as the photoionisation reaction
\begin{equation}\label{4.rec}
 p+e\longleftrightarrow\mathrm{H}+\gamma
\end{equation}
is able to maintain the equilibrium, the relative abundances of the electrons, protons and hydrogen will be fixed by the equation of chemical equilibrium. In this particular case, it is known as the Saha equation
\begin{equation}\label{4.saha}
 \frac{X_{\rm e}^2}{1-X_{\rm e}} =
 \left(\frac{m_{\rm e}T}{2\pi}\right)^{3/2}
 \frac{\hbox{e}^{-E_{\rm I}/T}}{n_{\rm b}}
\end{equation}
where $E_{\rm I}=m_{\rm e}+m_{\rm p}-m_{\rm H}=13.6$~eV is the hydrogen ionisation energy and where one introduces the ionisation fraction
\begin{equation}
 X_{\rm e} = \frac{n_{\rm e}}{n_{\rm p}+n_{\rm H}}\ ,
\end{equation}
where the denominator represents the total number of hydrogen nuclei, $n_{\rm b}=n_{\rm p}+n_{\rm H}$ [$n_{\rm e}=n_{\rm p}=X_{\rm e}n_{\rm b}$, $n_{\rm H}=(1-X_{\rm e})n_{\rm b}$] since electrical neutrality implies $n_{\rm e}=n_{\rm p}$.

The photon temperature is  $T= 2.725(1+z)$~K and the baryon density $n_{\rm b} = \eta n_{\gamma 0}(1+z)^3\,\mathrm{cm}^{-3}$. Notice first of all that when $T\sim E_{\rm I}$, the right hand side of Eq.~(\ref{4.saha}) is of order  of $10^{15}$ so that $X_{\rm e}(T\sim E_{\rm I})\sim1$. Recombination only happens for $T\ll E_{\rm I}$.  $X_{\rm e}$ varies abruptly between $z=1400$ and $z=1200$ and the recombination can be estimated to occur at a temperature between 3100 and 3800~K. Note that Eq.~(\ref{4.saha})
implies that
$$
 \frac{E_{\rm I}}{T} = \frac{3}{2}\ln\left(\frac{m_{\rm e}}{2\pi T}\right)
 -\ln\eta
 - \ln\left[\frac{2}{\pi^2}\zeta(3)\frac{X_{\rm e}^2}{1-X_{\rm e}}\right]\ ,
$$
which gives a rough estimates of the recombination temperature since the last term can be neglected; $T\sim3500$~K.

The electron density varies quickly at the time of recombination, thus the reaction rate $\Gamma_{\rm T} = n_{\rm e}\sigma_{\rm T}$, with $\sigma_{\rm T}$  the Thompson scattering cross-section, drops off rapidly so that the reaction (\ref{4.rec}) freezes out and the photons decouple soon after. An estimate of the decoupling time, $t_{\rm dec}$, can be obtained by the requirement
$\Gamma_{\rm T}(t_{\rm dec})=H(t_{\rm dec})$. It gives the decoupling redshift
\begin{equation}
 (1+z_{\rm dec})^{3/2} = \frac{280.01}{X_{\rm e}(\infty)}\left(\frac{\Omega_{\rm b 0}
 h^2}{0.02}\right)^{-1}\left(\frac{\Omega_{\rm m 0}
 h^2}{0.15}\right)^{1/2}\sqrt{1+\frac{1+z_{\rm dec}}{1+z_{\rm eq}}}\ .
\end{equation}
$X_{\rm e}(\infty)$ is the residual electron fraction, once recombination has ended. We can  estimate  that $X_{\rm e}(\infty)\sim7\times10^{-3}$.

This is indeed a simplified description since to describe the recombination, and to determine $X_{\rm e}(\infty)$, one should solve the Boltzmann equation and include the recombination of helium. A complete treatment, taking the hydrogen and helium contributions into account, is described in Ref.~\cite{jpu-book}.

%%----------------------------------------------------
\begin{figure}[htb!]
\centering
\includegraphics[width=.6\columnwidth,angle=270]{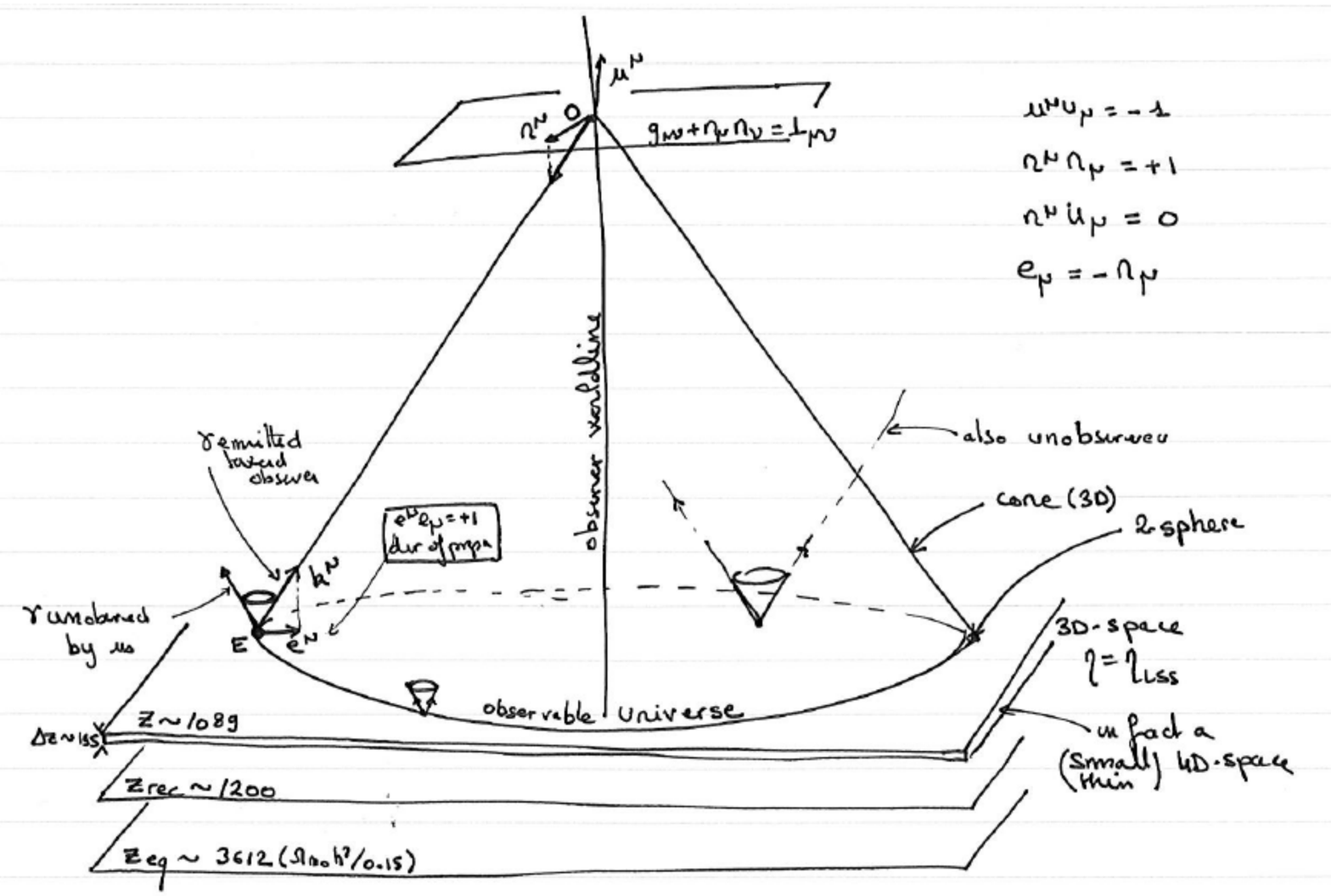}
\vskip-.25cm
 \caption{\footnotesize Spacetime diagram representing the equality hypersurface and the recombination followed by decoupling. Observed CMB photon are emitted on the intersection of this latter hypersurface and our past lightcone.} 
\label{fig-obs3}
\end{figure}
%%----------------------------------------------------

\paragraph{Last scattering surface}

The optical depth can be computed once the ionisation fraction as a function of the redshift around
decoupling is known,
\begin{equation}\label{4.eq34}
 \tau = \int n_{\rm e}X_{\rm e}\sigma_{\rm T}\dd\chi.
\end{equation}
It varies rapidly around $z\sim1000$ so that the visibility function $g(z)=\exp(-\tau)\dd\tau/\dd z$, which determines the probability for a photon to be scattered between $z$ and $z+\dd z$, is a highly peaked function. Its maximum defines the decoupling time, $z_{\rm dec}\simeq1057.3$ (see Fig.~\ref{ig-obs3}). This redshift defines the time at which the CMB photons last scatter; the universe then rapidly becomes transparent and these photons can propagate freely in all directions. The universe is then embedded in this homogeneous and isotropic radiation. The instant when the photons last interact is a space-like hypersurface, called the {\it last scattering surface}. Some of these relic photons can be observed. They come from the intersection of the last scattering surface with our past light cone. It is thus a 2-dimensional sphere, centered around us, with comoving radius  $\chi(z_{\rm dec})$. This sphere is not infinitely thin. If we  define its width as the zone where the visibility function is halved, we get  $\Delta z_{\rm dec}=185.7$; see Fig.~\ref{fig-obs3}.

\paragraph{Properties of the cosmic microwave background}

The temperature of the CMB, defined as the average of the temperature on the whole sky has been measured with precision by the FIRAS experiment on board of the COBE satellite~\cite{4.mather}
\begin{equation}
 T_0=2.725\pm0.001~{\rm K}
\end{equation}
at 2$\sigma$.  The observed spectrum is compatible with a black body spectrum
\begin{equation}
 I(\nu) \propto \frac{\nu^3}{\hbox{e}^{\nu/T} - 1}\ .
\end{equation} 
The fact that this spectrum is so close to a black body proves that the fossil radiation could have been thermalized, mainly thanks to interactions with electrons. However, for redshifts lower than $z\sim10^6$, the fossil radiation do not have time to be thermalized. Any energy injection at lower redshift would induce distortions in the Planck spectrum and can thus be constrained from these observations. It follows that
\begin{equation}
 n_{\gamma0} = 410.44\,\Theta_{2.7}^3\,\mathrm{cm}^{-3}
 \quad
 \rho_{\gamma0} = 4.6408\times10^{-34}\Theta_{2.7}^4\,\mathrm{g}\cdot\mathrm{cm}^{-3}
                \ 
\end{equation}
and
\begin{equation}\label{4.omgam0}
 \Omega_{\gamma0}h^2 = 2.4697\times10^{-5}\Theta_{2.7}^4\ .
\end{equation}
Note also that the temperature of the CMB can be measured at higher redshift to show that it scales as $(1+z)$ hence offering an independent proof of the cosmic expansion (see Fig.~\ref{fig-obs3}).

%%----------------------------------------------------
\begin{figure}[htb!]
\centering
\includegraphics[width=.4\columnwidth]{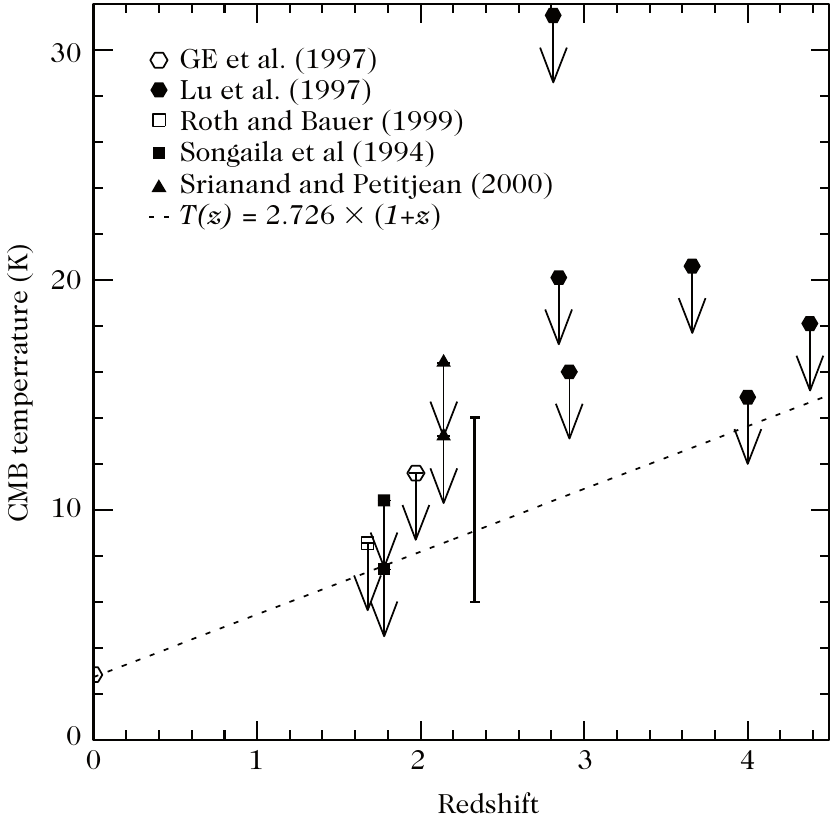}
 \caption{\footnotesize Measurements of the temperature of the CMB at different redshifts. The dashed line represents the prediction from the big-bang model. From Ref.~\cite{srianand}.} 
\label{fig-obs3}
\end{figure}
%%----------------------------------------------------

\paragraph{Residual fluctuations}

Once the monopole and dipole of the CMB have been removed, some temperature anisotropies remain, with relative amplitude $\sim10^{-5}$, which correspond to temperature fluctuations of the order of 30~$\mu$K. These anisotropies correspond to anisotropies in the cosmic microwave background at the time of recombination and redshifted by the cosmic expansion. They can be decomposed on a basis of spherical harmonics as
\begin{equation}\label{cmb1}
 \frac{\delta T}{T}(\vartheta,\varphi)=\sum_{\ell}\sum_{m=-\ell}^{m=+\ell}a_{\ell
  m}Y_{\ell m}(\vartheta,\varphi).
\end{equation}
The angular power spectrum multipole $C_\ell=\langle \vert a_{lm}\vert^2 \rangle$ is the coefficient of the decomposition of the angular correlation function on Legendre polynomials. 

This calls for an explanation to understand the origin of these fluctuations and their properties. They play a central role in modern cosmology that cannot be discussed in a smooth model of the universe.

\subsubsection{Existence of relics}

As the temperature of the universe drops, some interaction may freeze and let some thermal relics. This can be understood from Eq.~(\ref{4.nevo}) that can be rewritten as
\begin{eqnarray}\label{4.nevo3}
 \dot n_i + 3 H n_i &=& -\langle\sigma v\rangle
 \left(n_in_j - \frac{\bar n_i\bar n_j}{\bar n_k\bar n_l} n_k
 n_l\right)\ ,
\end{eqnarray}
where one sets $ n_i=\hbox{e}^{\mu_i/T}\bar n_i$, $\bar n_i\equiv n_i[\mu_i=0]$. $\langle\sigma v\rangle$ depends on the matrix elements of the reaction at hand.

As a simple example, consider a massive particle $X$ in thermodynamic equilibrium with its anti-particle $\bar X$ for temperatures larger than its mass. Assuming this particle is stable, then its density can only be modified by annihilation or inverse annihilation $ X +\bar{X}\longleftrightarrow l + \bar{l}$. If this particle had remained in thermodynamic equilibrium until today, its relic density, $n\propto (m/T)^{3/2}\exp(-m/T)$ would be completely negligible. The relic density of this massive particle, i.e. the residual density once the annihilation is no longer efficient, will actually be more important since, in an expanding space, annihilation cannot keep particles in equilibrium during the entire history of the universe. That particle is usually called a {\it relic}.

The equation~(\ref{4.nevo3}) takes the integrated form
\begin{equation}\label{4.nboltz}
 \dot n_X + 3 H n_X = -\langle\sigma v\rangle\left(n_X^2 - \bar
 n^2_X\right)\ ,
\end{equation}
that can be integrated once $\langle\sigma v\rangle$ is described. In the case of cold relics, the decoupling occurs when the particle is non-relativistic.  Assuming a form
\begin{equation}
 \langle\sigma v\rangle = \sigma_0 f(x)
\end{equation}
where $f(x)$ is a function of $x=m/T$ and $f(x)=x^{-n}$ one gets 
\begin{eqnarray}\label{4.cold}
 \Omega_{X0} h^2 &\simeq& 0.31\left[\frac{g_*(x_{\rm f})}{100}\right]^{-1/2}
 \left(n+1\right)x_{\rm f}^{n+1}\left(\frac{q_{*0}}{3.91}\right)\Theta_{2.7}^3
 \left(\frac{\sigma_0}{10^{-38}\,\mathrm{cm}^2}\right)^{-1}\ 
\end{eqnarray}
where $x_{\rm f}$ depends on the decoupling temperature.

This crude description shows that such relics can account for the dark matter. Any freeze-out may lead to the existence of relics which should not contribute too much to the matter budget of the universe, hence setting constraints on the microphysics (such as e.g. the problem with monopoles that can be produced in the early universe).

\subsubsection{Summary}

The formulation of the hot big-bang model answered partly the query by Lema\^{i}tre. It offered a way to discuss the origin of the way the Mendeleev table was populated and created a link with nuclear physics. It gives the historical pillars of the model that can be connected to observations: BBN, CMB, expansion of the universe. This demonstrates that the universe emerges from a hot and dense phase at thermal equilibrium.

Its main problems are (1) the lithium problem, (2) the origin of the homogeneity of the universe (which is a question of the peculiarity of is initial conditions, (3)  the existence of an initial singularity and (4) the fact that it describes no structure, contrary to an obvious observation.

\subsection{Large scale structure and dark matter}

\begin{figure}[htb!]
\centering
\vskip-1cm
\includegraphics[width=1\columnwidth]{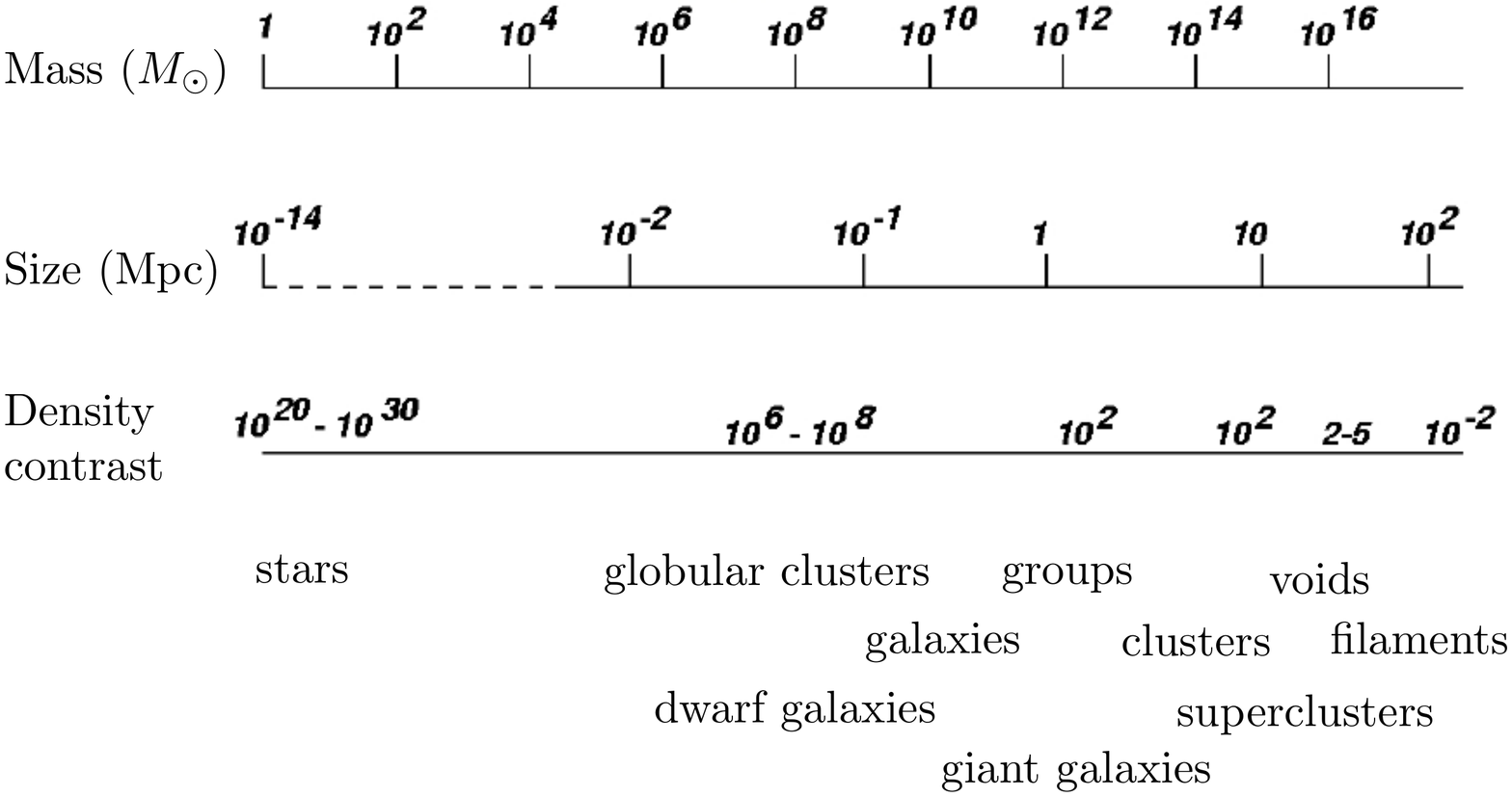}
\vskip-1.2cm
 \caption{\footnotesize  Scales associated with the different kinds of structures observed in the universe. Perturbation theory offers a tool to deal with the large scale structure only. (Courtesy of Yannick Mellier).} 
\label{fig-scale}
\end{figure}

The universe is indeed not smooth and a proper cosmological model needs to address the origin and variety of the large scale structure (see Fig.~\ref{fig-scale}). This means that it needs to model the evolution of density fluctuations and find a scenario for generating some initial density fluctuations. The first step is thus to develop a theory of cosmological perturbation, which can deal only with a part of the objects observed in the universe (see Fig.~\ref{fig-scale}).

%%----------------------------------------------------
%%----------------------------------------------------

\subsubsection{Perturbation theory}

In order to describe the deviation from homogeneity, one starts by considering the most general form of perturbed metric,
\begin{eqnarray}\label{5.metricpert}
 \dd s^2&=&a^2(\eta)\left[-(1+2A)\dd\eta^2 + 2B_i\dd x^i\dd\eta +(\gamma_{ij} +
h_{ij})\dd x^i \dd x^j\right],
\end{eqnarray}
where the small quantities $A$, $B_i$ and $h_{ij}$ are unknown functions of space and time to be determined from the Einstein
equations. 

The strategy is then to write down the Einstein equations for this metric taking into account a perturbed stress-energy tensor. There are two important technical points in this program. We refer to Ref.~\cite{pertbardeen} and to the chapters~5 and~8 of Ref.~\cite{jpu-book} or Ref.~\cite{dubook} for details.
\begin{itemize}
\item First, one can perform a Scalar-Vector-Tensor (SVT) decomposition of the perturbations and, at linear order, the modes will decouple. This decomposition is a generalization of the fact that any vector field can be decomposed as the sum of the gradient of a scalar and a divergenceless vector as
\begin{equation}\label{5.vecteur}
 B^i=D^iB+{\bar B}^i, \qquad
  h_{ij} = 2C\gamma_{ij} + 2D_iD_jE+ 2D_{(i}{\bar E}_{j)}+ 2{\bar E}_{ij},
\end{equation}
with $D^i {\bar B}_i=0$ and  $D_i{\bar E}^{ij}={\bar E}^i_i=0$. 
\item Second, one has to carefully look at how the perturbation variables change under a gauge transformation, $x^\mu\rightarrow x^\mu-\xi^\mu$, where $\xi^\mu$ is decomposed into 2 scalar degrees of freedom and 2 vector degrees of freedom ($\bar L^i$, which is divergenceless $D_i\bar L^i=0$) as
\begin{equation}\label{5.TJAUGE2}
  \xi^0= T,\qquad
  \xi^i= L^i = D^i L + {\bar L}^i.
\end{equation}
Under this change of coordinates, the metric and the scalar field transform as
$$
 g_{\mu\nu}\rightarrow g_{\mu\nu}+\pounds_\xi g_{\mu\nu}, \qquad
 \varphi\rightarrow \varphi+\pounds_\xi \varphi.
$$
In order to construct variables which remain unchanged under a gauge transformation, we introduce gauge invariant variables that ``absorb'' the components of $\xi^\mu$. We thus get the 7 variables
\begin{eqnarray}
  \Psi \equiv -C-{\cal H}(B-E') \quad \Phi\equiv A+{\cal H}(B-E')+(B-E')', \quad {\bar\Phi}^i\equiv{\bar E}^{i\prime}-{\bar B}^i,
\end{eqnarray} 
and it is obvious that the tensor modes ${\bar E}^{ij}$ were already gauge invariant.\\

The stress-energy tensor of matter now takes the general form $\bar T_{\mu\nu}+\delta T_{\mu\nu}$ with
\begin{equation}\label{5.deltatmunu}
 \delta T_{\mu\nu}=(\delta\rho+\delta P)\bar u_\mu \bar u_\nu + \delta P \bar  g_{\mu\nu}
 +2(\rho +P) \bar u_{(\mu}\delta u_{\nu)} + P\delta g_{\mu\nu} + a^2 P \pi_{\mu\nu},
\end{equation}
where $u^\mu=\bar u^\mu+\delta u^\mu$ is the four-velocity of a comoving observer, satisfying $u_\mu u^\mu=-1$.  The normalisation
condition of $\bar u^\mu$ (to zeroth order) implies $\bar u^\mu=a^{-1}\delta^\mu_0$, $\bar u_\mu=-a\delta_\mu^0$, and since the norm of  $\bar u^\mu + \delta u^\mu$ should also be equal to $-1$, we infer that  $2\bar u^\mu\delta u_\mu+\delta g_{\mu\nu} \bar u^\mu\bar u^\nu=0$ and thus that  $\delta u_0=-Aa$. We then write $\delta u^i\equiv v^i/a$ so that $\delta u^\mu = a^{-1}(-A,v^i)$ and decompose $v_i$ into a scalar and a tensor part according to $v_i=D_i v+\bar v_i$. The anisotropic stress tensor $\pi_{\mu\nu}$ then consists only of a spatial part, which can be decomposed into scalar, vector and tensor parts as $\pi_{ij}=\Delta_{ij}\bar pi + D_{(i}\bar pi_{j)} + \bar pi_{ij}$, where the operator $\Delta_{ij}$ is defined as $\Delta_{ij}\equiv D_iD_j -\frac{1}{3}\gamma_{ij}\Delta$.

As for the metric perturbations, one can define gauge invariant quantities. Different choices are possible, and we define
\begin{eqnarray}
&\deltanewt = \delta + \frac{\rho'}{\rho}(B-E'),\qquad \deltaflat = \delta - \frac{\rho'}{\rho}\frac{C}{\Hconf}, \qquad \deltacom  = \delta + \frac{\rho'}{\rho}(v+B), \label{5.deltacom}\\
&V          = v + E', \qquad   \bar V_i      = \bar v_i + \bar B_i. \label{5.bV}
\end{eqnarray}
The pressure perturbations  are defined in an identical way. In all these relations, we recall that $\delta=\delta\rho/\rho$ is the density contrast.\\

The Einstein and conservation equations provide 3 sets of independent equations: one for the scalar variables ($\Phi,\Psi,\delta,V$), one for the two vector variables ($\bar\Phi_i$ and $\bar V_i$) and one for the tensor mode $\bar E_{ij}$. To be solved, this system of equations needs to be closed by a description of the matter, that is by specifying an equation of state for the pressure and anisotropic stress.

We provide these equations for a single fluid but they can easily be derived for a mixture of fluids with non-gravitational integrations (see Chap.~5 of Ref.~\cite{jpu-book}). 

\paragraph{Scalar modes}

First of all, th Einstein equations provide two constraints,
\begin{eqnarray}\label{5.einsteinscalaire1}
 \left(\Delta + 3K\right) \Psi = \frac{\kappa}{2}a^2\rho
 \deltacom,
  \qquad\Psi-\Phi = \kappa a^2 P\bar\pi.\label{5.einsteinscalaire2}
\end{eqnarray}
The first equation takes the classical form of the Poisson equation when expressed in terms of $\deltacom$. The second equation tells us that the two gravitational potentials are equal if the scalar component of the anisotropic stress tensor vanishes.

The two other equations are
\begin{eqnarray}
&&\Psi'+\Hconf\Phi=-\frac{\kappa}{2}a^2\rho(1+w) V,
 \label{5.einsteinscalaire3}\\
&&  \Psi''+3\Hconf(1+c_s^2)\Psi' + \left[2\Hconf'+(\Hconf^2-K)(1+3c_s^2)
 \right]\Psi
 -c_s^2\Delta\Psi =-9c_s^2\Hconf^2\left(\Hconf^2+K\right)\bar\pi\nonumber\\
 &&\qquad\qquad-\left(\Hconf^2+2\Hconf'+K\right)\left[\frac{1}{2}\Gamma +
 (3\Hconf^2+2\Hconf')\bar\pi+\Hconf\bar\pi'
 +\frac{1}{3}\Delta\bar\pi\right]\
 \label{5.einsteinscalaire4}
\end{eqnarray}

The conservation of the stress-energy tensor gives a continuity and an Euler equations,
\begin{eqnarray}\label{5.deltared}
 &&\left(\frac{\deltanewt}{1+w}\right)' = -\left(\Delta V
 -3\Psi'\right)
    -3\Hconf \frac{w}{1+w}\Gamma,\\
&&    V' +\Hconf(1-3c_s^2)V=-\Phi - \frac{c_s^2}{1+w}\deltanewt -\frac{w}{1+w}\left[\Gamma +
\frac{2}{3}(\Delta +3K)\bar\pi\right].
\end{eqnarray}
These 2 equations are indeed redundant thanks to the Bianchi identities.

\paragraph{Vector modes}

There are two Einstein equations for the vector modes,
\begin{eqnarray}
 &\left(\Delta +2K\right) \bar \Phi_i = -2\kappa\rho a^2(1+w)\bar V_i,
 \label{5.einsteinvecteur1} \\
 &\bar \Phi_i'+2\Hconf\bar \Phi_i=\kappa P a^2\bar \pi_i,\label{5.einsteinvecteur2}
\end{eqnarray}
and here is a unique conservation equation 
\begin{equation}\label{5.consvec}
\bar V_i' +\Hconf\left(1-3c_s^2\right)\bar V_i =
 -\frac{1}{2}\frac{w}{1+w}\left(\Delta+2K\right)\bar\pi_i.
\end{equation}
It is clear from these equations that as long as there is no anisotropic stress source, the scalar modes decay away, $\bar V_i\propto a^{1-3c_s^2}$, $\bar\Phi_i\propto a^{-2}$. We shall thus not consider them so far even though they need to be included when topological defects, vector fields or magnetic fields are present.

\paragraph{Tensor modes.} Their evolution is described by a single wave equation
\begin{equation}\label{5.einsteintenseur}
\bar E_{kl}''+2\Hconf\bar E_{kl}'+\left(2K-\Delta\right)\bar E_{kl}=
 \kappa a^2 P {\bar \pi}_{kl}.
\end{equation}

\paragraph{Fourier modes}

As a consequence of the Copernican principle, these equations only involve the spatial Laplacian so they can easily be solved in Fourier space where they reduce to a set of coupled second order differential equations. (as a counter-example one may compare to the case of a Bianchi~I universe in which modes do not decouple).

Two regimes need to be distinquished according to the value of the wave-number $k$. For  $k\ll\Hconf$ the mode is said to be super-Hubble while for  $k\gg\Hconf$, is is said to be sub-Hubble 

\subsubsection{Link to the observed universe}\label{secpert}

%%----------------------------------------------------
\begin{figure}[htb!]
\centering
\includegraphics[width=.41\columnwidth]{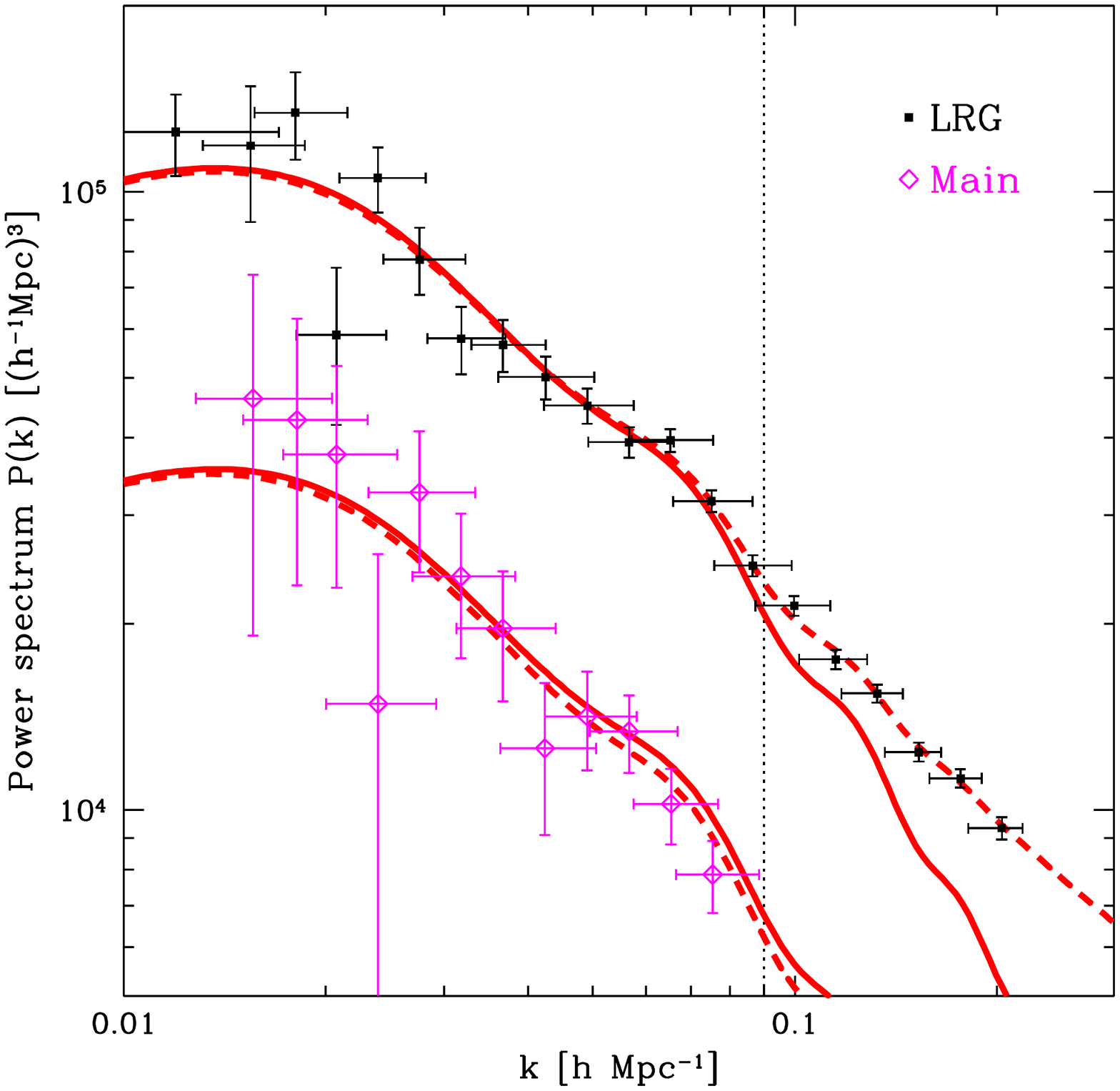}

\includegraphics[width=.41\columnwidth]{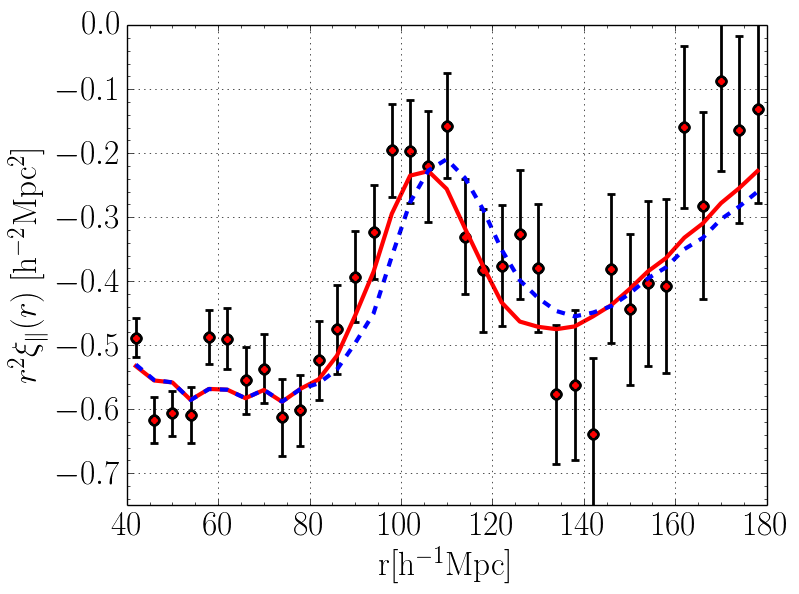}\includegraphics[width=.41\columnwidth]{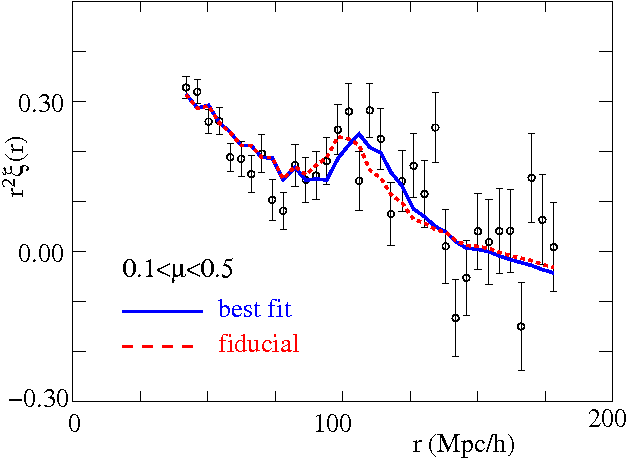}
 \caption{\footnotesize (left) Matter power spectra measured from the luminous red galaxy (LRG) sample and the main galaxy sample of the Sloan Digital Sky Survey (SDSS). Red solid lines indicate the predictions of the linear perturbation theory, while red dashed lines include nonlinear corrections. From Ref.~\cite{tegmark}. (right) Two-point correlation function for objects aligned with the line of sight (top), or orthogonal to the line of sight (bottom), measured with the BOSS quasars ($2.1\leq z\leq3.5$) and the integalactic medium traced by their Lyman-$\alpha$ forest, as function of comoving distance $r$. The effective redshift is $z=2.34$ here. From Ref.~\cite{debulac}.}
\label{fig-lss}
\end{figure}
%%----------------------------------------------------

These equations set the stage for all the analysis of the large scale structure. In simple (academic) cases, they can be solved analytically, which brings some insight on the growth of the large scale structure. Let us just point out that for a pressurlees fluid, neglecting $K$ and considering sub-Hubble modes, the scalar equations reduce to their Newtonian counterpart
\begin{equation}
\Delta\Phi = 4\pi G\rho a^2\delta,\qquad
\delta'=-\Delta V,\qquad
V'+\Hconf V = - \Phi
\end{equation}
where $\Psi=\Psi$ and the gauge can be forgotten. In this regime, we just get a close second order differential equations for $\delta$: $\delta''+2\Hconf\delta'-4\pi G\rho\delta=0$. It follows that the density perturbation can be split in terms of initial conditions and evolution as
\begin{equation}
 \delta({\bm k},a)=D_+({\bm k},a)\delta_+({\bm k},a_i) + D_-({\bm k},a)\delta_-({\bm k},a_i).
\end{equation}
It is then convenient to describe the evolution of the large scale structure in terms of a transfer function defined as
\begin{equation}
 \Phi(k,a) = T(k,a)\Phi(k,a_i).
\end{equation}
The observable modes today were initially super-Hubble and the theoretical models of the primordial universe (see below) allow us to predict the matter power spectrum of these modes, $P_\Phi(k,a_0)   = P_\Phi(k,a_i) T^2(k,a_0)$, and $P_\delta(k,a_0) = P_\delta(k,a_i) T^2(k,a_0)\left[\frac{D_+(a_0)}{D_+(a_i)}\right]^2$.\\

%%----------------------------------------------------
\begin{figure}[htb!]
\centering
\includegraphics[width=.32\columnwidth]{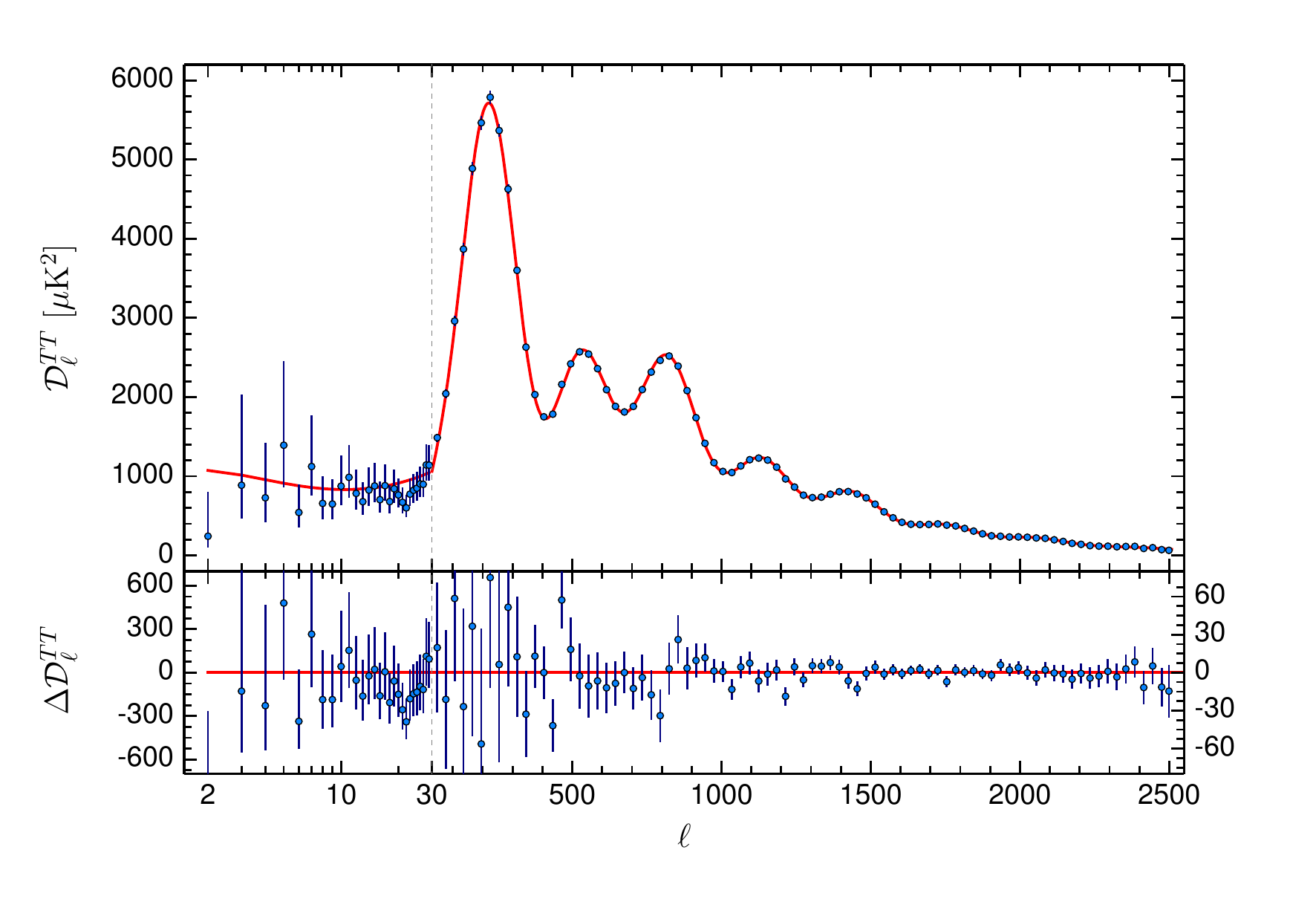}
\includegraphics[width=.31\columnwidth]{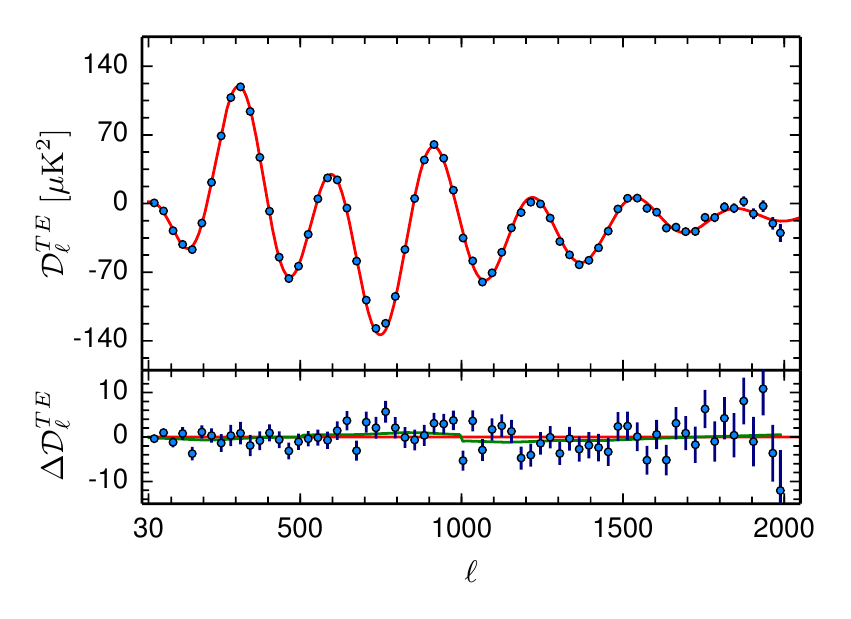}
\includegraphics[width=.32\columnwidth]{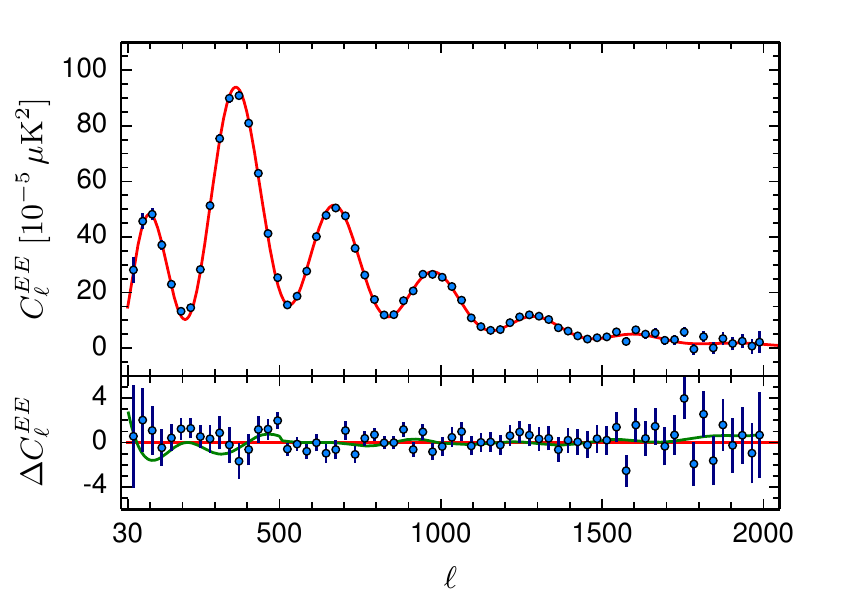}
 \caption{\footnotesize Angular power spectrum of the temperature (left), $E$-polarisation (right) anisotropies of the CMB, and their cross-correlation (middle) as measured by the Planck mission. From Ref.~\cite{planckpara}.}
 \label{fig-cmb}
\end{figure}
%%----------------------------------------------------

The general program to can then be summarized as followed. One needs to determine the {\em transfer function}. It will depend on the species that are assumed to exist in the universe, on their non-gravitational interaction (e.g. the Compton scattering between photons and electrons). It thus depends on the cosmological parameters as well on physical parameters (e.g. the mass of the neutrinos). Note that one needs to go beyond a fluid description to correctly describe the photon and neutrinos, for which a Boltzmann equation needs to be derived.

Then, one needs to specify the {\em initial conditions}, either on an ad hoc way or by specifying a model of the primordial universe, such as inflation. This will provide the linear power spectrum. Independently of any model of the early universe, one can use the observation of the cosmic microwave background to constrain the initial power spectrum. According to the recent study by Planck~\cite{planckpara}, it is well-described by an almost scale invariant spectrum (see the text by J.-L. Puget in this volume).

This provides the statistical properties of the distribution of the large scale structures. They can then be compared to the observed distribution. The developments of large scale surveys is a growing field. Fig.~\ref{fig-lss} depicts the matter power spectra measured from the luminous red galaxy (LRG) sample and the main galaxy sample of the Sloan Digital Sky Survey~\cite{tegmark}. In particular, it compares the predictions of the linear perturbation theory to  non-linear corrections. 

The comparison of the temperature anisotropies of the CMB to the distribution of the large scale structure confirms that the latter are formed through the gravitational collapse of initial density fluctuations of order of $10^{-5}$. It also shows that if any, modifications compared to Einstein gravity have to be small. The baryon acoustic oscillation, that explains the peak structure of  the CMB power spectrum corresponds to a rather large scale, which is weakly affected by the gravitational evolution of the universe between the epoch of recombination and today, contrary to small-scale inhomogeneities which tend to collapse and lose information about their initial conditions. As a consequence, this correlation within the distribution of baryonic matter in the universe has survived, only growing with cosmic expansion. They have been observed (see Fig.~\ref{fig-lss}) and demonstrate the consistency of this picture. Indeed, the comparison of these different observables allow one to sharpen the constraints on the transfer function and thus on the cosmological parameters.

One then needs to go beyond the linear regime. One can either work out the weakly non-linear regime perturbatively or resort on numerical simulation. The first approach can be performed either in a full-relativistic set-up or in the Newtonian regime while numerical simulations remain Newtonian. Both methods have their own limitations. In particular reaching sub-percent precision in small scales (typically $1-10 h$~Mpc) requires to control the validity of higher perturbation expansion. It was estimated~\cite{carlson} that the lowest mode $k$ at which linear perturbation theory deviates from a reference spectrum is $0.03$ at $z=0$ and reaches $0.09$ ate $z=1.5$ while with 1-loop (resp. 2-loop) correction is $0.08$ (resp. 0.04) at $z=0$ and $0.14$ (resp. $0.20$) at $z=1.5$. Various alternative techniques, e.g. based on the renormalisation group, have been developed in order to understand the effect of the small scales on the evolution of perturbations, hence leading to the idea that one may be able to construct an effective field theory of the large scale structure~\cite{eft}. Note also that most theoretical results are based on the idealized system of a collisionless cold matter fluid (described by point particles) that interacts only gravitationally. It is well-suited for cold dark matter, the mean free path of which is small but requires more attention when it comes to hot species, such as neutrinos which remain a challenge. And indeed, pressure has to be taken into account when one introduces baryons. We can conclude that we have a good understanding of the physcis at work but that high precision and realistic predictions still require progresses. 

\subsubsection{Dark matter and $\Lambda$}

The analysis of the large scale structure has led to a series of indications for the need of a dark sector composed of two components:
\begin{itemize}
 \item {\em dark matter} that can be assimilated to a non-interacting and non-baryonic dust component;
 \item{\em dark energy} that can be modeled by a smooth fluid with a negative pressure, the best candidate both from a phenomenological and observational point of view being the cosmological constant.
\end{itemize}
The evidence for the existence of dark matter can be traced back to the analysis by Zwicky of the dynamics of the Coma cluster~\cite{cdm1} in 1933, and then to the first rotation curves for the Andromeda nebula~\cite{cdm2} in 1939. The ideas that galaxies and clusters have a dark matter halo was not widely considered before the seventies, the change mostly trigerred by Refs.~\cite{cdm4,cdm5} and the arguments that massive halos are required to stabilize spiral galaxy disks~\cite{cdm6}. It was then followed by a formulation of the basis of a galaxy formation scenario~\cite{cdm7} in which a hierarchically merging population of dark matter halos provide the gravitational potential wells in which the intergalactic gas can cool and condense to form galaxies. In that model, the evolution of the distribution of the halo is described by the Press-Schechter analytic model~\cite{cdm11}. Note also that combining the growth rate of the perturbation, as obtained by Lifschitz~\cite{landau}, that shows that is scales as $a$, and the observational amplitude of density fluctuations from the CMB, $\delta\rho_{\rm b}/\rho_{\rm b}\sim10^{-5}$, the perturbation cannot become non-linear for a decoupling at $z\sim10^3$. This gives another indication for the need of dark matter, to form potential wells before decoupling. Many scenarios of structure formation were then investigated, mostly on the basis of numerical simulations, distinguishing between warm (i.e. relativistic) and cold (i.e. non-relativistic) dark matter. 

Among the potential candidates, a massive neutrino was the one to be known experimentally but it was shown not to offer a good cosmological solution. Today, it is important to remind that the need for dark matter in our cosmological model is one of the strong phenomenological arguments for the need to physics beyond the standard model. Many candidates such as axions or supersymmetric particles have been considered, and cosmology offers an important observational constraint on these models.

To summarize the proofs for the existence of dark matter comes from many observations: (1) the dynamics of gravitational bound objects (such as galaxies and clusters), (2) weak lensing effects, (3) the fact that baryonic matter interacts with radiation while dark matter does not (related to the amplitude of the peak structure of the CMB or the bullet cluster observations). Indeed several model tend to explain these observation by a modification of general relativity, usually in a low acceleration regime. It is important to say that these models also need to invoke new degrees of freedom that needs to be coupled to the standard matter. The main difference arise from the fact that in dark matter models, the energy density of the new field is an extra-source to the gravitational potential while in modified general relativity models, a new long-range interaction modifies the gravitational force. The current status is that none of these latter models can explain dark matter on all scales from galaxies to the CMB.\\

The observation of type~Ia supernovae (see Fig.~\ref{fig-hd}) shows that the cosmic expansion is accelerating today. From the Friedmann equations~(\ref{eq.fried2}) it can happen only if the matter content of the universe is dominated by a component such that $\rho+3P<0$, which has been named {\em dark energy}. The natural candidate is a cosmological constant, and no observation shows any deviation from this hypothesis (e.g. a variation with redshift of the equation of state of this component). The main problem arises from the amplitude of the associated energy density $\rho_\Lambda = \Lambda/8\pi G$ compared to what is inferred for the vacuum energy from quantum field theory in curved spacetime. This has split the approaches in two very different avenues. One the one hand, one can accept a cosmological constant and argue that the resolution of the cosmological constant lies in a multiverse hypothesis. On the other hand, one assumes the problem is solved (but no concrete way of the solution does actually exist) and invoke a new degrees of freedom. They can be either geometrical (i.e. by discussing that we have been fooled by assuming the validity of the FL geometry) or physical (i.e. new fields in the theory). In that latter case, as for dark matter, they can be associated to a modification of general relativity on large scale or not. It is also important to state that all of these models are phenomenological and of importance to discuss what  cosmological can detect or constrain but none of these models is on safe theoretical grounds. 

Today, we shall face that no deviation from general relativity+$\Lambda$ have been observationally detected and this is the best phenomenological model so far, hence that shall be considered as the reference model. Indeed, we know that general relativity needs to be extended to a quantum description, but it is hard to connect this to the required modification of general relativity to account for the acceleration of the universe, in terms of energy scales, and on the fact that it shall also solve the cosmological constant problem. See e.g. Refs.~\cite{jpu_cup,jpu-general relativityG,jpu-book} for further discussions and references.

\subsection{Modelling observations}

Most cosmological measurements rely, so far, on the observation of distant light sources, such as galaxies, supernovae, or quasars. Interpreting these observations requires to know the optical properties of the universe. In particular, the construction of a distance scale has been a long standing problem in astronomy.

\subsubsection{Light propagation and distances}

The interpretation of all cosmological observations relies on two equations that describe the propagation of a light ray and the evolution of the geodesic bundle, the first being more fundamental since it also applies in case of string lensing.

\paragraph{Fundamental equations} The first is the geodesic equation for the tangent vector, $k^\mu=\dd x^\mu/\dd\lambda$ to null geodesics $x^\mu(\lambda)$,
\begin{equation}
 k^\mu\nabla_\mu k^\nu=0,\qquad k^\mu k_\mu=0
\end{equation}
that derives from the Maxwell equation, $\nabla_\mu F^{\mu\nu}=0$ in the eikonal approximation. Its integration allows one to determine our past lightcone, that is to obtain the position of any object as its position on the sky $\lbrace \theta(\lambda),\varphi(\lambda)\rbrace$ and its redshift $z(\lambda)$, from which its distance needs to be determined, which means it is the byproduct of a model and not a direct observable.

In order to study object of finite size, one shall also describe the relative behavior of two neighboring geodesics, that is a bundle of geodesics, $x^\mu(\cdot,y^a)$ and $x^\mu(\cdot,y^a+\delta y^a)$, where $\xi^\mu =(\partial x^\mu/\partial y^a) \delta y^a$ is the connecting vector between two geodesics. If the origin $v=0$ of the affine parametrization of all rays is taken at $O$, then $k^\mu \xi_\mu = 0$ and then, the evolution of $\xi^\mu$ along the beam is governed by the geodesic deviation equation
\begin{equation}\label{GDE}
	k^\alpha k^\beta \nabla_\alpha \nabla_\beta \xi^\mu = {R^\mu}_{\nu\alpha\beta} k^\nu k^\alpha\xi^\beta,
\end{equation}
where ${R^\mu}_{\nu\alpha\beta}$ is the Riemann tensor.

Consider an observer, with four-velocity $u^\mu$ ($u_\mu u^\mu=-1$), who crosses the light beam. The tangent vecor to the beam can always be decomposed as $k^\mu =  -\omega (u^\mu + d^\mu)$ where
where $\omega =  u_\mu k^\mu$. The redshift $z$ in the observer's rest-frame  is thus given by
\begin{equation}\label{eq.a6}
	1+z = \frac{\nu_{\rm s}}{\nu_{\rm o}} 
	= \frac{u_{\rm s}^\mu k_\mu(v_{\rm s})}{u_{\rm o}^\mu k_\mu(v_{\rm o})} .
\end{equation}

These two equations allow to construct all observables. The difficult point lies in the fact that one needs to express them consistently in the cosmological framework, which in many cases is a difficult task.

\paragraph{Sachs formalism.} In order to measure shapes, the observer needs to define a screen (i.e. a 2-dimensional space on which the cross-section of the beam is projected) and spanned by 2 vectors $s_{A}^\mu$. Once projected on this basis, the geodesic equation reduces to the Sachs equation
\begin{equation}\label{eq:Sachs}
	\frac{\dd^2\xi_A}{\dd v^2} = \mathcal{R}_{AB} \, \xi^B,
\end{equation}
where $\xi_A = s_A^\mu \xi_\mu$ and $\mathcal{R}_{AB}={R}_{\mu\nu\alpha\beta}k^\nu k^\alpha s_A^\mu s_B^\beta$ are the screen-projected connecting vector  and Riemann tensor, usually called the {\em optical tidal matrix}. The properties of the Riemann tensor imply that this matrix is symmetric, $\mathcal{R}_{AB}=\mathcal{R}_{BA}$. Introducing the Jacobi matrix as
\begin{equation}
\xi^A(v) = {\cal D}{^A_B}(v) \dot{\xi}^B(0).
\end{equation}
that is as relating the physical separation~$\xi^A(v)$ of two neighbouring rays of a beam at $v$ to their observed separation~$\dot{\xi}^A(0)$, the Sachs equation takes the general form
\begin{equation}\label{jacobieq}
\ddot {\cal D}{^A_B}={\cal R}{^A_C}{\cal D}{^C_B},
\qquad
{\cal D}{^A_B}(0)=\delta^A_B,\quad
\dot{\cal D}{^A_B}(0)=0.
\end{equation}

\paragraph{Distances.} First of all, up to frequency factor $\omega_{\rm o}$ fixed to $1$ here, the determinant of the Jacobi matrix is related to the angular diameter distance as
\begin{equation}\label{eq.defDA}
D_{A}^2(v)=\frac{\text{area of the source at }v}{\text{observed angular size}}
= \det{\cal D}^A_B(v).
\end{equation}

The luminosity distance is then defined as
\begin{equation}\label{3.dlum_def}
 \phi_{\rm obs}=\frac{L_{\rm source}(\chi)}{4\pi D_{\rm L}^2}\ ,
\end{equation}
that relates the intrinsic luminosity of a source to the observed flux.

Interestingly, one can demonstrate a general theorem, that holds as long as the photon number is conserved, between the luminosity and angular distances, known as the distance duality relation
\begin{equation}
 D_L = (1+z)^2D_A(z).
\end{equation}
Such a relation can be tested observationally~\cite{jpu-duality} and allows one to rule out some dark energy models~\cite{refzal}.

\subsubsection{Background universe}

In a FL universe, these distances can be determined analytically and are given by Eq.~(\ref{eq.20}). This sets the basis of the interpretation of the Hubble diagram; see Fig.~\ref{fig-hd}. At small redshifts, it can be expanded to get
\begin{equation}
 D_{\rm A}(z)=  D_{{\rm H}_0}\left[1-\frac{1}{2}(q_0+3)z\right]z +{\mathcal O}(z^3)\ ,
\end{equation}
and 
\begin{equation}\label{3.374}
 D_{\rm L}(z)= D_{{\rm H}_0}\left[1-\frac{1}{2}(q_0-1)z\right]z + {\mathcal  O}(z^3)\ .
\end{equation}
Locally ($z\ll1$) all the distances reduce to $D_{{\rm H}_0}z$ and the deviations only depend on the value of the deceleration parameter. 

\subsubsection{Perturbation theory}

\paragraph{Generalities} When dealing with perturbation, one needs to solve the geodesic and Sachs equation at least at linear order in perturbation.

Each observable needs to be related to the expression of the gauge invariant perturbation variable. This requires to define clearly what is actually observed. At linear order, these computations are well-established but beyond linear order there are still some debates.

I just give a short overlook of two important examples: weak-lensing and the CMB anisotropies, in order to relate to the contributions by Yannick Mellier and Jean-Loup Puget. A full derivation can be obtained in chapters 6 and 7 of Ref.~\cite{jpu-book}.

\paragraph{Weak lensing} Usually, when dealing with weak lensing, treated as perturbation with respect to a FL spacetime, one uses that at background level both the shear and rotation vanish so that $\bar{\cal D}^C_B=\bar{D}_A\,\delta_B^C$. At linear order, it can be expanded to get the definition of the {\em amplification matrix} as
\begin{equation}
{\cal D}^A_B={\cal A}^A_C\bar{{\cal D}}^C_B + {\cal O}(2)
\end{equation}
with
\begin{equation}
{\cal A}^A_C
=
\begin{pmatrix}
1-\kappa-\gamma_1 & \gamma_2+\psi \\
\gamma_2-\psi& 1-\kappa+\gamma_1
\end{pmatrix},\label{eq:decomposition_amplification}
\end{equation}
where the {\em convergence} is defined as $\kappa= 1-\frac{1}{2}{\rm Tr}{\cal A}=\frac{D_{A}-\bar D_{A}}{\bar D_{A}} + \mathcal{O}(2)$. 

By expanding the Jacobi equation~(\ref{jacobieq}) to linear order, the amplification matrix in a cosmological space-time can thus be
expressed in terms of the gravitational potential as $\mathcal{A}_{ab}=\delta_{ab} -
\partial_{ab}\psi(\bm{\theta},\chi)$ with the lensing potential
\begin{equation}\label{7.81}
\psi(\bm{\theta},\chi) \equiv \frac{1}{c^2}
 \int_0^\chi
 \frac{f_K(\chi')f_K(\chi-\chi')}{f_K(\chi)}
 \left(\Phi+\Psi\right)[f_K(\chi')\bm{\theta},\chi']\dd\chi'.
\end{equation}
It follows that
\begin{eqnarray}
 \kappa(\bm{\theta},\chi)&=&\frac{1}{2}\left(\psi_{,11}+\psi_{22}\right),
                            \nonumber\\
 \gamma_1(\bm{\theta},\chi)&=&\frac{1}{2}\left(\psi_{,11}-\psi_{22}\right), \label{7.conv1}\\
 \gamma_2(\bm{\theta},\chi)&=&\psi_{,12}.
                            \nonumber
\end{eqnarray}
In order to be related to observation, one shall integrate over the source distribution, $p_z(z)\dd z=p_\chi(\chi)\dd\chi$, and the effective convergence is obtained by weighting the convergence (\ref{7.conv1}) with the source distribution
as
$$
\kappa(\bm{\theta}) = \int
 p_\chi(\chi)\kappa(\bm{\theta},\chi)\dd\chi.
$$
Splitting the perturbations as a transfer function, obtained by solving the perturbation equations, and initial conditions described in terms of an initial power spectrum, one can then work out the angular power spectrum of these observables (after a decomposition in spherical harmonics). It can be demonstrated that $P_\kappa=P_\gamma$. 

It follows that the key in the measurement of the cosmic shear relies on the measurement of the shape of background galaxies and on their ellipticity (see Refs.~\cite{7.bartel,7.mellier,7.witt} for details and problems). If the image of such a galaxy is represented by an ellipticity  $\bm{\varepsilon}=\varepsilon_1+i\varepsilon_2=(1-r)/(1+r)\exp(2i\phi)$,  $r=b/a$ being the ratio between the major and minor axes, then
$$
\left<\bm{\varepsilon}\right> =
\left<\frac{\bm{\gamma}}{1-\kappa}\right>.
$$
In the regime of weak distortion ($\kappa\ll1$), the ellipticities give access to the shear.

This techniques has witnessed tremendous progresses in the past 15 years. They are described in the contribution by Yannick Mellier. It is important to stress that weak lensing gives a way to directly measure the distribution of the gravitational potential. Combined with the observation of the matter distribution and of velocity fields, it gives a way to test the prediction of general relativity, as first pointed out in Ref.~\cite{ub2001} (see also Ref.~\cite{jpu-testRGcosmo} for a general review) and to reconstruct the distribution of dark matter.

\paragraph{CMB anisotropies} Any photon emitted during the decoupling and that is observed today has been redshifted due to the expansion of the universe but also because it propagates in an inhomogeneous spacetime. Besides, at emission, a denser region is hotter, the local velocity of the plasma is at the origin of a relative Doppler effect and the local gravitational potential induces an Einstein effect. The expression of the temperature fluctuation $\Theta=\delta T/T_{\rm CMB}$ obsrved here and today in a direction ${\bm e})$ can be derived by solving the perturbed geodesic equation~\cite{sachswolfe,panek} and leads to
\begin{eqnarray}\label{6.sachswolfe}
 \Theta({\bm x}_0,\eta_0,{\bm e}) &=& \left[\frac{1}{4}\deltanewt_\gamma
 + \Phi - e^i\left(D_i V_{\rm b} +\bar V_{{\rm b} i}+ \bar \Phi_i\right) \right]_{\rm
E},\bar\eta_{\rm E})\nonumber\\
&&\quad+ \int_{\rm E}^0
\left(\Phi'+\Psi'\right)[{\bm x}(\eta),\eta]\dd\eta
 - \int_{\rm E}^0e^i\bar \Phi'_i[{\bm x}(\eta),\eta]\dd\eta\nonumber\\
 &&\quad
 -\int_{\rm E}^0e^ie^j\bar E_{ij}'[{\bm x}(\eta),\eta]\dd \eta\ ,
\end{eqnarray}
where we integrate along the unperturbed geodesic that we chose to
parameterise as
\begin{equation}\label{6.geopara}
{\bm x} = {\bm x}_0 + {\bm e}(\eta_0-\eta)\ ,
\end{equation}
so that the integral now runs over the conformal time $\eta$.  This relation is known as the {\it Sachs-Wolfe equation}. It relates the observed temperature fluctuations to the cosmological perturbations. So $\Theta({\bm x}_0,\eta_0,{\bm e})$ is a stochastic variable that can be characterized by its angular correlation function, $C_\ell$ (see e.g. Refs~\cite{cmbtheory1} for early studies on the physcis of the CMB and Refs.~\cite{cmbtheory2} for more recent formalisms).

We shall not derive its expression, but the computation runs through decomposing all the perturbation in Fourier modes and then as a transfer function and initial conditions, the observables in spherical harmonics. The average then acts on the initial conditions so that $C_\ell$ depends directly on the initial power spectrum and on the transfer function. It follows that its computation requires to solve the perturbed equation prior to decoupling that is for the photon-baryon-electron plasma coupled through Compton scattering.

The description of radiation however requires to go beyond a fluid description and to rely on a Boltzmann equation, that needs to be derived at linear order in a gauge invariant form (see e.g. chapter~6 of Ref.~\cite{jpu-book} and references therein), but also beyond linear order~\cite{cmbcp}. This allows to determine, for any cosmological model, the angular correlation function of the temperature fluctuation, but also of the polarisation $E$- and $B$-modes, as well as the cross-correlation $TE$. Besides, the CMB photons are lens by the large scale structure of the universe, which offers a way to test the consistency of the whole model. This has recently been detected by the Planck experiment~\cite{plancklensing}.

\subsubsection{Beyond perturbation theory}\label{sectt}

As described above, the description of light propagation mostly relies on the cosmological perturbation theory~\cite{Futamase:1989hba}. At first order, it essentially introduces a dispersion of the distance-redshift relation with respect to the background FL prediction~\cite{Valageas:1999ch}, which can be partially corrected if a lensing map is known. This problem of determining the effect of inhomogeneities on light propagation has also been tackled in a non perturbative way, e.g. relying on toy models. The most common examples are Swiss-cheese models~\cite{SW1}. 

However, when it comes to narrow beams, such as those involved in supernova observations, the approximation of describing the cosmic matter by a fluid does no longer hold. The applicability of the perturbation theory in this regime, in particular, has been questioned in Ref.~\cite{Clarkson:2011br}. 

This specific issue of how the \emph{clumpiness} of the universe affects the interpretation of cosmological observables was first raised by Zel'dovich~\cite{Zeldo64} and Feynman~\cite{FeynmanColloquium}. The basic underlying idea is that in a clumpy medium, light mostly propagates through vacuum, and should therefore experience an underdense universe. This stimulated a corpus of seminal articles~\cite{DashevskiiZeldovich1965} to question why the use of a FL spacetime is a good approximation to interpret the Hubble diagram (see Fig.~\ref{fig-clump}. From a theoretical point it translates to the fact that in a FL universe, the Sachs equation is sourced only by the Ricci component (since the Weyl term vanishes) while in the true universe it is stress by the Weyl component (the Ricci term vanishing in vacuum).

%%----------------------------------------------------
\begin{figure}[h!]
\centering
\includegraphics[width=.53\columnwidth]{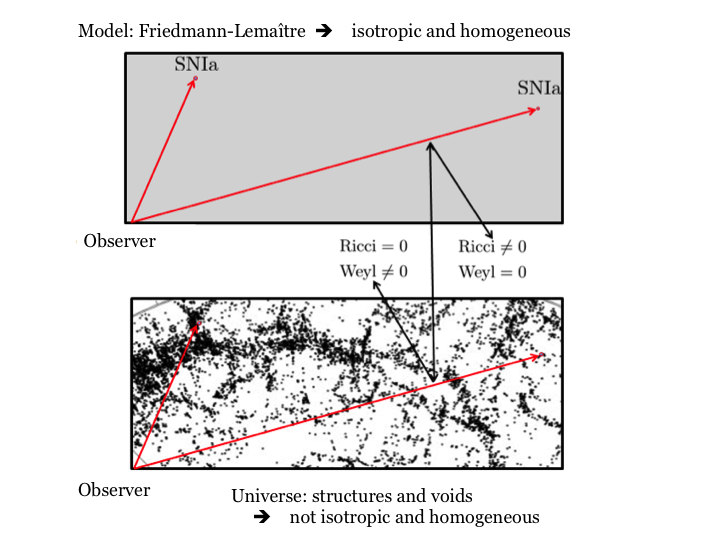}
\includegraphics[width=.39\columnwidth]{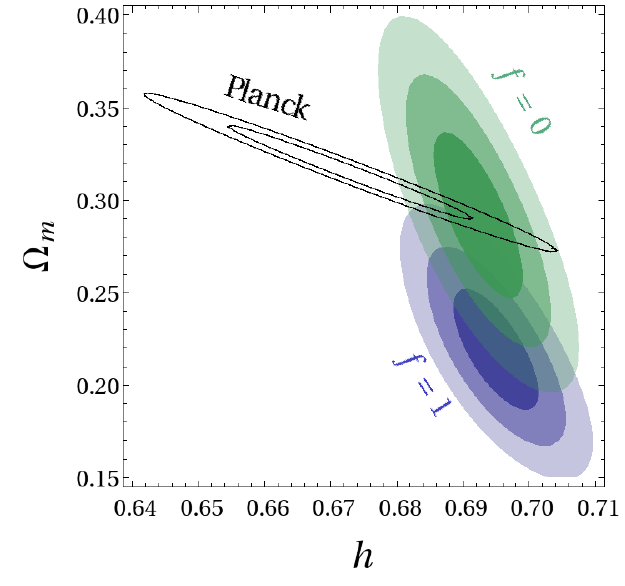}
 \caption{\footnotesize (left) The standard interpretation of SNe data assumes that light propagates in purely homogeneous and isotropic space~(top). However, thin light beams are expected to probe the inhomogeneous nature of the actual universe (bottom) down to a scale where the continuous limit is no longer valid. (right) Comparison of the constraints obtained by \textit{Planck} on  $(\Omega_{\rm m0}, H_0)$~\cite{planckpara} and from the analysis of the Hubble diagram constructed from the SNLS 3 catalog. The different contour plots correspond to different smoothness parameters $f$. For $f=1$, the geometry used to fit the data is  Friedmannian, as assumed in standard analysis. From Refs.~\cite{fleury1,fleury2}.} 
\label{fig-clump}
\end{figure}
%%----------------------------------------------------

When applied to the interpretation of SN data, two effects have to be described: (1) the fact that in average the beam probes an underdense media and (2) the fact that each line of sight is different so that one also expect a larger dispersion. The first aspect was described by the Dyer-Roeder approximation but  it does not tell anything about its dispersion, and a fortiori about its higher-order moments. The example of the analysis of Refs.~\cite{fleury1,fleury2} shows that taking into account the effects of the small scale structure improves the agreement between SN and CMB observations regarding the measurement of $\Omega_{\rm m}$ (see Fig.~\ref{fig-clump}).

Recently, a new approach was proposed~\cite{usnew} and offers an analytical and a priori non-perturbative framework for determining the statistical impact of small-scale structures on light propagation. Its main idea is that, on very small scales, the matter density field (i.e. the source of lensing) can be treated a \emph{white noise}, giving to lensing a diffusive behavior. The equations of geometric optics in curved spacetime then take the form of generalized Langevin equations, which come with the whole machinery of statistical physics. It then allows to write Fokker-Planck equation for the probability distribution function of the Jacobi matrix, which offers a very novel way to study the lensing in an inhomogeneous universe.

\subsubsection{Summary}

These examples show that relating observable to the perturbation variables and interpreting the effect of higher order correction consistently requires a deep theoretical insight. This steps cannot be avoided in order to determine how an observer in a given cosmological models sees his universe, which is what needs to be compared to the observations of our universe. Since all the predictions are from a statistical nature, one also needs to derive the variance of the observables.

This program is under control in a FL universe with linear perturbations, but many debates are open beyond linear order. One also needs to emphasize that as soon as the Copernican principle is not assumed, most of these computation, even at linear order, are difficult; see e.g. Ref.~\cite{usb1} for the theory of linear perturbation in a Bianchi universe and Ref.~\cite{usb2}  for the theory of weak-lensing.

\subsection{Summary: the $\Lambda$CDM model}\label{sec-back}

This description sets the basis of the standard cosmological model referred to as the $\Lambda$CDM model. In its minimal version it relies only on 6 cosmological parameters: 3 parameters ($\Omega_{\rm b}$, $\Omega_\Lambda$, $H_0$) to describe the evolution of the background universe, assumed to be a spatially Euclidean FL spacetime (so that $\Omega_K=0$, $\Omega_{\rm cdm}=1-\Omega_{\rm b}-\Omega_\Lambda$ and $\Omega_{\rm r}$ being given by the mean temperature of the CMB and neutrinos), 2 parameters to describe the initial perturbations (amplitude and index of the power spectrum) and 1 parameter to describe the reionization.

Such a model is an excellent phenomenological model since it is in agreement with all observations. Deviations from it have been constrained, see e.g. Ref.~\cite{planckpara}. Dark matter strengthen the connection, that was already established by BBN, with particle physics.

This description draws however many questions. First, the initial power spectrum is purely ad-hoc and a further step in the construction of our model is to find its origin in the dynamics of the primordial universe. The dark energy and dark matter components requires to identify the degrees of freedom to which they are associated. This is a very active part of research at the crossroad between particle physics and cosmology. The lithium problem still exhibits an inconsistency that requires attention. Last but not least, the description of the efficiency of assuming that the universe is well-described by a FL spacetime requires also some attention. Indeed, one has replaced the spacetime metric tensor by an average metric of large scale --see Eq.~(\ref{eq.fl})-- and that the matter stress-energy tensor has been smoothed to solve the Einstein equation. Since the Einstein equations are not linear, the Einstein tensor of the smooth metric differs from the smooth Einstein tensor. One shall then question the use of the Friedmann equations and try to define the smoothing procedure. This remains an open question~\cite{backreac}.

From a technical point of view, the use of perturbation theory to describe the large scale structure requires attention, in particular on its precision on small scales (see e.g. Ref.~\cite{carlson}) where linear perturbation theory cannot be used. The question of the resummation of a higher perturbation theory is under question, as well a the way to properly treat the physics (inclusion of baryons, neutrinos,...). The common approach is to use this techniques together with N-body numerical simulations, which have their own limitations, such as the way to incorporate baryons and neutrinos, and also all relativistic effects; see e.g. Ref.~\cite{refruth}. It is difficult to go below a 1\% precision, which becomes a limitation for the precision cosmology program.

Besides, even the interpretation of some data, such as those related to thin beams, may require some caution and it is not clear that one can use the same metric to interpret all the observation, specially on small scale where the fluid limit does not hold~\cite{fleury2}.

%%%%%%%%%%%%%%%%%%%%%%%%%%%%%%%%%%%%%
\section{The primordial universe}\label{section3}

\subsection{The inflationary paradigm}

{\em Inflation} is defined as a primordial phase of accelerated of accelerated expansion, which has to last long enough for the standard problems of the hot big-bang model to be solved. 

As explained above, it was initially proposed as a tentative solution to the problems of the hot big-bang model. In particular, it provides a simple explanation for the homogeneity and flatness of our universe.

Before the first models of inflation were proposed, precursor works appeared as early as 1965 by Erast Gliner~\cite{gliner}, who postulated a phase of exponential expansion. In 1978, Fran\c{c}ois Englert, Robert Brout and Egard Gunzig~\cite{EBG}, in an attempt to resolve the primordial singularity problem and to introduce the particles and the entropy contained in the universe, proposed a `fireball' hypothesis, whereby the universe itself would appear through a quantum effect in a state of negative pressure subject to a phase of exponential expansion. Alexei Starobinsky~\cite{staro}, in 1979, used quantum-gravity ideas to formulate the first semi-realistic rigorous model of an inflation era, although he did not aim to solve the cosmological problems. A simpler model, with transparent physical motivations, was then proposed by Alan Guth~\cite{guth} in 1981. This model, now called `old inflation', was the first to use inflation as a mean of solving cosmological problems. It was soon followed by Andrei Linde's `new inflation' proposal~\cite{lindenewinf}.

From the Einstein equations, an accelerated expansion requires that the matter dominating the dynamics of the universe satisfies a relation between its energy density and pressure, $\rho+3P<0$. A solution to implement this condition is to use a scalar field $\varphi$. As long as it is slow-rolling, it satisfies $P\sim-\rho$ hence leading to acceleration, while if it is fast-rolling it satisfies $P\sim+\rho$. As long as this scalar field is a minimally coupled canonical field, the only freedom is the choice of its potential $V(\varphi)$. 

Early models, known as {\em old inflation}, rely on a first-order phase-transition mechanism~\cite{oldinf}; see Fig.~\ref{fig2}. A scalar field is trapped in a local minimum of its potential, thus imposing a constant energy density, equivalent to the contribution of a cosmological constant. As long as the field remains in this configuration, the evolution of the universe is exponential and the universe can be described by a de Sitter spacetime. This configuration is metastable and the field can tunnel to its global minimum, $V(\varphi_f) = 0$, hence creating bubbles of true vacuum, which correspond to a non-inflationary universe. In models of {\em new inflation}~\cite{lindenewinf}, the scalar field exits its false vacuum by slowly rolling towards its true vacuum; see Fig.~\ref{fig2}. These models can only work if the potential has a very flat plateau around $\varphi=0$, which is artificial. In most of its versions, the inflaton cannot be in thermal equilibrium with other matter fields. The theory of cosmological phase transitions then does not apply and no satisfying realization of this model has been proposed so that it was progressively abandoned. The first models of inflation (see Ref.~\cite{revuelinde} for a review) were actually only incomplete modifications of the big-bang theory as they assumed that the universe was in a state of thermal equilibrium, homogeneous on large enough scales before the period of inflation. This problem was resolved by Linde with the proposition of chaotic inflation~\cite{lindechaos}. In this model, inflation can start from a Planckian density even if the universe is not in equilibrium. A new picture of the universe then appears. The homogeneity and isotropy of our observable universe would be only local properties, while the universe is very inhomogeneous on very large scales, with a fractal-like structure.

\begin{figure}[t!]
\begin{center}
\resizebox{12cm}{!}{\includegraphics[clip=true]{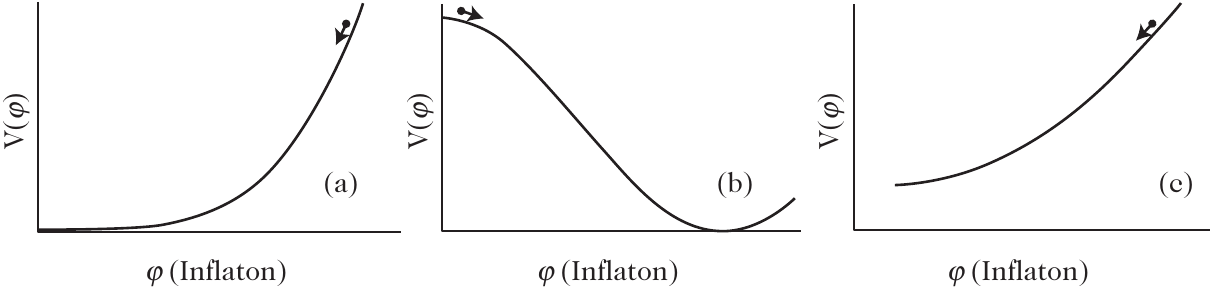}}
\end{center}
\vskip-0.5cm
 \caption{\footnotesize (left) The model of old inflation (left) is based on a first order phase transition from a local to a true minimum of the potential. The false vacuum is metastable and the field can tunnel to the true vacuum. In the models of new inflation (right) a scalar field slowly relaxes towards its vacuum. From Ref.~\cite{jpu-book}.
 (right) data analysis.}
\label{fig2}
\end{figure}

Following the study of quantum effects near black-holes and in de Sitter spacetimes, quantum effects were investigated during inflation (see Ref.~\cite{Lev} for an historical prospect). Two approaches were followed. The first originating in the 60s~\cite{dsschool}, uses the static form of the de Sitter metric so that an observer at the origin would detect thermal radiation from $R=1/H$ with a temperature $T=H/2\pi$, which corresponds to vacuum polarization of the de Sitter geometry. It is often used in the superstring community, in particular in the context of holography and the thermodynamics associated with horizons and the so-called ``hot tin can" picture \cite{susskind03}. The second approach is based on quantization of a scalar field in a time-dependent background described by an (almost) de Sitter spacetime. This led to the formulation by Viatcheslav Mukhanov and Gennady Chibisov of the theory of cosmological perturbations during inflation~\cite{chibi,mfb}, which links the origin of the large scale structure of the universe to quantum fluctuations, quickly followed by a series of works~\cite{timedep,otherQ}. The slow-roll phase is essential in this mechanism as it is during this period that the density fluctuations which lead to the currently observed large scale structure, are generated.

At the end of inflation, all classical inhomogeneities have been exponentially washed out, and one can consider them as non-existent at this stage. The universe has become very flat so that curvature terms can be neglected. Moreover, all the entropy has been diluted. If the inflaton potential has a minimum, the scalar field will oscillate around this minimum right after the end of inflation. Due to the Hubble expansion, these oscillations are damped and the scalar field decays into a large number of particles. During this phase, the inflationary universe (of low entropy and dominated by the coherent oscillations of the inflaton) becomes a hot universe (of high entropy and dominated by radiation). This reheating phase~\cite{reheating} connects inflation with the hot big-bang scenario and complete the picture. In principle, knowing the couplings of the inflaton to the standard matter field, one can determine the relative amount of all matter species and their distribution. This step is however very challenging theoretically and requires heavy numerical simulations.

\subsection{Early motivations: the problems of the hot big-bang model}

The different problems of the standard cosmological model are discussed in Chapters~3 and~5 of Ref.~\cite{jpu-book}. Let me start by the flatness and horizon problems in order to show how inflation can solve it. To that purpose, let me rewrite the equations of evolution in terms of the dimensionless parameter
$$
\Omega_K=-\frac{K}{a^2H^2}
$$
as
\begin{equation}\label{fdyn}
 \frac{\dd\ln\Omega_K}{\dd\ln a}=(1+3w)(1-\Omega_K)\Omega_K.
\end{equation}
It shows that $\Omega_K$ is a stable point of the dynamics which is not stable if $1+3w>0$. It means that with standard pressureless matter and radiation, the curvature term will tend to dominate the Friedmann equation at late time. But today, the spatial curvature of our universe is small $|\Omega_{K0}-1|<0.1$ (with the upper bound taken very generously). It implies that at the time of matter-radiation equality $|\Omega_{K}-1|<3\times10^{-5}$ and at Planck time $|\Omega_{K}-1|<10^{-60}$. The big-bang model does not give any explanation for such a small curvature at the beginning of the universe and this fine tuning is unnatural for an old universe like ours.

The cosmological principle, which imposes space to be homogeneous and isotropic, is at the heart of the FL solutions. By construction, these models cannot explain the origin of this homogeneity and isotropy. So it would be more satisfying to find a justification of this principle, at least on observable scales.  A simple way to grasp the problem is to estimate the number of initial cells, with an initial characteristic Planck size length, present today in the observable universe. This number is typically of order
$$
N\sim\left(\frac{1+z_p}{\ell_P H_0}\right)^3\sim10^{87}.
$$
The study of the CMB and of galaxies tends to show that their distribution is homogeneous on larger scales, so it is difficult to understand how initial conditions fixed on $10^{87}$ causally independent regions can appear so identical (at a $10^{-5}$ level!). This horizon problem is related to the state of thermodynamic equilibrium in which the universe is. The cosmological principle imposes a non-causal initial condition on spatial sections of the universe and in particular that the temperature of the thermal bath is the same at every point. The horizon problem is thus closely related to the cosmological principle and is therefore deeply rooted in the FL solutions.

\subsection{Inflation}

\subsubsection{The idea of inflation}

%%----------------------------------------------------
\begin{figure}[h!]
\centering
\begin{tabular}{ccc}
\includegraphics[width=.5\columnwidth]{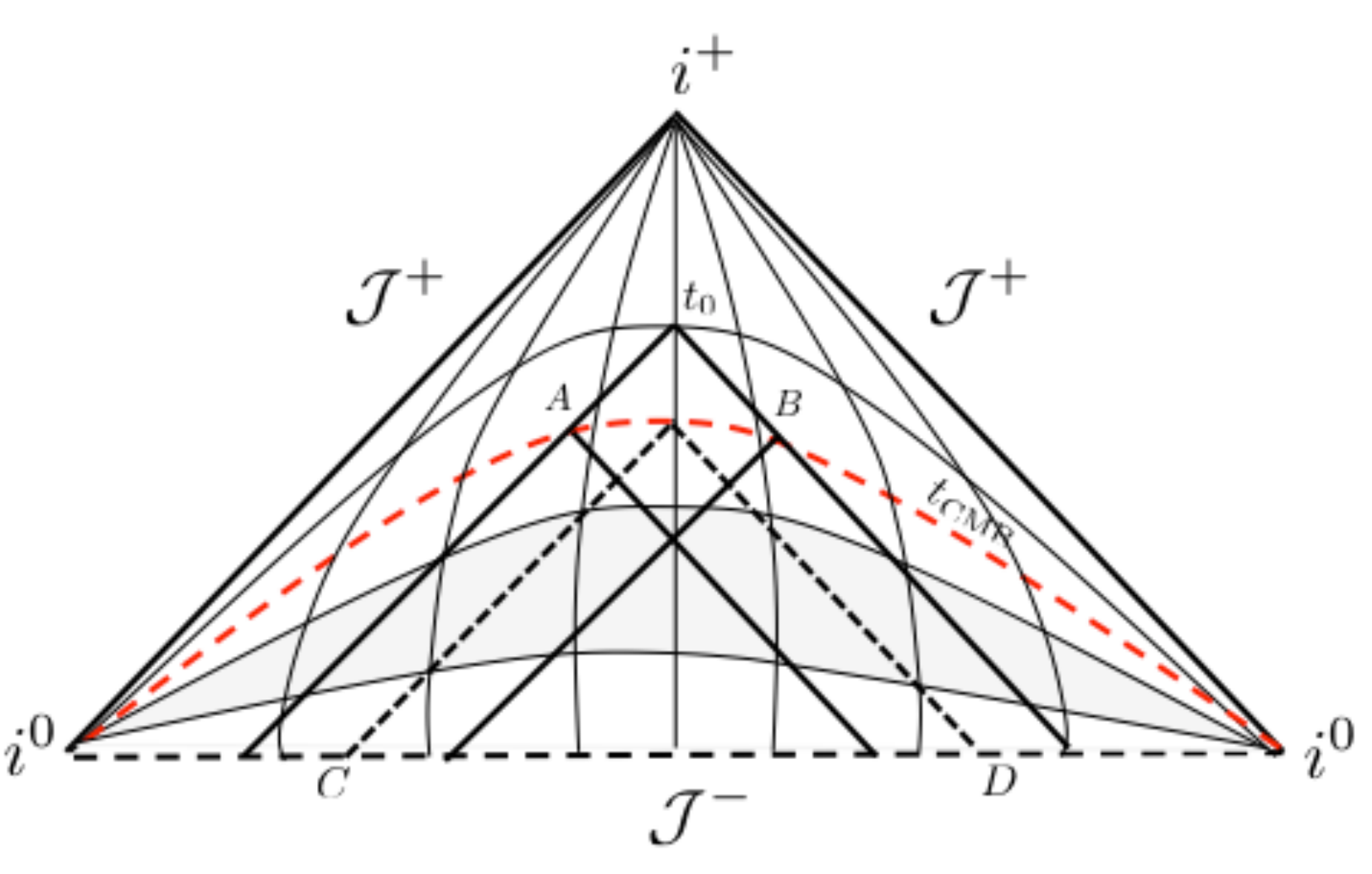}
\end{tabular}
 \caption{\footnotesize Penrose diagram of a universe with an intermediate stage of inflation (and no cosmological constant). It explains how such a period allow to solve the horizon problem. From Ref.~\cite{eu_cras}.} 
\label{fig-penrose1}
\end{figure}
%%----------------------------------------------------

Let us see how an phase of accelerated expansion can solve the flatness problem. Since the definition of $\Omega_K$ implies that $\Omega_K=-K/\dot a^2$, it is clear that it will (in absolute value) decrease to zero if $\dot a$ increases that is if $\ddot a>0$, i.e. during a phase of accelerated expansion. As discussed above, this requires $\rho+3P<0$ and cannot be achieved with ordinary matter. This can also be seen on the Friedmann equations under the form~(\ref{fdyn}). They clearly show that when $-1<w<-\frac{1}{3}$, the fixed point $\Omega_K=0$ is an attractor of the dynamics. 

So, the flatness problem can be resolved if the attraction toward  $\Omega_K=0$ during the period of inflation is sufficient to compensate its subsequent drift away from 0 during the hot big-bang, i.e. if inflation has lasted sufficiently long. To quantify the duration of the inflationary period, we define the quantity
\begin{equation}\label{9.defefold}
 N \equiv \ln\left(\frac{a_\mathrm{f}}{a_\mathrm{i}}\right)\ ,
\end{equation}
where $a_{\rm i}$ and $a_{\rm f}$ are the values of the scale factor at the beginning and at the end of inflation. This number measures the growth in the scale factor during the accelerating phase and is called the {\it number of ``$e$-folds''}. To give an estimate of the required minimum number of $e$-folds of inflation, note that, if we assume $H$ to be almost constant during inflation, then
$$
\left|\frac{\Omega_K(t_\mathrm{f})}{\Omega_K(t_\mathrm{i})}\right|
= \left(\frac{a_\mathrm{f}}{a_\mathrm{i}}\right)^{-2} =
\hbox{e}^{-2N}\ .
$$
In order to have $|\Omega_K(t_\mathrm{f})|\lesssim10^{-60}$ and
$\Omega_K(t_\mathrm{i})\sim\mathcal{O}(1)$, we thus need
\begin{equation}
 N \gtrsim 70\ .
\end{equation}

Similarly, in a accelerated universe, the comoving Hubble radius, ${\cal H}^{-1}=(aH)^{-1}$, decreases in time
\begin{equation}
 \frac{\dd}{\dd t}(aH)^{-1} < 0\ .
\end{equation}
Two points in causal contact at the beginning of inflation can thus be separated by a distance larger than the Hubble radius at the end of inflation. These points are indeed  still causally connected, but can {\it seem} to be causally disconnected if the inflationary period is omitted. So, inflation allows for the entire {\it observable} universe to emerge out of the same causal region before the onset of inflation. The horizon problem can also be solved if $N\gtrsim 70$.

The Penrose diagram of a universe with a finite number of $e$-folds is depicted on Fig.~\ref{fig-penrose1}.

\subsubsection{Dynamics of single-field inflationary models}

Most models of inflation rely on the introduction of a dynamical scalar field $\varphi$,  evolving in a potential, $V(\varphi)$ with an action given by
\begin{equation}
 S=-\int\sqrt{-g}\left[\frac{1}{2}\partial_\mu\varphi\partial^\mu\varphi
 + V(\varphi)\right]\dd^4 x\ ,
\end{equation}
so that its energy-momentum tensor takes the form
\begin{equation}\label{9.21}
 T_{\mu\nu}= \partial_\mu\varphi\partial_\nu\varphi  -\left(\frac{1}{2} \partial_\alpha\varphi\partial^\alpha\varphi +
 V\right)g_{\mu\nu}\ .
\end{equation}
It follows that the energy density and pressure of a homogenous scalar field are respectively given by
\begin{equation}\label{9.densphicosm}
\rho_\varphi=\frac{\dot\varphi^{2}}{2} + V(\varphi)\ , \qquad P_\varphi=\frac{\dot\varphi^2}{2} - V(\varphi)\ .
\end{equation}
These expressions show that $\rho_\varphi+ 3P_\varphi = 2(\dot\varphi^2-V)$. Since the Friedmann equations imply that the expansion is accelerated as soon as $\dot\varphi^2< V$. The expansion will be quasi-exponential if the scalar field is in slow-roll, i.e. if $\dot\varphi^2\ll V$. This clearly explains why this is a natural way of implementing inflation.

%%----------------------------------------------------
\begin{figure}[h!]
\centering
\includegraphics[width=.8\columnwidth]{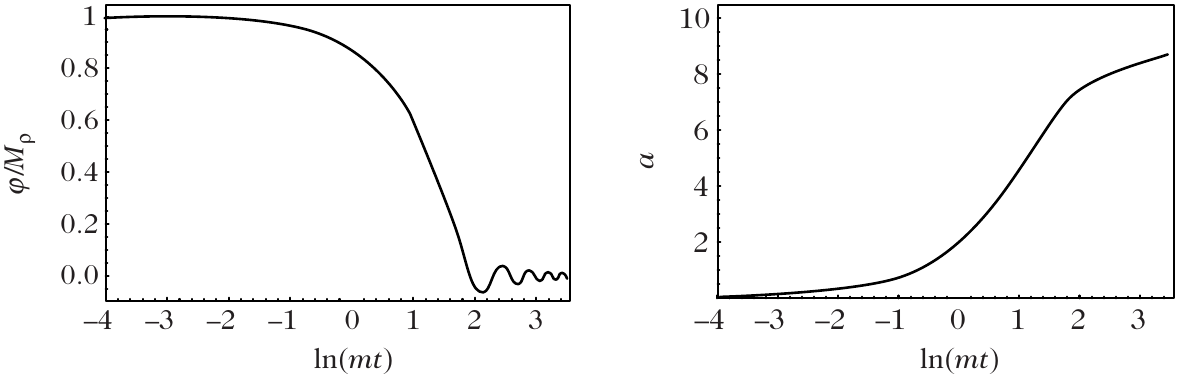}
 \caption{\footnotesize Evolution of the inflaton and scale factor during inflation with a potential~(\ref{9.potchaotic}). The scalar field is initially in a slow-roll regime and the expansion of the universe is accelerated. At the end of this regime, it starts oscillating at the minimum of its potential and it is equivalent to a pressureless fluid. From Ref.~\cite{jpu-book}.} 
\label{fig3}
\end{figure}
%%----------------------------------------------------

The Friedmann and Klein-Gordon equations take the form 
\begin{equation}
  H^2 = \frac{8\pi G}{3}\left(\frac{1}{2}\dot\varphi^{2} + V \right) -\frac{K}{a^2}\ ,
  \qquad
 \frac{\ddot a }{a} = \frac{8\pi G}{3}\left(V-\dot\varphi^{2}  \right)\ ,
 \qquad
 \ddot\varphi + 3H\dot\varphi + V_{,\varphi}=0\,.
\end{equation} 
Once the potential is chosen, the whole dynamics can be determined. As an example consider a free massive scalar
\begin{equation}\label{9.potchaotic}
 V(\varphi) = \frac{1}{2}m^2\varphi^2\ .
\end{equation}
The Klein-Gordon equation reduces to that of a damped harmonic oscillator. If $\varphi$ is initially large, then the Friedmann equation implies that $H$ is also very large. The friction term becomes important and dominates the dynamics so that the field must be in the slow-roll regime. The evolution equations then reduce to 
$$
 3H\dot\varphi + m^2\varphi = 0\ ,\qquad
 H^2 = \frac{4\pi}{3}\left(\frac{m}{M_p}\right)^2\varphi^2\ ,
$$
that  give
\begin{equation}
 \varphi(t) = \varphi_{\rm i} -\frac{mM_p}{\sqrt{12\pi}}t\ ,
 \qquad
 a(t) = a_{\rm i}\exp\left\{\frac{2\pi}{M_p^2} \left[ \varphi_{\rm i}^2 - \varphi^2(t)
 \right]\right\}\ ,
\end{equation}
where $\varphi_{\rm{i}}$ and $a_{\rm{i}}$ are the values of the field and the scale factor at $t_\mathrm{i}=0$ and $M_p$ is the Planck mass.  This can be compared to the numerical integration depicted in Fig.~\ref{fig3}, in which we see the subsequent (non-slow-rolling) phase with the damped oscillations at the bottom of the potential and the fact that $\ddot a$ changes sign to become negative. Figure~\ref{fig4} provides a phase space analysis of the dynamics, showing that the slow-roll trajectories are attractors of the dynamics. It follows that if inflation lasts long enough, the initial conditions on $(\varphi,\dot\varphi)$ become irrelevant.

%%----------------------------------------------------
\begin{figure}[h!]
\centering
\includegraphics[width=.4\columnwidth]{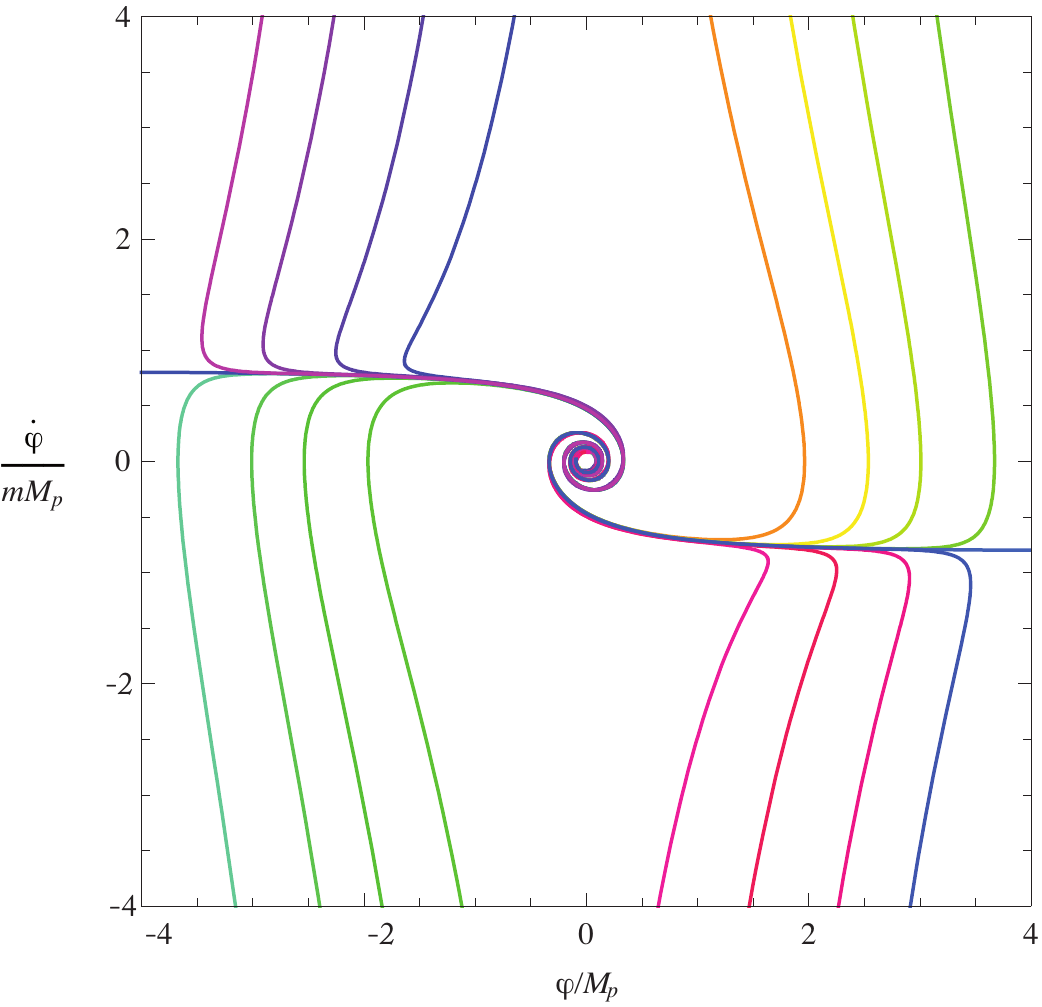}
 \caption{\footnotesize Phase portrait in the $(\varphi,\dot\varphi)$-plane of the dynamics of a scalar field with potential $V = m^2\varphi^2/2$, assuming $m=10^{-6}M_p$. This illustrates the mechanism of attraction toward the slow-roll solution. From Ref.~\cite{jpu-book}.} 
\label{fig4}
\end{figure}
%%----------------------------------------------------

\subsubsection{Slow-roll formalism}

As seen of the example of a massive scalar field, inflation occurs while the scalar field is slow-rolling. This has led to the development of a perturbative formalism to describe the dynamics. If the field is slow-rolling then $H$ is almost constant and it is convenient to define
\begin{equation}
\varepsilon = -\frac{\dot H}{H^2}\ ,\qquad
\delta =\varepsilon - \frac{\dot\varepsilon}{2H\varepsilon}\ ,\qquad
\xi = \frac{\dot\varepsilon-\dot\delta}{H}\,.
\end{equation}
These definitions depend only on the spacetime geometry. In the case of a single scalar field, they can be rewritten as
\begin{equation}
 \varepsilon = \frac{3}{2}\dot\varphi^2\left[\frac{1}{2}\dot\varphi^2 + V(\varphi) \right]^{-1}\ ,\qquad
 \delta = -\frac{\ddot\varphi}{H\dot\varphi}\ .
 \end{equation}
$\varepsilon$ can be used to rewrite the Friedmann equations as
\begin{equation}
 H^2\left(1-\frac{1}{3}\varepsilon\right) = \frac{8\pi G}{3} V\ ,\qquad
 \frac{\ddot a}{a} = H^2\left(1-\varepsilon\right)\ ,
\end{equation}
and the effective equation of state of the inflaton as $w_\varphi = -1 +\frac{2}{3}\varepsilon$ so that he condition for inflation reduces 
\begin{equation}
\ddot a>0 \Longleftrightarrow w<-1/3 \Longleftrightarrow
 \varepsilon <1\ .
\end{equation}

The number of $e$-folds between a time $t$ where the value of the inflaton is $\varphi$ and the end of inflation ($t=t_{\rm f}$ and $\varphi=\varphi_{\rm f}$, of the inflationary phase can be expressed as
\begin{equation}
 N(t,t_{\rm f}) = \int_t^{t_{\rm f}} H\dd t\ ,
\end{equation}
since, after integration, $a(t) = a_{\rm f}\, \hbox{e}^{-N}$. $N(t_{\rm i},t_{\rm f})$ corresponds to the duration of the inflationary phase, as defined earlier. $N$ can be expressed as
\begin{equation}\label{9.efold2}
  N(\varphi, \varphi_{\rm f})=
 \int_\varphi^{\varphi_{\rm f}} \frac{H}{\dot\varphi}\dd\varphi =
 -\sqrt{4\pi
 G}\int_\varphi^{\varphi_{\rm f}}\frac{\dd\varphi}{\sqrt{\varepsilon}}\ .
\end{equation}

As long as the slow-roll parameters are small, $H$ is almost constant. One can then develop the equations for the evolution of the background ($a,H,\varphi$) in terms of this small parameters. This allows to derive the observational predictions in term of these parameters and, interestingly they can be related to derivative of the inflationary potential~\cite{jpu-book,lidsey} as
$$
\varepsilon =  \frac{1}{16\pi G}\left(\frac{V_{,\varphi}}{V}\right)^2\ ,\qquad
 \delta = \frac{1}{8\pi G}\left(\frac{V_{,\varphi\varphi}}{V}\right)-\varepsilon.
$$

\subsubsection{Chaotic inflation}

In order to illustrate this formalism, let us come back to the massive scalar field. It is clear that $\varepsilon = \frac{M_p^2}{4\pi\varphi^2}$ and $\delta = 0$, so that the slow-roll regime lasts until $\varphi$ reaches $\varphi_{\rm f} = M_p/\sqrt{4\pi}$.  We infer that the total number of $e$-folds is
\begin{equation}
 N(\varphi_{\rm i}) =
 2\pi\left(\frac{\varphi_{\rm i}}{M_p}\right)^2 -\frac{1}{2}\ .
\end{equation}
In order to have $N\gtrsim70$, we need $\varphi_{\rm i}\gtrsim3M_p$. If $\varphi$ takes the largest possible value compatible with the classical description adopted here, i.e. $V(\varphi_{\rm i})\lesssim M_p^4$, we find that $\varphi_{\rm i}\sim M_p^2/m$. In this case, the maximal accessible number of $e$-folds would be $N_{\rm max}\sim2\pi M_p^2/m^2$. As can been deduced from the observations of the CMB impose that $m\sim10^{-6}M_p$, so that $N_{\rm max}\sim 10^{13}$. The maximal number of $e$-folds is thus very large compared to the minimum required for solving the cosmological problems.

Consequently, if the universe is initially composed of regions where the values of the scalar field are randomly distributed then the domains where the initial value of $\varphi$ is too small never inflate or only for a small number of $e$-folds.  The main contribution to the total physical volume of the universe at the end of inflation comes from regions that have inflated for a long time and that had an initially large value of $\varphi$. These domains produce extremely flat and homogeneous zones at the end of inflation with a very large size compared to that of the observable universe.

This is the idea of {\em chaotic inflation} proposed in Ref.~\cite{lindechaos} and which predicts, if we are typical observers, that we shall not be surprised to observe an extremely Euclidean, homogeneous and isotropic universe. This conclusion has to be contrasted with the early version of the big-bang model and to the discussion on its problems.

\subsubsection{End of the inflationary phases}

Inflation ends when ${\rm max}(\varepsilon,\delta)\sim1$. At the end of inflation, all classical inhomogeneities have been exponentially washed out, and one can consider them as non-existent at this stage (still the fact that we neglect them from the start of the analysis has motivated many works to carefully understand the homogeneization ans isotropization of the universe, both at the background and perturbative levels; see e.g. Ref.~\cite{Keffect} for a discussion of the curvature and Ref.~\cite{anisotropy} for a discussion on the isotropization and their effects on the maximum number of $e$-folds) so that one can set $K = 0$ during all the late stages of the primordial phase. 

During inflation, all the energy is concentrated in the inflaton. Shortly after the end of inflation, the universe is cold and ``frozen'' in a state of low entropy where the field oscillates around the minimum of its potential. The coherent oscillations of the inflaton can be considered as a collection of independent scalar particles. If they couple to other particles, the inflaton can decay perturbatively to produce light particles. The interaction of the inflaton should therefore give rise to an effective decay rate, $\Gamma_\varphi$, and reheating would only occur after the expansion rate had decreased to a value $H\sim\Gamma_\varphi$. This also implies that during the first ${\rm m}hcal{O}(m/\Gamma_\varphi)$ oscillations of the inflaton, nothing happens.

To illustrate this consider an inflaton with potential $V\propto\varphi^n$. During the oscillatory phase, $H<m$ and the inflaton undergoes several oscillations during a time $H^{-1}$. It is thus reasonable to use the mean value of the pressure and density over several oscillations.  It follows that $\langle\dot\varphi^2/2\rangle = (n/2)\langle V(\varphi)\rangle$, so that $\langle P_\varphi\rangle= (n-2)/(n+2)\langle \rho_\varphi\rangle$. The scalar field thus behaves as a dust fluid for $n=2$ and as a radiation fluid for $n=4$.

In order for the inflaton to decay, it should be coupled to other fields. Including quantum corrections, the Klein-Gordon equation becomes
\begin{equation}
 \ddot\varphi + 3H\dot\varphi + \left[m^2 + \Pi(m)\right]\varphi = 0\ ,
\end{equation}
where $\Pi(m)$ is the polarisation operator of the inflaton. The real part of $\Pi(m)$ corresponds to the mass correction. $\Pi$ has an imaginary part $\hbox{Im}[\Pi(m)] = m\Gamma_\varphi$. Since $m$ is much larger than both $\Gamma_\varphi$ and $H$ at the end of inflation, we can solve the Klein-Gordon equation by assuming that both these quantities are constant during an oscillation.  It follows that
\begin{equation}\label{9.PhiSol}
 \varphi = \Phi(t)\sin m t\ ,\qquad
 \Phi = \varphi_0\exp\left[-\frac{1}{2}\int(3H+\Gamma_\varphi)\dd
 t\right]\ .
\end{equation}
As long as $3H>\Gamma_\varphi$, the decrease in the inflaton energy caused by the expansion (Hubble friction) dominates over particle decay. Thus,
\begin{equation}
 \Phi = \varphi_{\rm f}\frac{t_{\rm f}}{t} =
 \frac{ M_p}{m}\frac{1}{\sqrt{3\pi }t}
\end{equation}
where we have used $\varphi_{\rm f}=  M_p/\sqrt{4\pi}$, $t_{\rm f}=2/3H_{\rm f}$ and $H_{\rm f}^2=(4\pi/3)(m/ M_p)^2\varphi_{\rm f}^2$. Reheating occurs in the regime $\Gamma_\varphi\gtrsim 3H$ and
\begin{equation}
 \Phi = \frac{ M_p}{m}\frac{1}{\sqrt{3\pi t}}
 \hbox{e}^{-\Gamma_\varphi t/2}\ .
\end{equation}
We define the time of reheating, $t_{\rm reh}$, by $\Gamma_\varphi = 3H$ so that the energy density at that time is $\rho_{\rm reh} = \frac{\Gamma_\varphi^2 M_p^2}{24\pi}$. If this energy is rapidly converted into radiation, its temperature is $\rho_{\rm reh} = \frac{\pi^2}{30}g_* T_{\rm reh}^4$. This defines the reheating temperature ias
\begin{equation}\label{9.Treh}
 T_{\rm reh} = \left(\frac{5}{4\pi^3
 g_*}\right)^{1/4}\sqrt{\Gamma_\varphi  M_p}
 \simeq 0.14\left(\frac{100}{g_*}\right)^{1/4}\sqrt{\Gamma_\varphi  M_p}
 \ll
 10^{15}\,{\rm GeV}\ ,
\end{equation}
where the upper bound was obtained from the constraint $\Gamma_\varphi \ll m\sim10^{-6} M_p$, assuming $g_*\gtrsim100$ for the effective number of relativistic particles. 

This description of perturbative reheating is simple and intuitive in many aspects. However, the decay of the inflaton can start much earlier in a phase of {\it preheating} (parametric reheating) where particles are produced by parametric resonance. The preheating process can be decomposed into three stages: (1) non perturbative production of particles, (2) perturbative stage and (3) thermalization of the produced particles. We refer to Ref.~\cite{reheating} for details on this stage that connects inflation to the standard hot big-bang model.

\subsection{A scenario for the origin of the large scale structure}

The previous section has described a homogeneous scalar field. As any matter field, $\varphi$ has quantum fluctuations and cannot be considered as strictly homogeneous. This drives us to study the effects of these perturbations. Any fluctuation of the scalar field will generate metric perturbations since they are coupled by the Einstein field equations. We should thus study the coupled inflaton-gravity system to understand. 

In terms of the variables to consider, we start from 10 perturbations for the metric and 1 for the scalar field, to which we have to subtract 4 gauge freedoms and 4 constraint equations (2 scalars and 2 vectors). We thus expect to identify 3 independent degrees of freedom to describe the full dynamics: one scalar mode and a tensor mode (counting for 2 degrees of freedom, one per polarization).

The following presentation relies on the extended description of Ref.~\cite{jpu-book} (chapter 8).

\subsubsection{Perturbation theory during inflation}

Starting from the perturbation of the scalar field as $\varphi = \varphi(t) + \delta\varphi({\bm x},t)$, its stress-energy tensor takes the form
\begin{eqnarray}
 \delta T_{\mu\nu} &=&
 2\partial_{(\nu}\varphi\partial_{\mu)}\delta\varphi
 -\left(\frac{1}{2}g^{\alpha\beta}\partial_\alpha\varphi\partial_\beta\varphi
 +V\right)\delta g_{\mu\nu}\nonumber\\ &&\qquad
 -g_{\mu\nu}\left(\frac{1}{2}\delta g^{\alpha\beta}
 \partial_\alpha\varphi\partial_\beta\varphi
 +g^{\alpha\beta}\partial_\alpha\delta\varphi\partial_\beta\varphi +
 V'\delta\varphi \right)\ .
\end{eqnarray}
One can then use the standard theory of gauge invariant cosmological perturbation described above. To that purpose, one needs introduce the two following gauge invariant variables for the scalar field
\begin{eqnarray}
   \chi = \delta\varphi + \varphi'(B-E')\ ,\quad\hbox{or}\qquad
   Q = \delta\varphi - \varphi'\frac{C}{\cal H}\ ,
\end{eqnarray}
They are related by $Q = \chi + \varphi'{\Psi}/{\cal H}$. and $\chi$ is the perturbation of the scalar field in Newtonian gauge while $Q$, often called the Mukhanov-Sasaki variable is the one in flat slicing gauge.
\end{itemize}

Forgetting about the technicalities of the derivation, one ends up with the following equations of evolution. We were left with $7=2+2+3$ degrees of freedom respectively fo the T, V and S modes. For simplicity we assume $K=0$.
\begin{itemize}
\item{\it Vector modes}. Since a scalar field does not contain any vector sources, there are no vector equations associated with the Klein-Gordon equation and there is only one Einstein equation for these modes and a constraint equation,
$$
 \Delta {\bar\Phi}_i = 0 \ .
$$
We can conclude that ${\bar\Phi}_i=0$ independently of any model of inflation, that the vector modes are completely absent at the end of inflation. The 2 constraints equation kill the vector degrees of freedom.
\item{\it Tensor modes}. There is only one tensor equation which can be obtained from the Einstein equations
\begin{equation}\label{9.GWevo}
 \bar E_{ij}'' + 2{\cal H}{\bar E}_{ij}'-\Delta{\bar E}_{ij} = 0\ .
\end{equation}
It describes the evolution of the two polarisations of a gravity wave as a damped harmonic oscillator. Shifting to Fourier space and decomposing the gravity waves on a polarization tensor as ${\bar E}_{ij}({\bm k},,\eta)=\sum_\lambda{\bar E}_\lambda({\bm k},\eta)
 \varepsilon_{ij}^\lambda({\bm k})$ and defining
 $$
  \mu_\lambda({\bm k},\eta) =\sqrt{\frac{M_p^2}{8\pi}} a(\eta){\bar E}_\lambda({\bm k},\eta)
 $$
the equation of evolution takes the form
\begin{equation}\label{eq.mu}
 \mu''_\lambda + \left(k^2-\frac{a''}{a}\right)\mu_\lambda=0
\end{equation}
for each of the two polarizations.
\item{\it Scalar modes}. The Einstein equations provide 2 constraint equations and an equation of evolution. This implies that we shall be able to derive a master equation for the only propagating degree of freedom. This can be achieved by defining~\cite{mfb,jpu-book}
$$
 v({\bm k},\eta)= a(\eta)Q({\bm k},\eta), \qquad z(\eta)= \frac{a\varphi'}{{\cal H}}
$$
that can be shown to satisfy 
\begin{equation}\label{eq.v}
 v'' + \left(k^2-\frac{z''}{z}\right)v=0.
\end{equation}
Bote that $v$ is a composite variable of the field and metric perturbations. As expected from general relativity, it is not possible to separate them.
\end{itemize}
The set of equations~(\ref{eq.mu}-\ref{eq.v}) for the tensor and scalar modes are second order differential equations in Fourier space. In order for them to be predictive, one needs to fix their initial conditions. It is striking that these equations look like Schr\"odinger equations with a time dependent mass.

\subsubsection{Setting the initial conditions}

\begin{figure}[h!]
\centering
\includegraphics[width=.8\columnwidth]{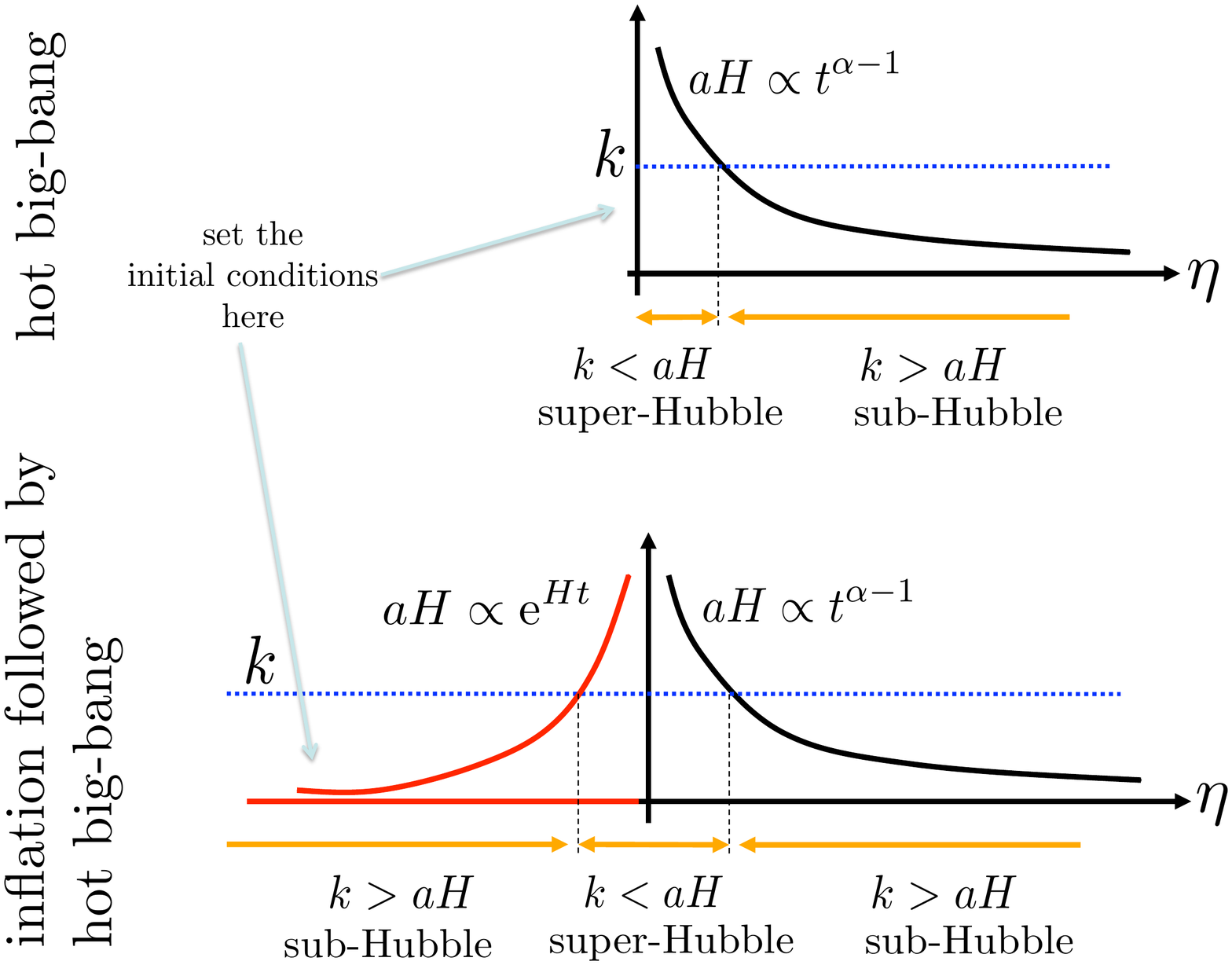}
 \caption{\footnotesize Evolution of the comoving Hubble radius and of a comoving mode of wavenumber $k$ with conformal time $\eta$ in the standard hot big-bang model (top) and with an inflationary phase (bottom). Without inflation, the mode is always super-Hubble in the past and becomes sub-Hubble as the universe expands. This implies that initial conditions have to be set on super-Hubble scales. The existence of super-Hubble correlation are thus thought to be acausal. With an inflationary phase, the mode was sub-Hubble deep in the inflationary era, which means that initial conditions have to be set on sub-Hubble scales.} 
\label{fig5}
\end{figure}

Let us first compare the evolution of a mode with comoving wave-number $k$ in the standard big-bang model and during inflation. Any equation of evolution can be shown to have two regime whether the mode is super-Hubble ($k<{\cal H}$) or sub-Hubble ($k>{\cal H}$); see Fig.~\ref{fig5}.

%%----------------------------------------------------
%%----------------------------------------------------

In the post-inflationary era, the expansion is decelerating so that ${\cal H}$ is a decreasing function of $\eta$ while $k$ remains constant (since it is comoving). It follows that the super-Hubble modes become sub-Hubble while the universe expands. All observable modes are sub-Hubble today and one needs to set initial conditions in the early universe while they were super-Hubble. Actually, asymptotically toward the big-bang all modes where super-Hubble. A difficulty arises from the observation of the CMB and large scale structure that show that there shall exist super-Hubble correlations. It seems unnatural to set correlated initial conditions on super-Hubble scales since it would appear as acausal. Besides, there was no natural procedure to fix the initial conditions so that they were completely free and reconstructed to get an agreement with the observations.

During inflation, the picture is different because the expansion is accelerated, which means that ${\cal H}$ is increasing with $\eta$ while $k$ remains constant. It follows that any super-Hubble modes was sub-Hubble deep in the inflationary era. If inflation lasts long enough then all the observable modes can be sub-Hubble where $v$ and $\mu_\lambda$ behave as harmonic oscillator. We shall thus find a mechanism to set initial conditions in this regime since we are interested only in modes that are observable today, at least as far as confronting inflation to observation is concerned.\\

The idea of Ref.~\cite{chibi} was to take seriously the fact that $Q$, and thus $v$, enjoys quantum fluctuations. They demonstrated that when expanded at second order in perturbation the Einstein Hilbert + scalar field action reduces (see Refs.~\cite{mfb,anisotropy,dubook}) to
\begin{equation}\label{9.q1}
 \delta^{(2)}S=\frac{1}{2}\int\dd\eta\dd^3{\bm x}\left[ \left(v'\right)^2
 - \delta^{ij}\partial_i v\partial_jv +
 \frac{z''}{z}v^2\right]\equiv\int{\mathcal L}\dd^4x\ ,
\end{equation}
up to terms involving a total derivative and that do not contribute to the equations of motion. The variation of this action indeed gives the equation~(\ref{eq.v}) but it tells us more on the structure of the theory: we recognize the action of a canonical scalar field with a time-dependent mass, $m^2=-z''/z$ in Minkowski spacetime. It was thus proposed that the variable to quantize is $v$ and we shall quantize it as we quantize any canonical scalar field evolving in an time-dependent exterior field~\cite{9.grib,mwbook}, where here the time-dependence would find its origin in the spacetime dynamics~\cite{9.bd}. 

The procedure then goes as follows
\begin{enumerate}
\item  $v$ is promoted to the status of quantum operator in second quantization in Heisenberg representation
\begin{equation}\label{9.DEC}
 \hat v({\bm k},\eta) = \int\frac{\dd^3{\bm k}}{(2\pi)^{3/2}}\left[
 v_k(\eta)\hbox{e}^{i{\bm k}\cdot{\bm x}}\hat a_{\bm k} +
 v_k^*(\eta)\hbox{e}^{-i{\bm k}\cdot{\bm x}}\hat a_{\bm k}^\dag
 \right]\ ,
\end{equation}
where $\hat a_{\bm k}$ and $\hat a_{\bm k}^\dag$ are creation and annihilation operators.
\item One introduces he conjugate momentum of $v$,
\begin{equation}
 \pi = \frac{\delta{\cal{L}}}{\delta v'} = v'\ ,
\end{equation}
which is also promoted to the status of operator,  $\hat\pi$.
\item One can then get the Hamiltonian
\begin{equation}\label{9.HHH}
 H = \int\left(v'\pi-\cal{L}\right)\dd^4x =
     \frac{1}{2}
     \int\left(\pi^2 + \delta^{ij}\partial_{i}v\partial_jv -
 \frac{z''}{z}v^2\right)\dd^4x\ .
\end{equation}
The equation of evolution for $\hat v$ is indeed Eq.~(\ref{eq.v}) which is equivalent to the Heisenberg equations  $\hat v' = i \left[\hat H,\hat v\right]$ and $\hat \pi' = i \left[\hat H,\hat \pi\right]$.
\item The operators $\hat v$ and $\hat\pi$ have to satisfy canonical commutation relations on constant time hypersurfaces
\begin{equation}\label{9.comm}
 [\hat v({\bm x},\eta),\hat v({\bm y},\eta)] =  [\hat \pi({\bm x},\eta),\hat
 \pi({\bm y},\eta)]= 0\ ,\quad
 [\hat v({\bm x},\eta),\hat \pi({\bm y},\eta)] = i\delta({\bm x}-{\bm y})\ .
\end{equation}
\item As for quantization in Minkowski spacetime, the creation and annihilation operators appearing in the decomposition~(\ref{9.DEC}) satisfy the standard commutation rules
\begin{equation}
 [\hat a_{\bm k},\hat a_{\bm p}] =  [\hat a_{\bm k}^\dag,\hat a_{\bm p}^\dag]= 0\ ,\qquad
 [\hat a_{\bm k},\hat a_{\bm p}^\dag] = \delta({\bm k}-{\bm p})\ .
\end{equation}
They are consistent with the commutation rules~(\ref{9.comm}) only if  $v_k$ is normalized
according to
\begin{equation}\label{9.wronsk}
 W(k)\equiv v_k {v'}^{*}_k - v^*_k v'_k =i\ ,
\end{equation}
This determines the normalisation of their Wronskian, $W$.
\item The vacuum state $\left|0\right\rangle$ is then defined by the condition that it is annihilated by all the operators
$\hat a_{\bm k}$, $\forall\,{\bm k}$, $\hat a_{\bm k}\left|0\right\rangle = 0$. The sub-Hubble modes, i.e. the high-frequency modes compared to the expansion of the universe, must behave as in a flat spacetime. Thus, one decides to pick up the solution that corresponds adiabatically to the usual Minkowski vacuum
\begin{equation}\label{9.condinitv}
 v_k(\eta) \rightarrow
 \frac{1}{\sqrt{2k}}\hbox{e}^{-ik\eta}\ ,\qquad
 k\eta\rightarrow-\infty\ .
\end{equation}
This choice is called the {\em Bunch-Davies vacuum}. Let us note that in an expanding spacetime, the notion of time is fixed by the background evolution which provides a preferred direction. The notion of positive and negative frequency is not time invariant, which implies that during the evolution positive frequencies will be generated.
\end{enumerate}

Note that this quantization procedure amounts to treating gravity in a quantum way at linear order since the gravitational potential is promoted to the status of operator. 

It can be seen that it completely fixes the initial conditions, i.e. the two free functions of integration that appear when solving Eq.~(\ref{eq.v}). This can be seen on the simple example of a de Sitter phase ($H=$~constant, $a(\eta)=-1/H\eta$ with a test scalar field. Then $z''/z=2/\eta$ and the equation~(\ref{eq.v}) simplifies to $v_k'' + \left(k^2 -\frac{2}{\eta^2}\right)v_k=0$ for which the solutions are given by
\begin{equation}
 v_k(\eta) = A(k)\hbox{e}^{-ik\eta}\left(1
 +\frac{1}{ik\eta}\right) + B(k)\hbox{e}^{ik\eta}\left(1
 -\frac{1}{ik\eta}\right).
\end{equation}
The condition~(\ref{9.condinitv}) sets $B(k)=0$ and fixes the solution to be (for $Q=v/a=H\eta v$)
\begin{equation}\label{9.chisol}
 Q_k = \frac{H\eta}{\sqrt{2k}}\left(1 +\frac{1}{ik\eta}\right)\hbox{e}^{-ik\eta}\ .
\end{equation}
Notice that on sub-Hubble $Q_k\propto 1/\sqrt{k}$ as imposed by quantum mechanics but when the mode becomes super-Hubble ($k\eta\ll 1$) it shifts to $Q_k\propto 1/\sqrt{k^3}$, which corresponds to a scale invariant power spectrum. Hence the scale invariance on super-Hubble scales is inherited from the small scale properties fixed by quantum mechanics and no tuning is at work.

On super-Hubble scales, the field $Q$ acquires a constant amplitude  $\left|Q_k\right|(k\eta\ll1) = \frac{H}{\sqrt{2k^3}}$. In this limit
$$
 \hat Q \rightarrow \int\frac{\dd^3{\bm k}}{(2\pi)^{3/2}}\,
 \hat \chi_{\bm k}\,\hbox{e}^{i{\bm k}\cdot{\bm x}}=
 \int\frac{\dd^3{\bm k}}{(2\pi)^{3/2}}\,
 \frac{H}{\sqrt{2k^3}}
 \left( \hat a_{\bm k}+ \hat a_{-{\bm k}}^\dag\right)\,\hbox{e}^{i{\bm k}.{\bm x}}\
$$
All the modes being proportional to $( \hat a_{\bm k}+ \hat a_{-{\bm k}}^\dag)$, $\hat Q$ has the same statistical properties as a Gaussian classical stochastic field and its power spectrum defined by
\begin{equation}
 \langle Q_{\bm k} Q_{{\bm k}'}^* \rangle = P_Q(k)\, \delta^{(3)}({\bm k}-{\bm k}')\ ,
\end{equation}
from which one easily deduces that
\begin{equation}\label{e.spec}
 P_Q(k) = \frac{2\pi^2}{k^3} {{\cal P}}_Q(k) = |Q_k|^2=\frac{|v_k|^2}{a^2}\ .
\end{equation}
It is thus a scale invariant power spectrum since ${\cal{P}}_Q(k) = \left(\frac{H}{2\pi}\right)^2$.

\subsection{Generic predictions and status}

The analysis described in the previous section applies in any model of inflation and the conclusions can be computed analytically in the slow-roll regime. The departure from a pure de Sitter phase reflects itself in a spectral index in Eq.~(\ref{e.spec}). They also apply identically for the gravity waves, the only difference being that $v$ and $\mu_\lambda$ have different ``mass terms'' and that they are related to $Q$ and ${\bar E}_\lambda$ respectively by $z$ and $a$. It implies that they will have different spectral indices and amplitudes.

%%----------------------------------------------------
\begin{figure}[h!]
\centering
\includegraphics[width=.8\columnwidth]{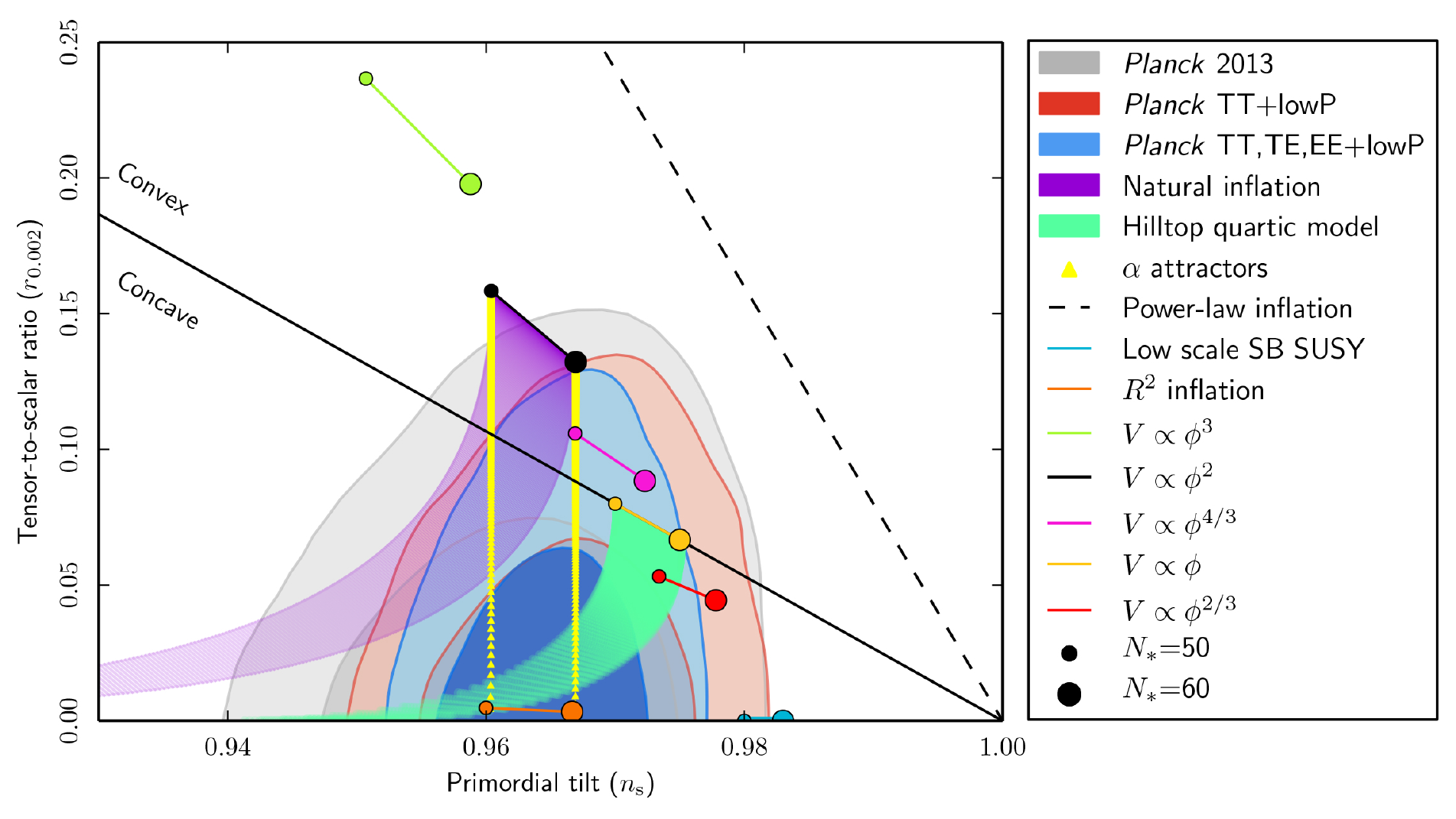}
 \caption{\footnotesize  Constraints on inflationary models from Planck: marginalized joint 68\% and 95\% confidence level regions for the spectral index $n_s$ and scalar-to-tensor ratio $r_{0.002}$ from Planck in combination with other data sets, compared to the theoretical predictions of selected inflationary models. From Ref.~\cite{planckinflation}.} 
\label{fig_infl}
\end{figure}
%%----------------------------------------------------

in conclusion, single field inflationary models have robust predictions that are independent of their specific implementation.
\begin{itemize}
 \item{The observable universe must be homogeneous and isotropic}. Inflation erases any classical inhomogeneities. The observable universe can thus be described by a FL spacetime.
 \item{The universe must be spatially Euclidean}. During inflation, the curvature of the universe is exponentially suppressed. Inflation therefore predicts that $\Omega_K=0$ up to the amplitude of the super-Hubble density perturbations, i.e. up to about $10^{-5}$.
 \item{Scalar perturbations are generated}. Inflation finds the origin of the density perturbations in the quantum fluctuations of  the inflaton which are amplified and redshifted to macroscopic scales. All the modes corresponding to observable scales today are super-Hubble at the end of inflation. Inflation predicts that these perturbations are adiabatic, have a Gaussian statistics and have an almost scale-invariant power spectrum. The spectral index can vary slightly with wavelength. These perturbations are coherent, which is translated into a structure of  acoustic peaks in the angular power spectrum of the  CMB's temperature anisotropies.
 \item{There are no vector perturbations}. 
 \item{Gravitational waves are generated}. In the same way as for scalar modes, the gravitational waves have a quantum origin and are  produced through parametric amplification. They also have Gaussian statistics and an almost scale invariant power spectrum. 
 \item There is a  consistency relation between their spectral index and the relative amplitude of the scalar and tensor modes.
  \item The quantum fluctuations of any light field  ($M<H$) are amplified. This field develops super-Hubble  fluctuations of amplitude $H/2\pi$.
\end{itemize}
Note that this can be viewed as a no-hair theorem for inflation since any classical perturbations prior to inflation are erased while quantum fluctuations survive.

More generally, inflationary models radically change our vision of cosmology in at least three respects. These predictions are in full agreement with observations, and CMB in particular (see Fig.~\ref{fig_infl}).
\begin{itemize}
\item Predictions are by essence probabilist. One can only predict spectra, correlations so that only the statistical properties of the galaxy distribution can be inferred, not their exact position.
 \item{Particles are produced by the preheating mechanism}. The  inflaton decay allows us to explain the production of particles at
 the end of inflation. It may affect the standard predictions~\cite{modulated}.
 \item{Inflation is eternal}. Placing inflationary models in the context of chaotic initial conditions, we obtain a very different  picture of the universe. The latter should be in eternal inflation and gives rise to island universes.
\end{itemize}

\subsection{Extrapolation}

\subsubsection{Eternal inflation}

%%----------------------------------------------------
\begin{figure}[h!]
\centering
\begin{tabular}{ccc}
\includegraphics[width=.49\columnwidth]{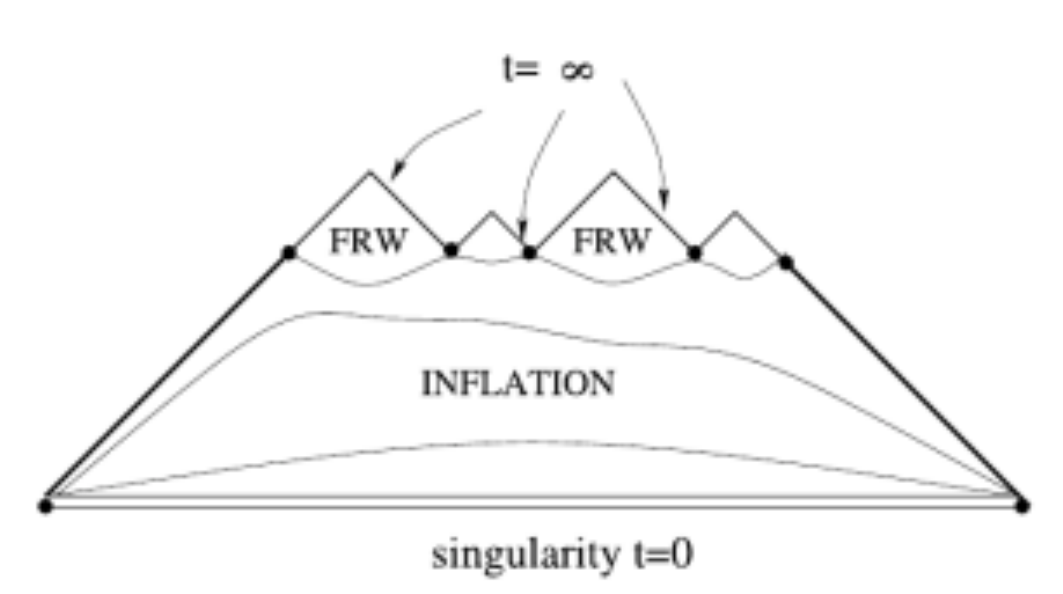}
\hskip1cm\includegraphics[width=.65\columnwidth]{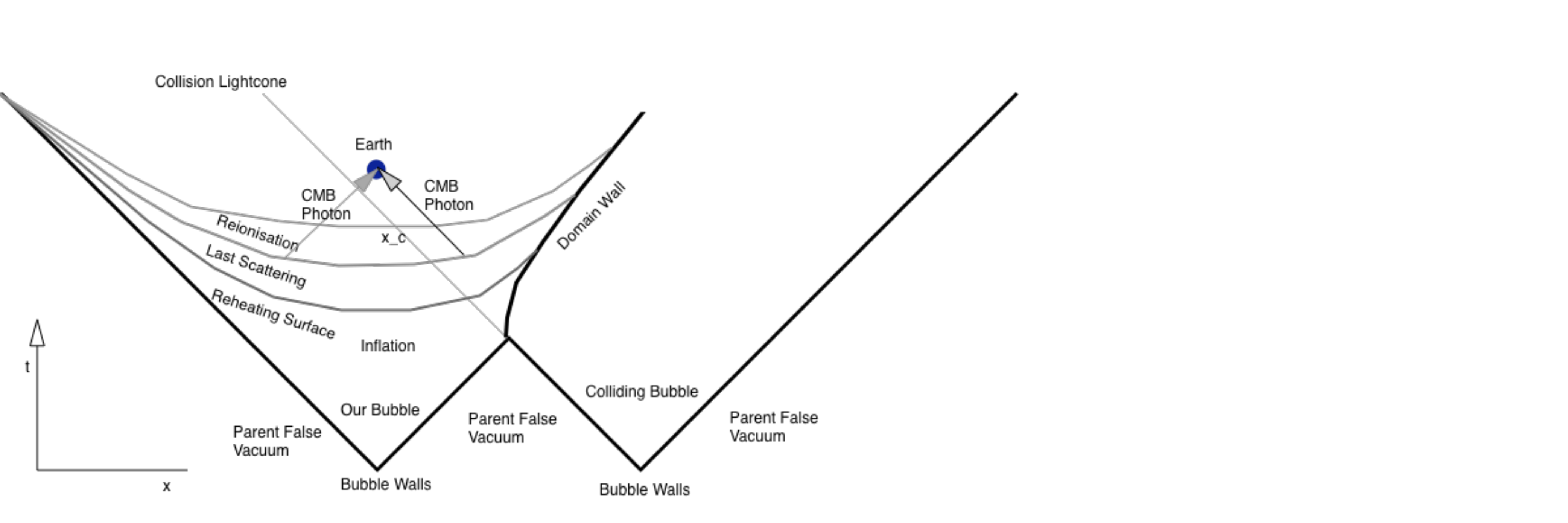}
\end{tabular}
 \caption{\footnotesize Penrose diagrams for a universe with eternal inflation (left) and with nucleation of bubble universes (right). Respectively from Ref.~\cite{Lev} and Ref.~\cite{kleban}.} 
\label{fig-penrose2}
\end{figure}
%%----------------------------------------------------

To finish, let us describe the process of eternal inflation that plays a central role in the discussion of the multiverse. The discovery of the self-reproduction process is a major development for inflation and cosmology. This mechanism was known for the models of old and new inflation and then extended to chaotic inflation~\cite{eternal}.  In inflationary models with large values of the inflaton, quantum fluctuations can locally increase the value of the inflaton. The expansion of these regions is then faster and their own quantum fluctuations generate new inflationary domains. This process naturally leads to a self-reproducing universe where there are always inflating regions. 

We just sketch a heuristic description here~\cite{guthreview}. Regions separated by distances greater than  $H^{-1}$ can be considered to be evolving independently. So any region of size $H^{-1}$ will be considered as an independent universe decoupled from other regions. Consider such a region of size $H^{-1}$ for which the scalar field is homogeneous enough and has a value $\varphi\gg  M_p$ and consider a massive field. In a time interval $\Delta t\sim H^{-1}$, the field classically decreases by $\Delta\varphi\sim\dot\varphi\Delta t\sim\dot\varphi/H$. The Klein-Gordon equation in the slow-roll regime then implies that $\Delta\varphi\sim - M_p^2/4\pi\varphi$. This value should be compared with the typical amplitude of quantum fluctuations, $|\delta\varphi|\sim H/2\pi\sim m\varphi/\sqrt{3\pi} M_p$ (see next section).

Classical and quantum fluctuations have the same amplitude for
\begin{equation}
 |\Delta\varphi|\sim|\delta\varphi|\Longleftrightarrow
 \varphi\sim\varphi_*\equiv
 \frac{ M_p}{2}\sqrt{\frac{ M_p}{m}}\ .
\end{equation}
We can thus distinguish between three phases in the evolution of the inflaton: a phase during which the quantum fluctuations are of the same order (or larger) as the classical field variation, a phase in which the field is in classical slow-roll towards its minimum and a phase when the field oscillates around its minimum. Note that $V(\varphi_*) =(m/ M_p) M_p^4/8\lesssim 10^{-6} M_p^4$ so that the
first regime can be reached even at energies small compared to $ M_p$.

When $\varphi\gg\varphi_*$, $\delta\varphi\gg\Delta\varphi$. The characteristic length of the fluctuations $\delta\varphi$ generated during the time $\Delta t$ is of the order of $H^{-1}$ so that the initial volume is divided into $(\exp{H\Delta t})^3\sim20$ independent volumes of radius $H^{-1}$. Statistically the value of the field in half of these regions is $\varphi+\Delta\varphi-\delta\varphi$ and $\varphi+\Delta\varphi+\delta\varphi$ in the other half. So the physical volume of the regions where the field has a value greater than $\varphi_*$ is ten times larger
 \begin{equation}
 V_{t+\Delta t}(\varphi>\varphi_*) \sim\frac{1}{2} \left(\exp{H\Delta t}\right)^3
 V_{t}(\varphi>\varphi_*)\sim 10 V_{t}(\varphi>\varphi_*)\ .
 \end{equation}
The physical volume where the space is inflating therefore grows exponentially in time. The zones where the field becomes lower than  $\varphi_*$ enter a slow-roll phase. So they become inflationary universes, decoupled from the rest of the universe, with a slow-roll phase, a reheating phase and a hot big-bang. These zones are {\it island universes} (or pocket universes) and our observable universe would be only a tiny part of such an island universe.

This scenario has important consequences for cosmology. In chaotic inflation, the universe has a very inhomogeneous structure on scales larger than $H^{-1}$ with regions undergoing eternal inflation continuously giving rise to new zones themselves undergoing inflation. On large scales, the universe has a fractal structure with continuous production of island universes. Each one of these island universes then undergoes a phase of ``classical'' inflation, with a large number of $e$-folds and is thus composed of many regions of the size of our observable universe. This simplistic model assumes only one scalar field and a potential with a unique minimum. Realistic models in high energy physics on the other hand involve many scalar fields. The potential of these fields can be very complex and have many flat directions and minima. So the same theory can have different vacua which correspond to different schemes of symmetry breaking. Each of these vacua can lead, at low energy, to physically different laws. Due to the exploration of this {\it landscape} by quantum fluctuations, the universe would find itself divided into many regions with different low-energy physical laws (for instance different values of the fundamental constants). If this vision of the primordial universe is correct, physics alone cannot provide a complete explanation for all the properties of our observable universe since the same physical theory can generate vast regions with very different low energy properties. So our observable universe would have the properties it has not because the other possibilities are impossible or improbable, but simply because a universe with such properties allows for a life similar to ours to appear.

Eternal inflation thus offers a framework to apply the anthropic principle since the self-reproduction mechanism makes it possible to generate universes with different properties and to explore all possible vacua  of a theory. This approach is used more and more to address the question of the value of the fundamental constants and to address of the cosmological constant problem. This framework also allows us to address questions beyond the origin of the properties of our universe thus defining the limitation of what we will be capable of explaining.

The Penrose diagram in eternal inflation is depicted on Fig.~\ref{fig-penrose2} and compared to another scenario based on bubble nucleation. In the case the ${\cal I}^+$ hypersurface has a fractal structure. As observers, we live in one of the late time asymptotic FL pockets.

%%%%%%%%%%%%%%%%%%%%%%%%%%%%%%%%%%%%%
\section{Conclusions}\label{section5}

Our current cosmological model can be considered as a simple and efficient model to explain the dynamics of the universe and of the structures it contains. It relies on 4 main hypothesis on the laws of nature and the symmetries of the geometry of the cosmological solutions. The Copernican principle thus plays a central role in this construction. The model is structured in different layers, that have historically been developed one after the others: (1) cosmological solution of general relativity, (2) the hot big-bang model, (3) the description of the large scale structure and (4) a scenario of the primordial universe and of the origin of the large scale structures. At each stage, it has been connected to some observations.

The main paradigm lies on the dynamics of a universe that cools while expanding, the quantum fluctuations of a scalar field in the early universe, gravitational instability and then a progressive structuration that depends on the value of the cosmological parameters, the nature of the dark matter and the dissipative processes that affect baryons. Among the main successes of the model, we can recall: the origin of the variety of atoms (during BBN), the origin of matter (during reheating), the origin of the large scale structure (during inflation). This description is robust to our knowledge of the laws of physics since the energies involved since 1~s after the big-bang are smaller than 100~MeV, that is energies well-tested in the laboratory. It relies on nothing speculative and is an almost direct consequence of general relativity and known laboratory physics. We can state that the main uncertainty lies in the cosmological hypotheses. Only at earlier times  are there doubts about the extrapolations used. This robustness is balanced by all the limitations we have discussed (one observable universe, observed from one point, the existence of horizons, the need for abduction, etc...)

The matter contained in the model includes photons (0.01\%), baryons and electrons (5\%), neutrinos (0.1-2\%), non-baryonic dark matter (25\%), a cosmological constant (or dark energy) (70\%). This means that the minimal number of cosmological parameters boils down to 6: $\Omega_{\rm baryon}$, $\Omega_{\gamma}$, $\Omega_{\nu}$, $\Omega_{\rm cdm}$, $\Omega_{\Lambda}$ (which reduces to 3 since $\Omega_{\gamma}$, $\Omega_{\nu}$ are determined and the sum of all $\Omega$s is 1), the Hubble constant $H_0$, the amplitude $A_S$ and spectral index $n_S$ of the initial spectrum and the reionisation parameter $\tau_e$. It can be trivially extended to include a spatial curvature ($\Omega_K$), a dynamical dark energy component ($w,...$) and extra-neutrinos ($N_\nu$). These 6 parameters are today well measured and all the extra-parameters are well constrained.

The model nevertheless suffers from a series of problems and questions.
\begin{itemize}
\item {\em The lithium problem}, discussed in \S~\ref{secbbn}.
\item {\em The understanding of the form in which baryons are.} The baryonic density is well determined from CMB analysis but the baryon budget at low redshift is more difficult to establish: intergalactic gas (cold or hot), stars, planets...
\item {\em The understanding of the end of the dark age}: formation of the first stars, origin of reionization, history of galaxies. Do we see all galaxies (in particular are there low brightness galaxies) and to which extent is the distribution of the galaxies a good tracer of the distribution of matter in the universe? What is the role of black holes in the formation of galaxies?...
\item {\em Validity and precision of the perturbation theory}. See \S~\ref{secpert}.
\item {\em The understanding of the nature and properties of dark matter}. This component of matter is not included in the model of particle physics as we know it but there exist many candidates in various of its extension (such as the lightest supersymmetric particle). Cosmology hence provides a strong indication for the need of an extension of the standard model of particle physics. It can be optimistically thought that the conjonction of astrophysical observations, direct searches and accelerator experiments shall characterize this particle within the coming ten years. An alternative to dark matter is the formulation of modified theories for gravity, among which MOND, but this route is not favored.. 
\item {\em The cosmological constant problem or the understanding of the nature of dark energy}. This component is required to explain the late time acceleration of the universe. Dark energy confronts us with a compatibility problem since, in order to ``save the phenomena'' of the observations, we have to include new ingredients (constant, matter fields or interactions) beyond those of our established physical theories~\cite{jpu-general relativityG}. The important conclusion here is that we either have to accept a cosmological constant, as favored by all observations, or introduce new degrees of freedom in the model, either as geometric degrees of freedom of the cosmological solution, or as fundamental degrees of freedom (i.e. new physical fields), that can act as a new matter component or as a modification of general relativity if they are responsible of a long-range interaction due to their coupling with standard matter. See e.g. Ref.~\cite{jain} for the observational prospective on the understanding of dark energy.
\item {\em The microphysics behind inflation}. For example (but far fom being limitative), To which extent can we reconstruct the potential of the inflaton from cosmological observations? How is the inflaton connected to the standard model of particle physics (can it be the Higgs~\cite{uee}, is single field inflation stable to the existence of other heavier fields~\cite{seb})? Can one construct models without scalar field? How robust are the predictions and can one construct radically different models (as the now-ruled out opponent theory of topological defects)? Can one relate inflation to supersymmetry or string theory in a satisfying way? What happens to other fields? Should we be worried by trans-Planckian modes? How robust is the hypothesis of a Bunch-Davies vacuum~\cite{usb2}? Can one detect features (non-Gaussianity, correlations) of the initial conditions that would reveal the physics at work? Can one constrain the number of $e$-folds? Can one detect the primordial gravity waves? What is the structure of spacetime in eternal inflation?... 
\item {\em Connecting inflation to the hot big-bang}. The physics of reheating needs to be better investigated, in particular to include the standard model of particle physics and the transition from the quantum perturbation to classical stochastic perturbations also needs to be better understood.
\item {\em The backreaction question} discussed in \S~\ref{sec-back}.
\item {\em The effect of small scale structure} on the observation of the universe and to which extent it biases our observations; see \S~\ref{sectt}.
\item{\em Emergence of complexity} The fact that we can understand the universe and its laws has also a strong implication on the structure of the physical theories. At each step in our construction of physical theories, we have been dealing with phenomena below a typical energy scale, for technological constraints, and it turned out (experimentally) that we have always been able to design a consistent theory valid in such a restricted regime. This is not expected in general and is deeply rooted in the mathematical structure of the theories that describe nature.  We can call such a property  a {\it scale decoupling principle} and it refers to the fact that there exist energy scales below which effective theories are sufficient to understand a set of physical phenomena that can be observed. {\it Effective theories} are the most fundamental concepts in the scientific approach to the understanding of nature and they always come with a domain of validity inside which they are efficient to describe all related phenomena. They are a successful explanation at a given level of complexity based on concepts of that particular levels. This implies that the structure of the theories are such that there is a kind of stability and independence of higher levels with respect to more fundamental ones. This lead to the idea of a hierarchy of physical theories of increasing complexity~\cite{ellisc}, which affects our understanding of causality to take into account contextuality, selection effects (in the Darwinian sense), emergence etc.
\item {\em Fine tuning.} A related issues lies in the observational fact that we (as a complex biological system) are here observing the universe. It means that the fundamental laws of physics, which set the space of possibilities of higher complexity theories (such as chemistry, biology etc.) need to allow for the existence of complexity. The study of the variation of the fundamental constants have taught us that small variation in the fine structure constant or other parameters can forbid the emergence of such a complexity, one of the more critical parameter being the cosmological constant. How is our universe fine-tuned (and according to which measure)? is an essential question. It can be answered for some physical system, e.g. the production of carbon-12 in population~III stars require a tuning at the level of $10^{-3}$~\cite{c12}. This line of arguments is an essential motivation to promote the idea of a multiverse.  It still requires a better understanding.
\item{\em Primordial singularity.} The cosmological model exhibits a spacelike singularity when extrapolated in the past. This big-bang is not a physical event but a limit in a given theoretical model. It is important to understand how this conclusion may be changed by quantum theories of gravity. It also appears as an explicit limit of our model that motivates many questions on the origin of the universe, that goes beyond the purpose of the model itself.
\item{\em Links with quantum theories of gravity}. In the early universe quantum gravity may be important to understand the dynamics of the universe. A large activity focuses on the development of a realistic phenomenology of these theories, and in particular string theory. 
\end{itemize}
Whatever the problem or the discrepancy in the model, one can always look for 3 kinds of solutions: astrophysical (i.e. an astrophysical effect that would affect and alter the observation), cosmological (i.e. the fact that the cosmological hypotheses are two strong and need to be relaxed, hence modifying our interpretation of the data) or physical (i.e. a need to modify the fundamental laws of physics).

An important step of the past decade has been the development of many tests of the hypothesis of the model, to be contrasted with the mainstream quest for a better precision of the measurement of the cosmological parameters. This includes tests of the field equations of general relativity on large scale~\cite{jpu_cup,jpu-general relativityG,jpu-testRGcosmo,ub2001}, tests of the equivalence principle by the constancy of fundamental constants~\cite{uzancst1,uzancst2} (but also tests of some well-defined extensions of general relativity such as the scalar-tensor theories~\cite{dampich,msu}), tests of the Copernican principle (both homogeneity~\cite{testCP} and isotropy~\cite{usb2}), tests of the distance duality relation~\cite{jpu-duality}, and constraints on the topology of the universe~\cite{topogen,topocmb,topolimite}. No deviations from the standard model, despite the need of a dark sector, have been detected so far.\\

The field of cosmology can be seen as following two routes, mostly independent. 

Observational cosmology has been booming, with the development of new techniques and of large surveys in all possible wavelengths. This has led to the idea of {\em precision cosmology}. One has to stress that measuring with higher accuracy does not mean that we get a better understanding of our model, and in particular one needs to be aware that the definition of the cosmological parameters depends on the cosmological model. One also needs to control the validity of the model and of its predictions with a similar accuracy. As discussed above, the precision of the observations may lead to question the validity of the use of a FL spacetime on all scales~\cite{fleury2}. Again, this justify the need to test the hypothesis of the model. A trend has been to introduce arbitrary (or badly motivated) extensions of the model (and thus some extra-cosmological parameters) and to constrain them, backed up by the development of computational techniques. While it gives an idea of the validity of the minimal model, it does not teach much on physics. Note also that the interpretation of data also requires some deep investigation on the understanding of the non-linear regime, in particular by including all relativistic effects, and on the relation between perturbation theory and N-body simulations, which are essentially Newtonian. The precision on some parameters (such as the curvature parameter, the spectral index or its running) are indeed important while a high precision on the value of the cosmological constant does not improve our understanding of the physics behind.

Primordial cosmology tends to obtain a better description of the early phases of our the universe. It deals with many extension of the laws of nature as we know them. It is the playground of phenomenology for theories of quantum gravity and of many speculations, often out of reach of any observational testability, such as the multiverse. The important question is the ``realism" of this phenomenology, which is often based on simplified version of the underlying theories. The guide here is mostly the search for a consistent picture and this is where one needs to be careful with the distinction between cosmology and Cosmology.\\

One century after the formulation of general relativity and ninety-eight years after the first relativistic cosmological model, cosmology has made significant progresses and managed to connect theory with a blooming observational activity. It offers new views of our universe. 

Today, we can state that there is no need, from an observational and phenomenological point of view, to doubt general relativity (including a cosmological constant). Indeed the main problems here are theoretical (cosmological constant problem and the need for quantum gravity at high energy) but we may question that there exist some intermediate scale at which general relativity needs to be modified in a way that it would imprint phenomena in our observable universe. The trend to introduce often badly motivated extension of general relativity (which in particular do not solve the issue of the cosmological constant or of the quantum aspects) has mostly been induced by the need of a phenomenology for comparing to observations. This has indeed taught us the effects of many of these extensions, but they have to be considered only as toy models.

The next step in our understanding of gravity and of our universe lies probably in our ability to detect gravitational waves, which will open a new observational window on our universe, and inevitably lead to a new picture.

\section*{Acknowledgements}
%%%%%%%%%%%%%%%%%%%%%%%%%%%%%%%%%%%%%%%%%%%%%%%%%%%%%%%%%%%%%%%%

This work made in the ILP LABEX (under reference ANR-10-LABX-63) was supported by French state funds managed by the ANR within the Investissements d'Avenir programme under reference ANR-11-IDEX-0004-02. I thank George Ellis, Pierre Fleury and Cyril Pitrou for their comments.

%%%%%%%%%%%%%%%%%%%%%%%%%%%%%%%%%%%%%

\end{document}